\documentclass[aps,prb,twocolumn,superscriptaddress,floatfix,morefloats]{revtex4-1}
\usepackage{longtable}
\usepackage{subfigure}
\usepackage{stmaryrd}
\usepackage{amsmath,amssymb,amsfonts,bm}
\usepackage{graphicx}
\usepackage{ulem}
\usepackage{color}
\usepackage{hyperref}

\def\KeyWord#1{$\backslash$\IfColor{$\!\!$\textRed{#1}\textBlack}{#1}$\!\!$}

\newcommand{\maC}{\mathcal{C}}
\newcommand{\proj}{\mathcal{P} }

\newcommand{\be}{\begin{equation} }
\newcommand{\ee}{\end{equation} }
\newcommand{\ba}{\begin{eqnarray} }
\newcommand{\ea}{\end{eqnarray} }
\newcommand{\n}{\nonumber \\ }

\newcommand{\db}[1]{\left( #1\right)}
\newcommand{\mac}{\mathcal}

\newcommand{\bit}{\begin{itemize}}
\newcommand{\eit}{\end{itemize}}

\newcommand{\eqnref}[1]{Eq.~\eqref{#1}}
\newcommand{\figref}[1]{Fig.~\ref{#1}}
\newcommand{\tabref}[1]{Table~\ref{#1}}
\newcommand{\secref}[1]{Sec.~\ref{#1}}
\newcommand{\appref}[1]{Appendix~\ref{#1}}

\newcommand{\ket}[1]{\mid#1\rangle}

\newcommand{\tr}{\operatorname{Tr}}
\renewcommand{\vec}[1]{\mathbf{#1}}

\newcommand{\punc}[1]{\,#1}

\newcommand{\expecval}[1]{\langle#1\rangle}
\newcommand{\lev}{\epsilon^{\mu \nu \rho \sigma}}

\newcommand{\wt}[1]{\widetilde{#1}}

\def\note#1{$\color{red}{\bullet}$}
\def\okay#1{$\color{green}{\bullet}$}

\begin{document}

\title{Three-dimensional topological lattice models with surface anyons}

\author{C.W.~von Keyserlingk}
\author{F.J.~Burnell}
\author{S.H.~Simon}
\affiliation{Rudolf Peierls Centre for Theoretical Physics, 1 Keble Road, Oxford,
OX1 3NP, United Kingdom}
\date{\today}

\begin{abstract}
We study  a class of three dimensional exactly solvable models of topological matter first put forward by Walker and Wang [arXiv:1104.2632v2]. While these are not models of interacting fermions, they may well capture the topological behavior of some strongly correlated systems. In this work we give a full pedagogical treatment of a special simple case of these models, which we call the 3D semion model: We calculate its ground state degeneracies for a variety of boundary conditions, and classify its low-lying excitations. While point defects in the bulk are confined in pairs connected by energetic strings, the surface excitations are more interesting: the model has deconfined point defects pinned to the boundary of the lattice, and these exhibit semionic braiding statistics. The surface physics is reminiscent of a $\nu=1/2$ bosonic fractional quantum Hall effect in its topological limit, and these considerations help motivate an effective field theoretic description for the lattice models as variants of $bF$ theories. Our special example of the 3D semion model captures much of the behavior of more general `confined Walker-Wang models'. We contrast the 3D semion model with the closely related 3D version of the toric code (a lattice gauge theory) which has deconfined point excitations in the bulk and we discuss how more general models may have some confined and some deconfined excitations.  Having seen that there exist lattice models whose surfaces have the same topological order as a bosonic fractional quantum Hall effect on a confining bulk, we construct a lattice model whose surface has similar topological order to a fermionic quantum hall effect. We find that in these models a fermion is always deconfined in the three dimensional bulk.
\end{abstract}


\maketitle

\include{epsfx}

\section{Introduction}

It has been known for decades that two is a privileged dimension for point particles. In greater than two dimensions, point particles are restricted to be either bosons or fermions; but for $D=2$, a vast number of other {\it anyonic} statistics are possible -- some of which are believed to be realized by fractional quantum Hall states\cite{Arovas84}. The study of the 2D {\it topological phases} of matter that possess such anyonic excitations has grown into a substantial enterprise in condensed matter physics, spurred on by the ultimate goal of creating topologically protected qubits\cite{TQCrefs}.

Our substantial understanding of topological phases in two dimensions naturally leads to the question of whether any of their defining features can also be realized in  three dimensional systems. Though statistics between point particles in $D=3$ must be fermionic or bosonic, interesting statistical interactions can still exist between vortex defects, between point particles and vortex defects (as in discrete gauge theories), or between point-like defects with additional structure\cite{TeoKane,MicrosoftProjectiveStatistics}.
Thus one class of 3D topological phases contains particles that have unconventional statistics, as they are not strictly point-like.
A second category of 3D phases that share some of the characteristics of 2D topological phases are fractional topological insulators\cite{Levin11,Maciejko10,Swingle11}. When time reversal symmetry is present, the surfaces of these systems have protected gapless excitations. If time reversal is broken, however, the surfaces will exhibit a fractional Hall effect. These gapped surfaces might thus be expected to support chiral anyonic excitations.

In the present work, we focus on a third class of 3D topological phases, which can be realized by a family of Hamiltonians introduced by Walker and Wang\cite{Walker11}.  Specifically, we discuss phases realized by a subset of these models, which we will call confined Walker-Wang models. These are reminiscent of fractional topological insulators in that the 2D surfaces of these 3D systems display the physics of a chiral anyon model (whose topological properties are those of a fractional quantum Hall state).  
 They differ, however, in that a fractional topological insulator must preserve time-reversal symmetry in the bulk-- which requires a topological ground-state degeneracy in periodic boundary conditions\cite{Swingle11}. The confined Walker-Wang models have non-degenerate ground states on a system without boundaries, and explicitly broken time reversal symmetry in the bulk. This breaking of time-reversal ensures that unlike fractional topological insulators\cite{Levin11}, our models do not have protected gapless boundary modes. Rather, they describe a system resembling a confined phase of the fractional topological insulator, in which time-reversal invariance has been lost, and the surface modes have been gapped such that they are always found in their fractional Hall state.

The exactly solvable Walker-Wang Hamiltonians we study here are 3D cousins of the 2D string-net models introduced by Levin and Wen\cite{Levin05}, and operate on the same underlying principles. These families of models are not intended to describe any known physical system; rather, they are interesting because they can be solved exactly, and capture the long-wavelength physics of certain topological states of matter. They thus provide a framework in which to study the physical properties of these systems, including their ground state degeneracy and the properties of their low-lying excitations.  

Our key findings are as follows. As mentioned above, and unlike both discrete gauge theories and true fractional topological insulators, the confined Walker-Wang models that we consider have a unique ground state  on any {\it closed} 3D lattice (i.e. a lattice that has no boundary). Thus in the absence of boundaries, the models are not topologically ordered in the usual sense. The possible excitations in the bulk are line defects with an energy cost per unit length: in other words, the bulk admits no deconfined point particles.
Unlike vortex lines in 3D systems, which must form closed loops, these line defects can end at points in the bulk.
The name ``confined" Walker-Wang model stems from the fact that these point particles are confined --- that is they must sit at the end-points of line defects  that have a fixed energy cost per unit length (analogous to flux-tubes in a confining gauge theory).
If the lattice does have a boundary, however, the system has the more conventional hallmarks of topological order: with appropriate boundary conditions, the surfaces admit deconfined anyonic excitations, and the ground state is degenerate.
Though much of our work focuses on the simplest (abelian) example of the family of confined Walker-Wang models, we will show that these properties also apply to models that have deconfined non-abelian anyons at their surfaces.

Readers well-versed in the quantum Hall effect should note that
surface states of the confined Walker-Wang models always have the topological properties of a bosonic quantum Hall state.  However, Walker-Wang models with fermionic surface states  also exist; in addition to the surface topological order, these have a deconfined fermion and $\mac{Z}_2$ topological order in the bulk, which we will also discuss in some detail.

Unlike topological insulators, whose single surface Dirac cone cannot be realized in any 2D model with the same symmetries, the  surface states of confined Walker-Wang models are no different in their long-wavelength properties from phases that can be realized by purely $2$D systems.  Indeed, by adding an appropriate 2D layer to the boundary of a confined Walker-Wang model, it is possible to destroy the topological order resulting from these surface states.
By this metric, they do not represent genuine $3$D topological phases of matter.

 The appearance of a chiral anyon model at the surface of an exactly solvable $3$D Hamiltonian is nonetheless striking.
 Simulating a general  chiral Chern-Simons theory {\it locally} on a lattice is long-outstanding challenge\cite{}.
That is, though a chiral Chern-Simons theory can emerge as the effective description of a purely $2$D system, there is no ``fixed-point'' lattice model (in the sense of Ref. \onlinecite{Levin05}) describing the limit in which the correlation length is less than the lattice spacing.   Indeed, a chiral Chern-Simons theory has topologically protected gapless boundary modes, meaning that when the $2$D system has a boundary, the correlation length at the boundary must be infinite. Thus it is believed\cite{WenPersonal} that such fixed-point Hamiltonians do not exist for chiral 2D Chern-Simons theories (or, more generally, chiral $2$D anyon models).
Confined Walker-Wang models, whose Hamiltonians are exactly solvable, realize chiral anyon models as surface states of fixed-point Hamiltonians (whose correlation length is $0$). The possibility of constructing a theory in terms of the boundary of a higher dimensional model is analogous to the domain-wall solution of the fermion doubling problem\cite{Kaplan92}.

The outline of the remainder of this work is as follows.
The bulk of the paper will focus on the simplest topological models -- those that can be described as loop gases.
 In \secref{s:2DLM} we review the 2D loop gases: the toric code\cite{Kitaev03} and the 2D doubled semion (DSem) model in the form introduced by Levin and Wen\cite{Levin05}.  (More expert readers may wish to skip this introductory section).
 The generalization of these quantum loop gas models to three dimensions is the content of much of this paper.   
 In \secref{WW} we present the 3D generalizations of these loop gas models: the $3$D Toric code\cite{Hamma05} (essentially a 3D $\mathbb{Z}_2$ lattice gauge theory) and a 3D analogue of the semion model (3DSem), which is our paradigm for confined Walker-Wang models. After describing these Hamiltonians, we discuss their low-energy physics for periodic boundary conditions (i.e., if the system is on a three-torus, $\mathbb{T}^3$), showing that whereas the toric code has deconfined point particles and a degenerate ground state, the 3DSem model has a unique ground state and no deconfined point-like excitations, just like a confined fractional topological insulator.

In \secref{sec:DsemD^2xS^1} we study the low-energy physics of these models with open boundaries. We find that for 3DSem, boundaries introduce new ground state sectors, and that deconfined anyonic excitations can exist on the boundary. \secref{s:DSemFT} discusses the field theory that captures the properties of the ground states and (confined) excitations of 3DSem.

A large part of this paper focuses on the 3D quantum loop gases (the 3D Toric code, and the 3DSem model). However, these constitute only a small subset of the possible Walker-Wang models. In \secref{s:MTCGS} we discuss how our results extend to the more general case. We show that 3DSem is one member of a large family of confined Walker-Wang models for which all point defects are confined in the bulk, and the ground state degeneracy is completely determined by the topology of the boundary of the system. Most generally, the surface excitations of confined Walker-Wang models may be non-abelian anyons. We discuss the effective field theory for these lattice models in \secref{ss:bFbbMTC}.  In \secref{QHSect}, we study one family of the even more general case: Walker-Wang models that describe fermionic systems, whose surface states have the topological order of fermionic quantum Hall systems and whose bulk has a deconfined fermionic point excitation.

A number of technical aspects of the models are addressed in the appendices.  Some readers will find \appref{ExplicitDSem},  in which we give exact expressions for the Hamiltonian and quasi-particle creation operators in the doubled semion model in the presence of arbitrary defects, of particular interest.

Since the Hamiltonians we discuss were introduced by Walker and Wang\cite{Walker11}, it is worth emphasizing what is new in the present work. Our main contribution is to treat in detail the excitations of these Hamiltonians, particularly on their two dimensional surfaces. This allows us to describe the full spectrum of ``topological" features of the systems in question, and make more explicit the connection to field theories suggested in Ref.~\onlinecite{Walker11}.  We also discuss the possible boundary conditions of these models, and under what circumstances the topological order associated with the surface can be destroyed.

\section{2D Lattice models}\label{s:2DLM}
While 3D lattice models are the main focus of this paper, it is useful to first review the topological order found in their 2D analogues --- starting with some very simple examples. Perhaps the simplest exactly solvable topological model is the 2D toric code: a lattice model where each bond takes one of two possible states (which we can call spin-down and spin-up). It turns out that a very slight modification of the 2D toric code Hamiltonian\cite{Kitaev03}  gives a different 2D topological phase of matter known as the doubled semion (DSem) model\cite{Shtengel04,Levin05}. Both the toric code and the doubled semion model can be thought of as quantum loop gases in that the bonds with down spins must form closed loops. While in this aspect the two models are quite similar, the resulting low energy theories are quite different: The toric code is the simplest example of a discrete gauge theory, whereas the doubled semion model is the simplest example of a doubled Chern-Simons theory\cite{Shtengel04} --- that is, a theory whose low energy spectrum is two identical anyon models with opposite chiralities.  (In the doubled semion model, the anyon theory contains one species of anyon, which has ``semionic" statistics in which exchanging two particles induces a complex phase $\pm i$ to the wave function, whence the name).

In \secref{ss:Tcintro} we introduce the well-known toric code\cite{Kitaev03,Levin05}, and then calculate its ground state degeneracies and classify its low-lying excitations; experts may wish to skip this section. Then in \secref{ss:DSemintro} we do the same for the doubled semion model\cite{Levin05}, paying particular attention to the types of defects that arise and how they differ from those found in the toric code. The notation and techniques introduced in this section provide a firm foundation for our study of the 3D models in \secref{WW}.
\subsection{ The toric code}\label{ss:Tcintro}

\begin{figure}
 \includegraphics[width=1\linewidth]{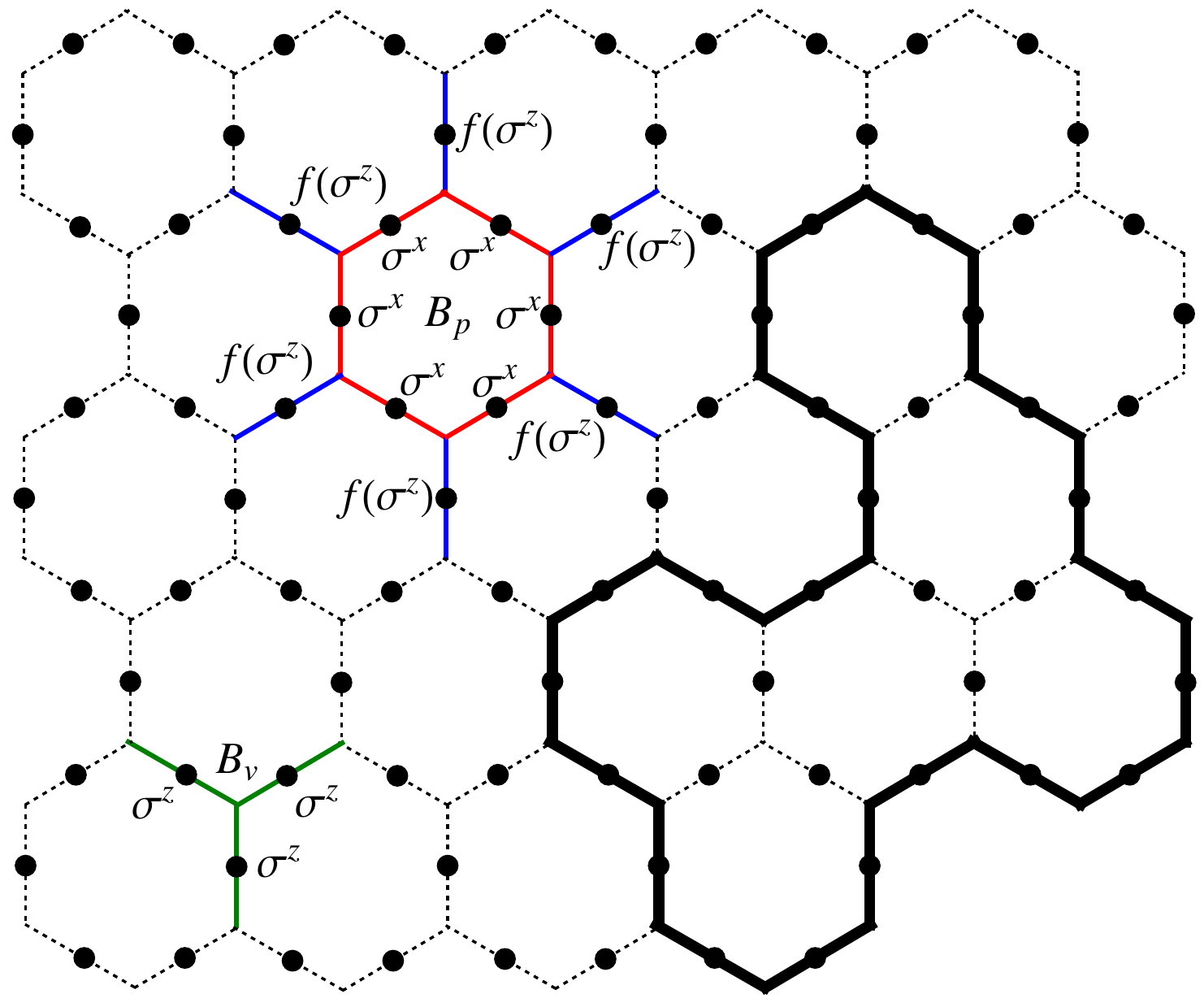}
 \caption{(Color online) This figure shows a region of the lattice for the 2D models, where we have represented the spin degree of freedom on each edge with a black dot. Bold black edges represent $\sigma^{z}=-1$ spin configurations, while dashed edges represent $\sigma^{z}=+1$. The figure also indicates the edges involved in the definition of $B_v$; the three green edges in $s(v)$ are acted on with $\sigma^z$ matrices as shown. The edges involved in $B_p$ differ between the toric code and the DSem model. In the toric code $B_p$ acts on the six red edges in $\partial p$ with $\sigma^x$ matrices as shown, and does not act on the six blue edges in $s(p)$ i.e. $f(\sigma^z)=1$. In the DSem model, in addition to acting on the red edges with $\sigma^x$, the plaquette operator also acts on the six blue edges in $s(p)$ with $f\db{\sigma^z} = i^{(1 - \sigma^z)/2}$. }
 \label{toricFig1}
\end{figure}

The Hilbert space of the toric code consists of a two-state system $\sigma^z = \pm 1$ on each edge of a honeycomb lattice as shown in \figref{toricFig1}. The Hamiltonian is given by
\be \label{HTC}
H =  - \sum_v \underbrace{ \prod_{i\in s(v)} \sigma^z_i }_{B_v} - \sum_p \underbrace{\prod_{i\in \partial p} \sigma^x_i }_{B_p} \punc{,}
\ee
where $\sum_v$ runs over all vertices, and $\sum_p$ over all plaquettes, $s(v)$ is the set of three edges attached to $v$ and $\partial p$ is the set of six edges bounding plaquette $p$ (see \figref{toricFig1}).
The vertex operator $B_v$ takes value $\pm1$ depending on whether there are an even/odd number of down-spins on the edges coming into vertex $v$. The plaquette operator $B_p$ flips the spins on every edge of a plaquette $p$ and commutes with $B_v$ because it flips a {\it pair} of spins at each vertex $v$. Since clearly $\left[ B_v,B_{v'}\right]=\left[ B_p,B_{p'}\right]=0$, the Hamiltonian is a sum of commuting operators, allowing us to write down its ground states exactly.

 \begin{figure}
 \includegraphics[width=1\linewidth]{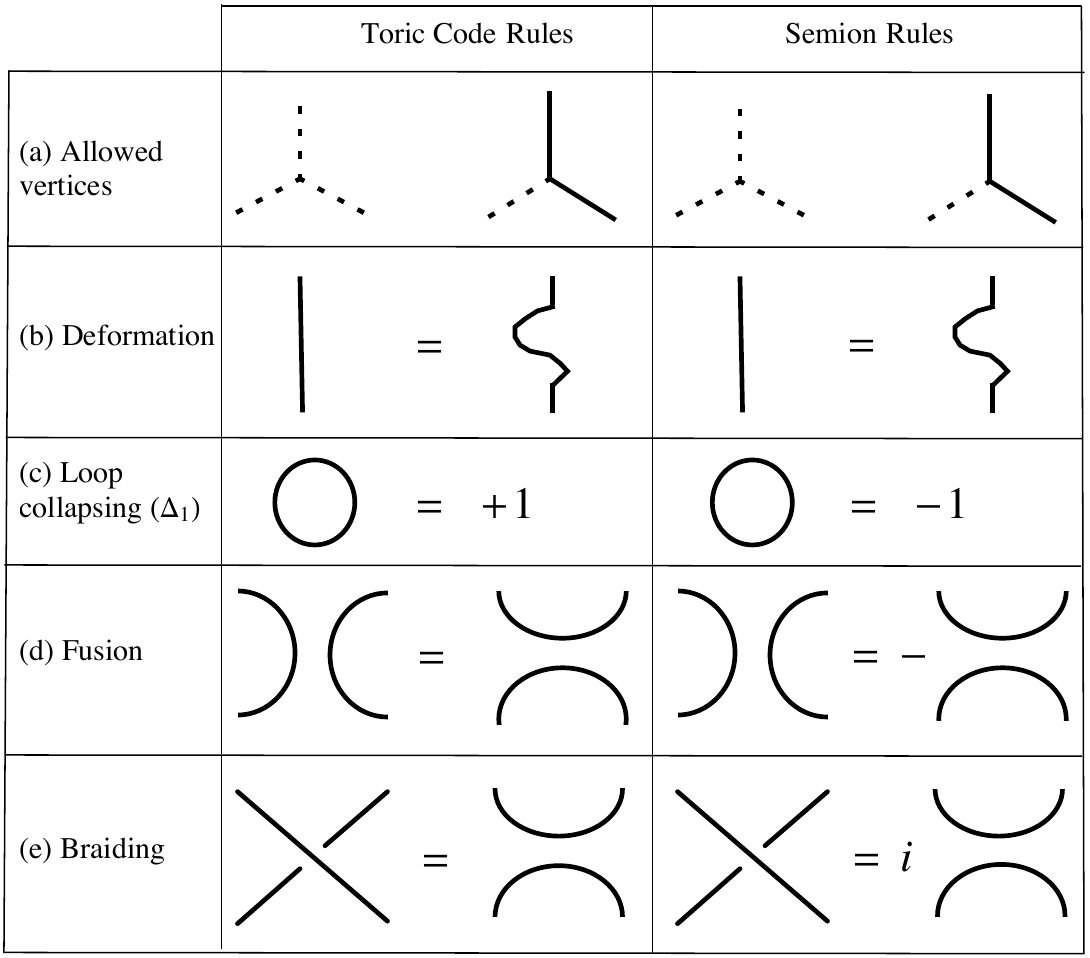}
 \caption{This figure shows the graphical rules for the toric code and DSem models. Row (a) represents the fact that the ground state only involves vertices with $B_v=1$. The diagrams in (b)-(d) serve two purposes. Firstly, they tell us the relative amplitudes of loop gas configurations in the ground state e.g. row (c) tells us that configurations related by removing a closed loop occur with the same amplitude in the toric code ground state, but with a relative minus sign in the DSem ground state. Second, these diagrams provide a neat graphical mnemonic for the definitions of string operators.}
\label{tcdsem}
 \end{figure}

 \subsubsection{Ground States}\label{tcGSdisc}
\begin{figure}
  \includegraphics[width=.9\linewidth]{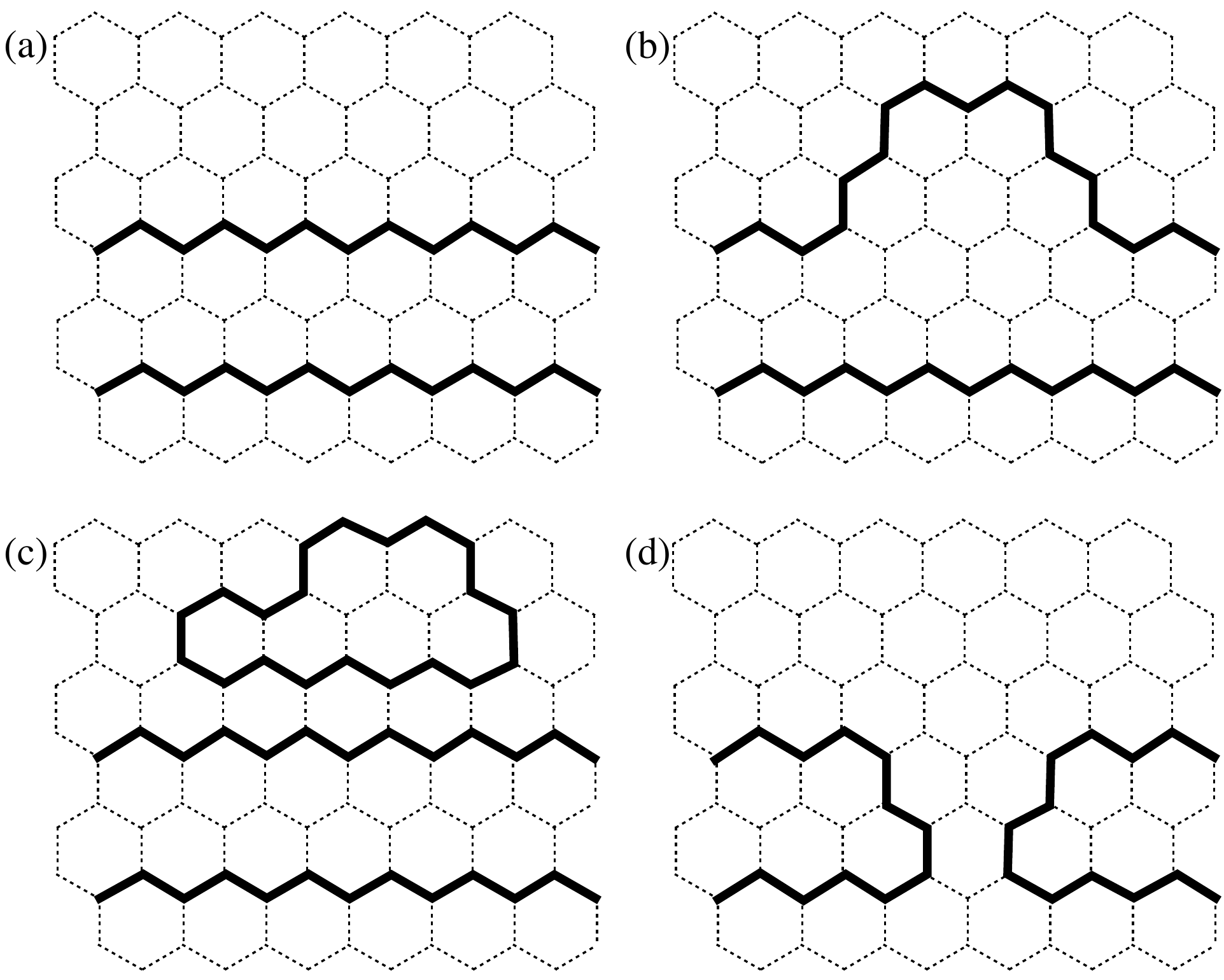}
 \caption{This diagram shows a specific small region of the lattice for four distinct spin configurations. The configurations are taken to be the same everywhere outside this region. For the toric code, all of these configurations occur in the ground state wave function with the same amplitude. This is because (b), (c), and (d) can all be made to look like (a) using loop deformation, loop collapsing and fusion respectively. In the DSem model, (a) and (b) occur with the same phase because they are related by deforming a loop. (a) and (c) occur with a relative minus sign because they differ by the presence of a single closed loop. (a) and (d) also occur with a relative minus sign because they are related by a single fusion (\figref{tcdsem}(d)).}
\label{gskets}
 \end{figure}

Using the fact that all $B_v$ and $B_p$ have eigenvalues $\pm 1$ we see that a lower bound on the ground state energy is obtained when

\be\label{GSconstraints}
\prod_{i \in s\db{v}} \sigma^z_i = 1 \text{ for all } v \ \ \ \ \ \ \prod_{i \in \partial p} \sigma^x_i=1 \text{ for all } p\punc{.}
\ee
Any state which satisfies all of these conditions is automatically a ground state because it saturates a lower bound on the Hamiltonian. Remarkably such states exist.
Henceforth we work with a basis of states for which the spins on the edges of the lattice are $\sigma^z$ eigenstates, and we choose a `loop gas' representation of these states as diagrams with only down-spin edges colored. The first condition in \eqnref{GSconstraints} ensures that there are an even number of down-spins on the edges attached directly to any vertex, so the ground state must be made of a superposition of spin configurations which all involve only non-intersecting closed loops. The second condition in \eqnref{GSconstraints} tells us that any two spin configurations that are related to each other by flipping all the edges of a plaquette appear with the same coefficient in the ground state. This has three important consequences: Given two spin configurations obeying the first constraint in \eqnref{GSconstraints}, if the loop gas pictures of these spin configurations can be related by deforming loops over plaquettes (as shown in Figure \ref{tcdsem}b), by removing contractible loops (as shown in Figure \ref{tcdsem}c), or by fusing loops together (as shown in Figure \ref{tcdsem}d), then the spin configurations appear in the ground state with the same amplitude. This is because the actions of deforming loops, removing contractible loops and fusion can all be implemented by applying a sequence of $B_p$ operators, and $B_p=+1$ on the ground state. As a concrete example, if four configurations differ only in a small region in the manner shown in \figref{gskets}(a)-(d) then they all appear in the ground state with the same amplitude. Loop deformation relates \figref{gskets}(a) and (b), loop collapsing relates \figref{gskets}(a) and (c) and loop fusion relates \figref{gskets}(a) and (d).

\begin{figure}
 \includegraphics[width=0.95\linewidth]{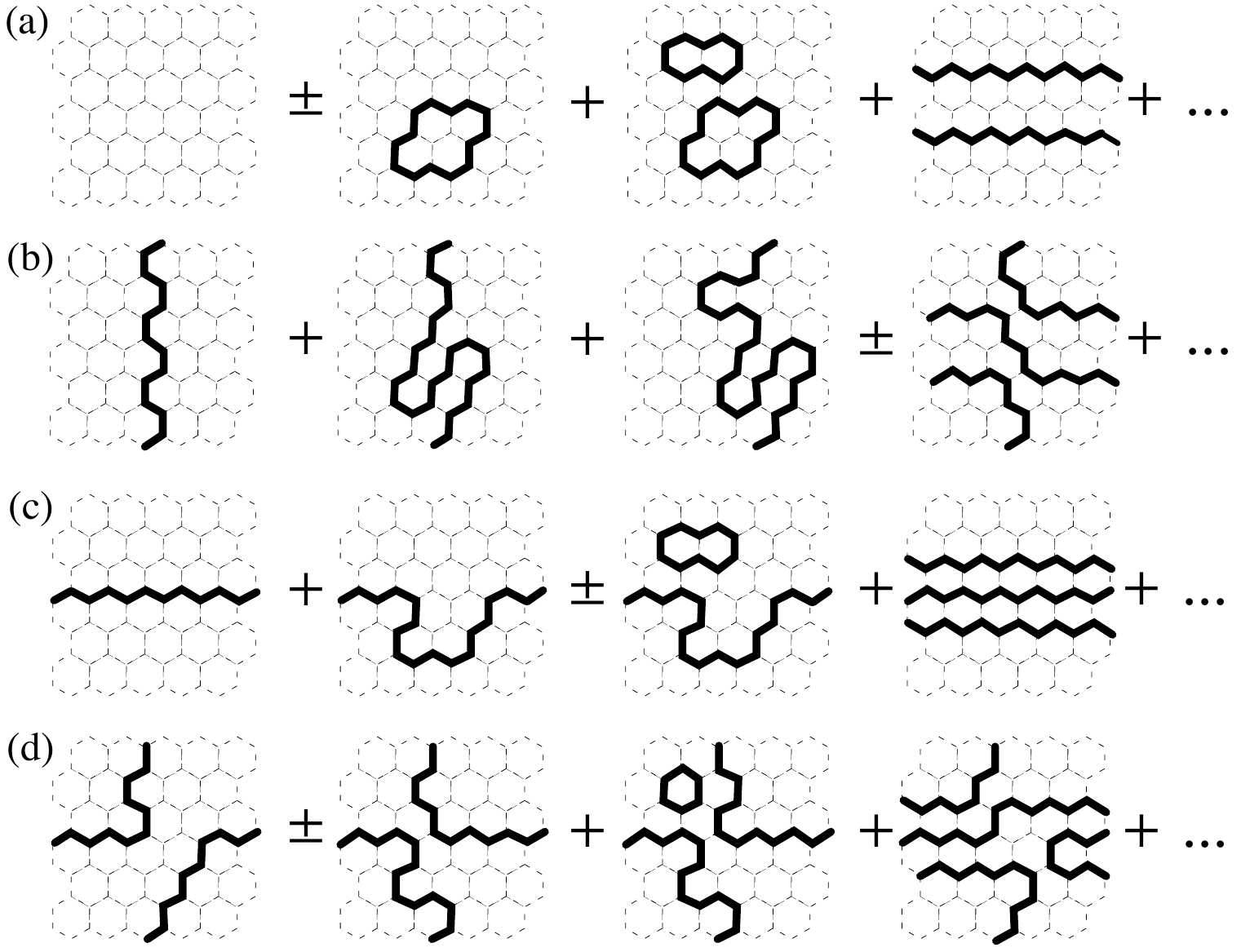}
 \caption{This figure shows the 4 independent ground state sectors in the toric code with periodic boundary conditions in both directions (i.e., on a torus). The 4 ``canonical" kets can be taken to be the four spin configurations on the far left side labeled (a)-(d). Each for the 4 ground states is a superposition of all spin configurations which can be made to look like the canonical ket by using the equivalence rules shown in Fig.~\ref{tcdsem}b-d. For the toric code the relative coefficient is always $+$. The ground state structure for the DSem model is similar except the kets marked with a $\pm$ appear with negative coefficients, a fact that can easily be checked using the rules in \figref{tcdsem}. }
 \label{explicitgs}
\end{figure}

Now that we have concrete relations (\figref{tcdsem}(b)-(d)) between the coefficients of different spin configurations in the ground state we can deduce the ground state degeneracy for periodic boundary conditions in both directions (i.e.  on the torus $\mathbb{T}^2$). Starting with any configuration of closed loops one can show that it appears in the ground state with the same amplitude as one of the four `canonical' configurations labelled (a)-(d) in \figref{explicitgs}.   These are distinguished by the winding number parities
\be
 P_{x}=\prod_{i \in c_{x}}\sigma^z_i \ \ \ \ \ \ \ \ \ P_{y}=\prod_{i \in c_{y}}\sigma^z_i \ \ ,
 \ee
   where $c_{x}$ and $c_{y}$ are cuts along the $x$ and $y$ directions respectively. Each distinct ground state is an equal superposition of all configurations related to a canonical ket by the equivalences in \figref{tcdsem}(b)-(d); some of the kets in each superposition are shown in \figref{explicitgs}.

%
%
%


\subsubsection{Excitations}\label{excitationstc}
The Hamiltonian \eqnref{HTC} has two types of excitations: vertex defects where $B_v=-1$, and plaquette defects where $B_p=-1$. These defects are exact eigenstates because both $B_p$ and $B_v$ are conserved (that is, they commute with the Hamiltonian). Two further features of these excitations will be important in what follows. Firstly, an excitation necessarily involves a pair of vertex or plaquette defects. Secondly, both types of excitations are `deconfined', meaning the energy of a pair of vertex (or plaquette) defects is independent of their relative separation.

\begin{figure}
 \includegraphics[width=1\linewidth]{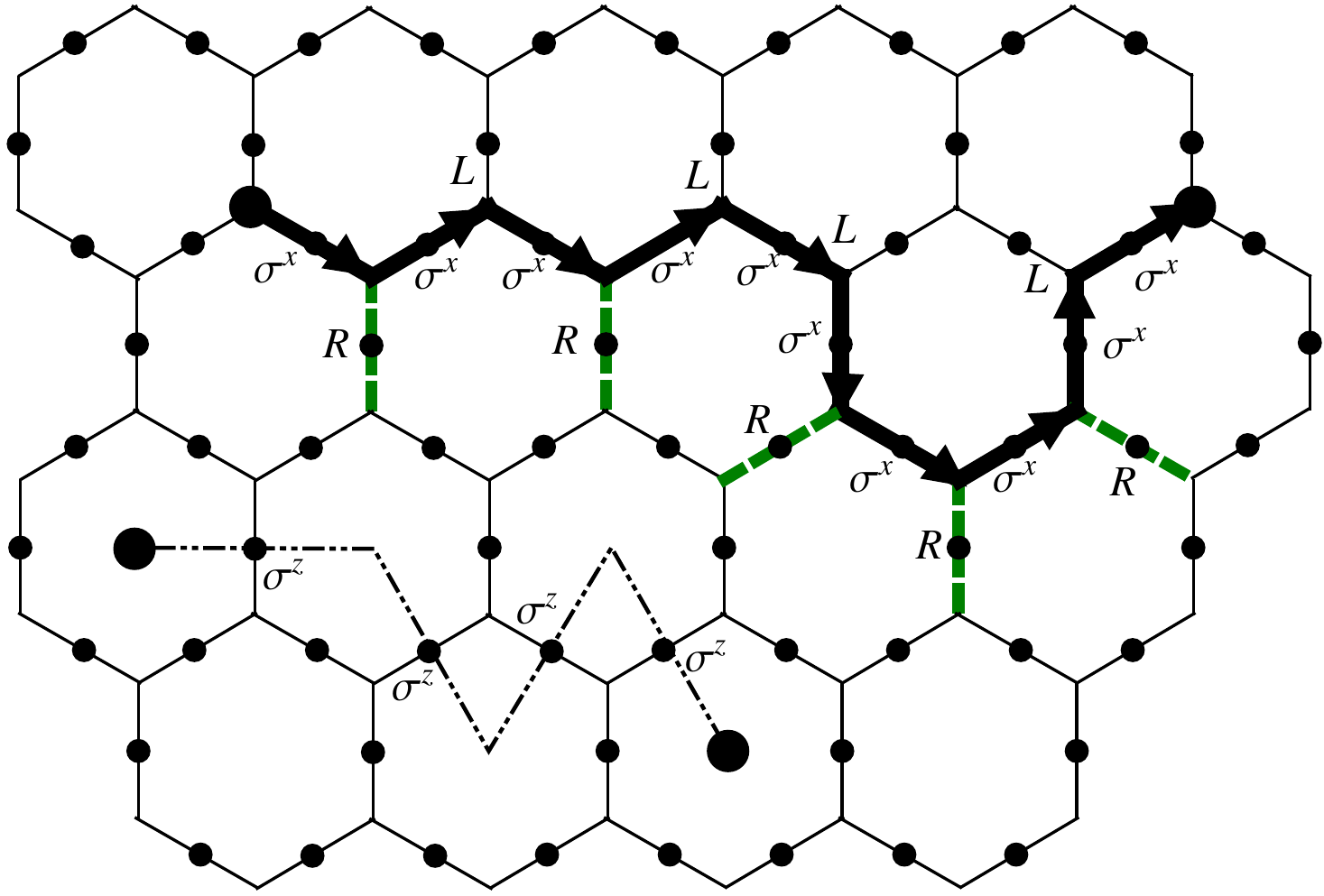}
 \caption{(Color online) This figure shows the different types of string operators in the simple 2D lattice models. For the toric code, the vertex type string operator simply flips the spin on every edge of a path with $\sigma^x$. The plaquette type string operator is represented as a line living on the dual lattice, and every edge that crosses the line is acted on with $\sigma^z$. For the DSem model, the plaquette type string operator is precisely the same. The vertex type string operator, however, includes additional phases which depend on the spins touching the path. A phase $(\pm i)^{(1-\sigma^z)/2}$ is associated with the dashed (green) edges labelled $R$ which lie just to the right of the path; the choice of $\pm$ determines the chirality of the string. Vertices labelled $L$ are attached to an edge lying on the left of the path, and these vertices are associated with a phase $(-1)^{\frac{1}{4}(1-\sigma^z_i)(1+\sigma^z_j)}$ where $i/j$ are the edges on the path just before/after the L-vertex.}
 \label{stringoperators}
\end{figure}

We can understand both of these features by considering how defects are created. To create a vertex defect at $v$, one must flip an edge spin in $s(v)$. This, however, creates a violation on the adjacent vertex $v_2$ sharing this edge. We can remove this violation by flipping a second edge spin in $s(v_2)$. However, this just moves the violation to a third vertex $v_3$, and so on. Hence we can never create a lone vertex defect but we can create a pair of vertex defects by flipping the spins along a continuous path $\mathcal{C}$ of edges by acting with the operator $\hat{W}_V(\maC)=\prod_{i\in \mathcal{C}}\sigma^{x}_i$, as shown in \figref{stringoperators}. We call this a vertex-type string operator. Crucially the string operator commutes with all plaquette operators along its length (because it only involves $\sigma^{x}$'s), so only has an energy cost from the vertex defects at its endpoints. This tells us that vertex excitations are deconfined, and that closed strings without end-points have no energy cost. Indeed,
 taking $\mathcal{C} = \partial P$ gives precisely the plaquette operator $B_p$, which commutes with the Hamiltonian by construction.  A closed vertex type string operator that wraps several plaquettes is just the product of $B_p$'s for the enclosed plaquettes. However, interestingly, one can also construct non-contractible vertex type string operators which flip spins along paths $\maC_x, \maC_y$ which wind completely around the system in the $x$ or $y$ direction respectively. The resulting operators $\hat{W}_V(\maC_x), \hat{W}_V(\maC_y)$ flip the parities $P_y$ or $P_x$ respectively and so have the effect of toggling between the four different ground state sectors in \figref{explicitgs}.

One can similarly define plaquette-type string-operators that create pairs of plaquette defects. These take the form $\hat{W}_P(\maC')=\prod_{i\in \mathcal{C}'}\sigma^{z}_i$, where $\mathcal{C}'$ is a string on the dual lattice as shown in \figref{stringoperators}. Vertex defects have nontrivial relative statistics with plaquette defects: Moving a vertex defect all the way around a plaquette defect gives a Berry phase of $-1$, and vice versa. These statistics are a defining element of topological order, and will allow us to distinguish the toric code from the doubled semion model which we discuss in the next section.

\subsubsection{Graphical representation of operators as strings}\label{tcgrep}
\begin{figure}
 \includegraphics[width=0.95\linewidth]{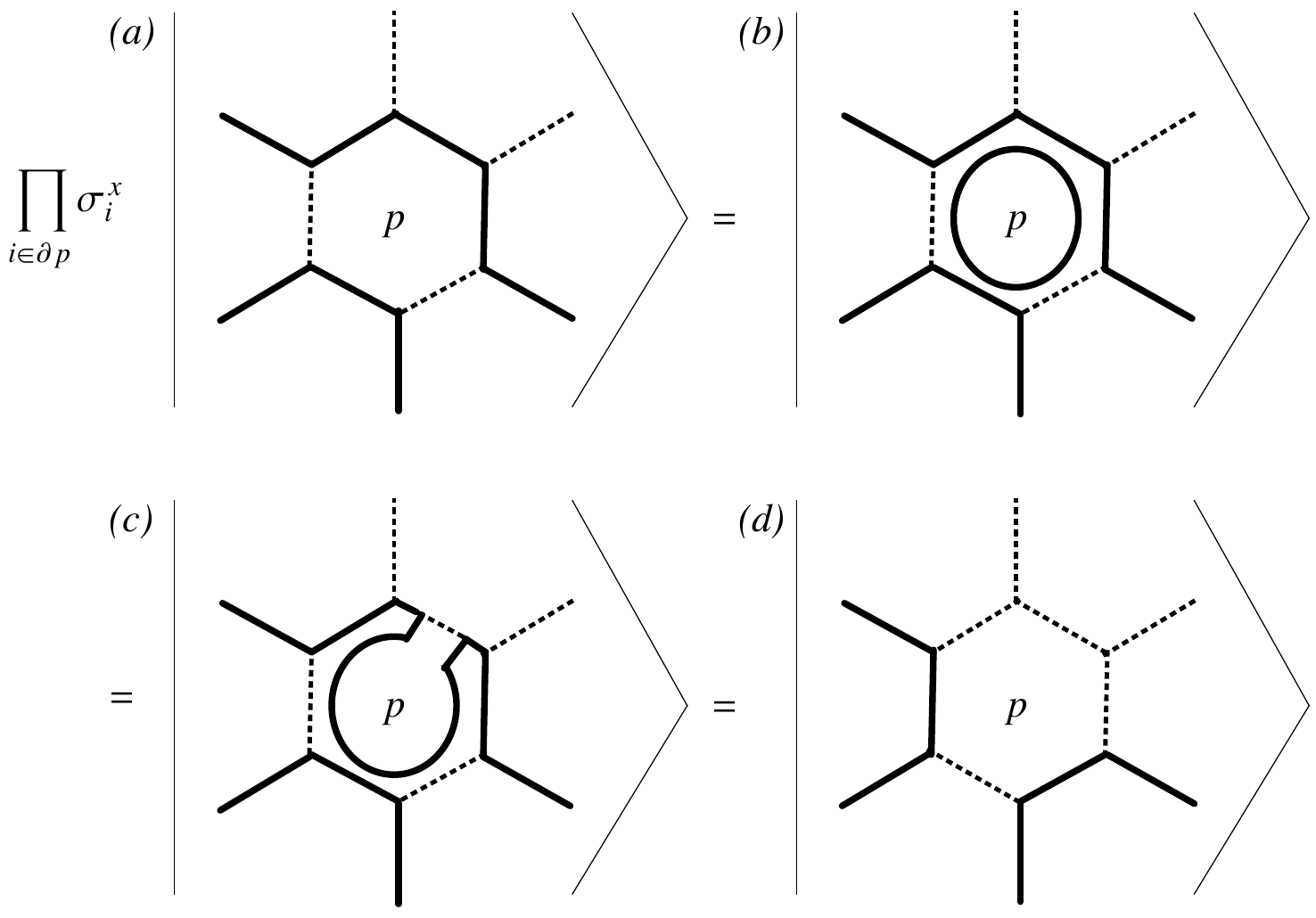}
 \caption{This diagram shows how we can use the fusion rules to graphically define a plaquette flipping term (a) using a string that lives off lattice shown in (b). The key step occurs in going from (b)-(c) where we use \figref{tcdsem}(d) to fuse the string into the edge. The end result (d) is that all edges on the plaquette are flipped, which is precisely the action of $\prod_{i\in\partial p} \sigma^x_i$.}
 \label{plaquettestring}
\end{figure}

Before discussing our next 2D topological lattice model, it is useful to review the connection between the rules in \figref{tcdsem} and the lattice model. Thus far, we have established that the rules in \figref{tcdsem} (b)-(d) can be used to evaluate the relative amplitudes of different kets in a particular ground state. However, they can also be used to depict the action of the string operators defined in the previous section. An example is shown in \figref{plaquettestring}: the plaquette operator $B_p$ can be represented by a string running along the inside of the plaquette $p$. To determine the action of this operator on a particular ket, we ``fuse" this string with the relevant edge labels of $p$ using \figref{tcdsem}(d).

\begin{figure}
 \includegraphics[width=1\linewidth]{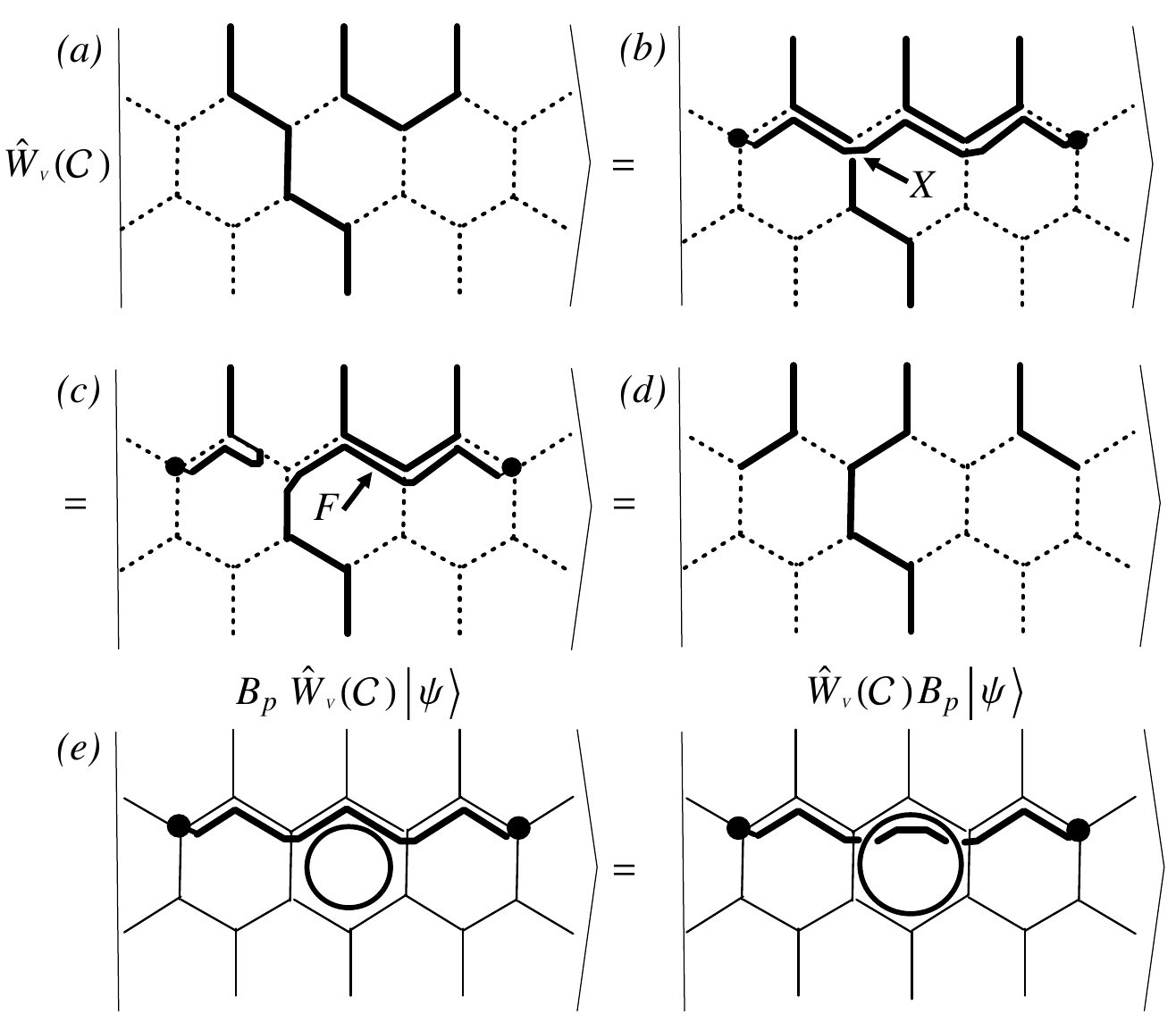}
 \caption{This diagram illustrates how we can use an off-lattice string to define a string-operator $\hat{W}\db{C}$. Starting with some ket in (a) we define the string operator using the diagram in (b), where the black dots denote the end-points of the string operator. In order to evaluate this diagram we need to use the fusion rule \figref{tcdsem}(d) at the point labelled $F$ and the braiding rule \figref{tcdsem}(e) at point $X$ to resolve the string under-crossing the lattice edge. (e) Represents a graphical proof that the plaquette operator commutes with the string operator along its length; on the left hand side the string acts first, on the right hand side the plaquette acts first and we can move between these two diagrams by deforming the string locally which is allowed as long as string operator does not move through the centre of a plaquette.}
 \label{stringdefn}
\end{figure}

Vertex string operators $\prod_{i\in \mathcal{C}}\sigma^{x}_i$ are defined using a similar procedure: we depict the string operator graphically as a string running along the path $\mathcal{C}$. The action of the operator on a ket is given by fusing this string into the edges along $\mathcal{C}$ using \figref{tcdsem}(d) and (e). (For the Toric code the rule (e) is superfluous; however in other models it will be necessary as we will assign a definite meaning to strings drawn crossing over, as opposed to under, the edges in its path (\figref{stringdefn}).)

Using these graphical representations of states and operators, it is possible to describe many more 2D topological lattice models than the ones we describe here\cite{Levin05}. A careful discussion of the conditions under which these graphical procedures give a consistent and unambiguous action for the operators in the theory can be found in Refs. \onlinecite{Levin05,Kitaev06}.

 \subsection{The doubled semion model}\label{ss:DSemintro}
The toric code model shows that it is possible to write a simple and exactly solvable spin Hamiltonian exhibiting topological order. Many generalizations are known but here we describe one of the simplest, a second loop model known as the Ôdoubled semionÕ or DSem model first discussed in Refs.~[\onlinecite{Levin05}] and [\onlinecite{Shtengel04}]. The Hamiltonian takes the form
\be\label{DSemplaq}
H = - \sum_v \underbrace{ \prod_{s(v)} \sigma^z_i }_{B_v} + \sum_p \underbrace{(\prod_{i\in \partial p} \sigma^x_i ) \prod_{j\in s(p)} i^{(1 - \sigma^z_{j})/2}}_{B_p} \punc{,}
\ee
where $s(p)$ is the set of six legs radiating from plaquette $p$, as shown in \figref{toricFig1}, and as above $s(v)$ and $\partial p$ denote the three edges entering vertex $v$, and the six edges bordering plaquette $p$ respectively. Comparing \eqnref{DSemplaq} with the toric code Hamiltonian in \eqnref{HTC}, we see that $B_v$ is unchanged but the plaquette terms differ: in addition to flipping the edges of $p$, the new $B_p$ operator includes a phase which depends on the spins in $s(p)$. Furthermore, the plaquette part of the DSem Hamiltonian appears with a total $+$ sign rather than a $-$ sign.

For any state $|\Psi_{loop} \rangle$ obeying $B_v|\Psi_{loop} \rangle=|\Psi_{loop} \rangle$ (a `loop gas state'), it is easy to show that
\be
\left[B_p, B_v \right]|\Psi_{loop} \rangle =\left[B_p, B_{p'} \right]|\Psi_{loop} \rangle=0
\ee
and furthermore that $B_p$ has eigenvalues $\pm1$ on these loop gas states. Thus if we restrict our Hilbert space to the loop gas states, the Hamiltonian is a sum of commuting projectors and is thus exactly solvable. This fact will allow us to solve exactly for the ground states of the Hamiltonian, because we will find that these only involve closed loop configurations. Though it will not be necessary for our purposes, with a more complex definition of $B_p$ it is possible to write a Hamiltonian that is exactly solvable on the full Hilbert space; we direct the interested reader to \appref{ExplicitDSem}.

\begin{figure}
 \includegraphics[width=.95\linewidth]{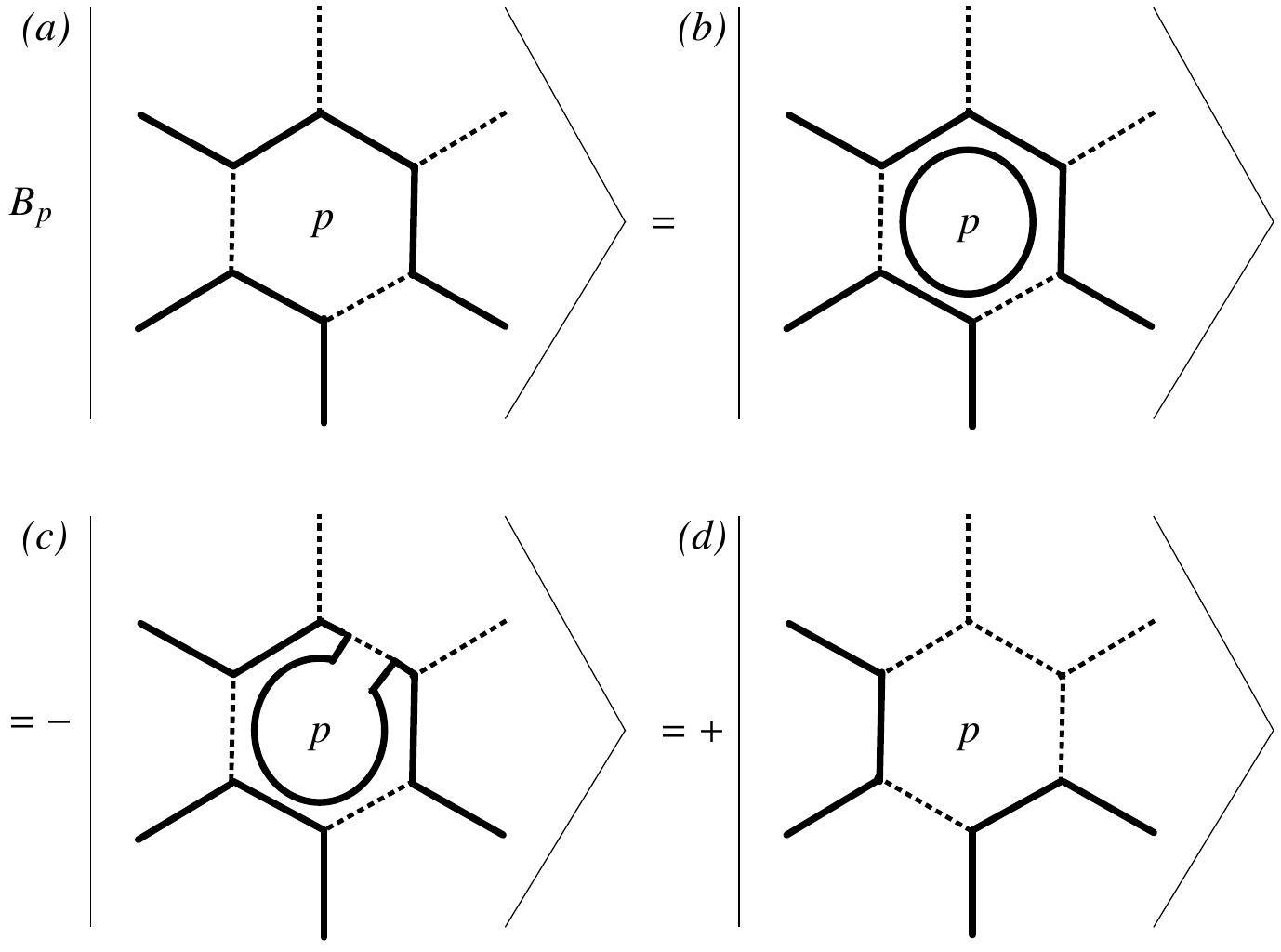}
\caption{This figure shows the graphical definition of a $B_p$ on a ket in the DSem model. (b) shows the definition in terms of a string operator. The step (b)-(c) requires the use of a fusion rule \figref{tcdsem}(d), so produces a minus sign. Similarly, step (c)-(d) requires a fusion, giving a net plus sign. The original ket and (d) are related by a single application of $B_p$, therefore they appear with opposite sign in the ground state because $B_p=-1$ on the ground state. }
\label{plaquetteDSem}
 \end{figure}

			\subsubsection{Ground States}\label{sss:DSemGS}

A lower bound on the ground state energy is obtained when
\be\label{DSemGSconstraints}
B_v= 1 \text{ for all } v \ \ \ \ \ \ B_p=-1 \text{ for all } p\punc{.}
\ee
Any state which satisfies all of these conditions is automatically a ground state because it saturates a lower bound on the Hamiltonian. Once again such states exist. Since the vertex condition is the same, the ground states of the doubled semion model are also made of a superposition of spin configurations which involve only non-intersecting closed loops. However, as we will see below, the coefficients in the superposition are different from those of the toric code.

The condition $B_p=-1$ implies that the two spin configurations involved in the equation

\be
\mathord{\includegraphics[height=14ex]{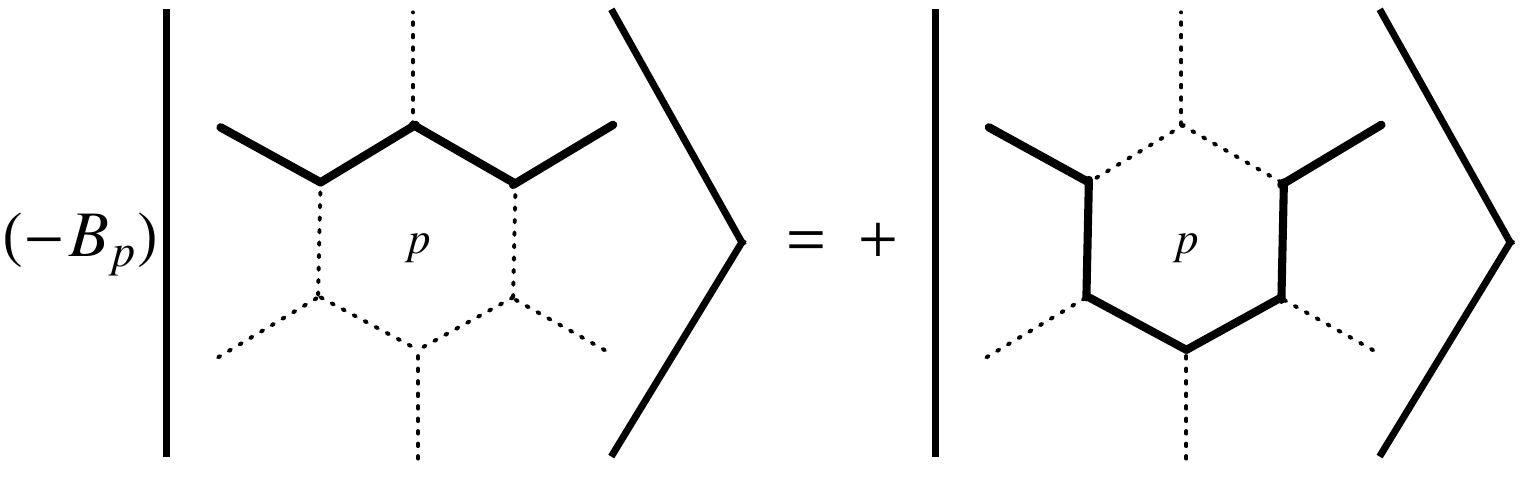}}\punc{,}
\ee
occur with the same coefficient in the ground state superposition. Therefore, similar to the toric code, configurations related by deforming a loop over a plaquette occur with precisely the same coefficient in the ground state (as indicated by \figref{tcdsem}(b)). However, the $B_p=-1$ condition implies that the two kets involved in
\be
\mathord{\includegraphics[height=14ex]{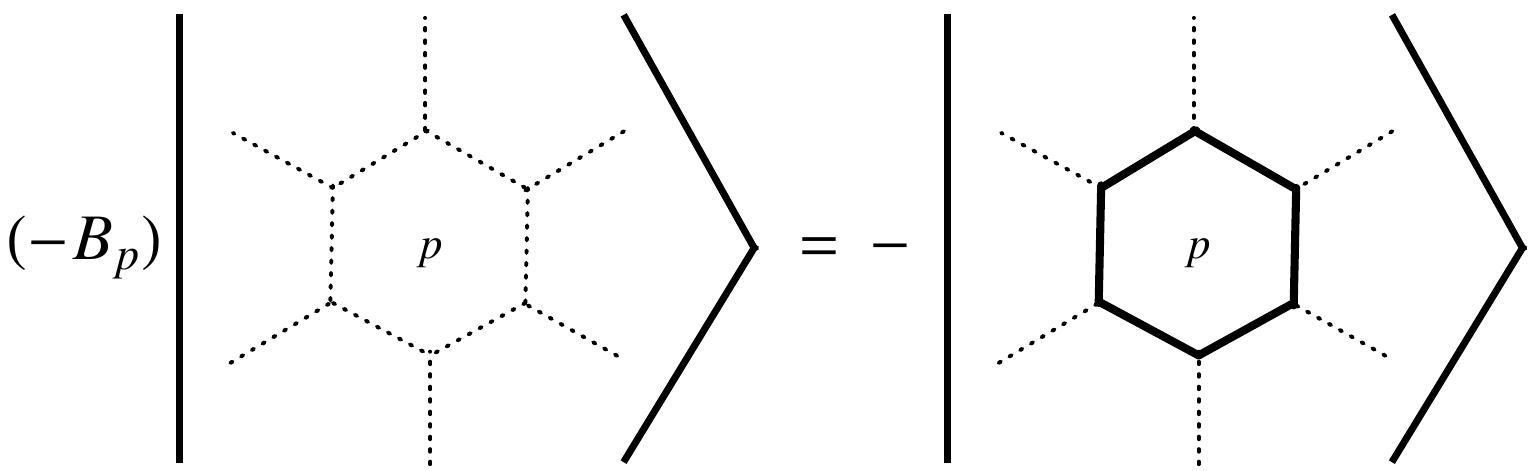}}\punc{,}
\ee
occur with a relative minus sign in the ground state superposition. Therefore spin configurations related by the creation or elimination of a closed contractible loop have a relative minus sign in the ground state superposition (as indicated by the rule in \figref{tcdsem}(c)). In the same way one can show that configurations related by the fusion rule \figref{tcdsem}(d) occur with a relative minus sign.
Thus the doubled semion graphical rules \figref{tcdsem}(b)-(d) give linear relations between the coefficients of spin configurations in the ground state as illustrated in \figref{gskets}.

In the ground state space, the amplitude of any loop configuration in can be related, by using the DSem graphical rules \figref{tcdsem}(b)-(d), to the amplitude of one of the canonical kets labelled (a)-(d) in \figref{explicitgs}. As these new rules have minus signs, the ground states are similar to those in the toric code except some configurations appear with a $-1$ coefficient, as shown in the superposition in \figref{explicitgs}. Analogous to the toric code, the ground state space is thus split into $2^2$ orthogonal sectors distinguished by parity operators $P_{x},P_{y}$.


\subsubsection{Excitations}\label{sss:DSemExcitations}

The low lying excitations of the doubled semion Hamiltonian appear as pairs of vertex defects where $B_v=-1$, or pairs of plaquette defects where $B_p=+1$. As for the Toric code, the operators that create pairs of defects are `string operators' acting on all edges along some path $C$ joining the pair. Here we will show how to construct string operators $\hat{W}_P, (\hat{W}_V)$ that violate exactly two plaquettes (vertices) at their end-points, and no other terms in the Hamiltonian. It follows that both vertex and plaquette excitations are deconfined: the energy cost of creating a pair of either of these is independent of their separation.

The plaquette type string operator is identical to that in the toric code, taking the form $\hat{W}_P(\maC')=\prod_{i\in \mathcal{C}'}\sigma^{z}_i$, where $\mathcal{C}'$ is a string on the dual lattice (see \figref{stringoperators}). Such an operator manifestly commutes with every $B_v$, and only induces plaquette defects at its endpoints, where it anti-commutes with the $\prod \sigma^x$ part of the $B_p$ operators. On the other hand, the vertex type string operator from the toric code model, $\prod_{i\in \mathcal{C}}\sigma^{x}_i$, fails to commute with the modified plaquette operators bordering on $\mathcal{C}$ because of the $i^{(1 - \sigma^z_{j})/2}$ factor in the new definition of $B_p$. To write down a vertex type string operator that produces vertex violations at its endpoints but commutes with plaquettes along its length, $\prod_{i\in \mathcal{C}}\sigma^{x}_i$ must be combined with phases depending on the spins on edges touching the path to form the following operator $\hat{W}^{\pm}_V(\mathcal{C})$:
\be\label{DSemstring}
\prod_{i\in \mathcal{C}} \sigma^x_i \prod_{k\in\text{$L$ vertices}} (-1)^{\frac{1}{4}(1-\sigma^z_i)(1+\sigma^z_j)} \prod_{l\in \text{$R$ legs}} (\pm i)^{(1-\sigma^z_l)/2}\punc{.}
\ee
To specify the phases such that $\hat{W}^{\pm}_V(\mathcal{C})$ commutes with $B_p$ everywhere, we have had to choose an orientation for our string operator (\figref{stringoperators}), and separate the vertices into two types: L-vertices are attached to an edge on the left of $\mathcal{C}$, while R-vertices are attached to an edge on the right of $\mathcal{C}$ which we call an R-leg. Each L-vertex is also attached to two edges on $\mathcal{C}$, and these are denoted by $i,j$ where $i$ occurs before $j$ on the directed path $\mathcal{C}$. The choice of $\pm$ determines whether the string has $\pm$ `chirality'. The phases in the definitions of the string operators imply that two vertex string operators of the same chirality anti-commute if their paths cross. This indicates a sensitivity to the order in which particle pairs are created and destroyed, and can be used to show that vertex defects of the same chirality have semionic statistics\cite{Levin05}: exchanging a pair of these induces a phase of $\pm i$ in the wave function.

\begin{figure}
 \includegraphics[width=1\linewidth]{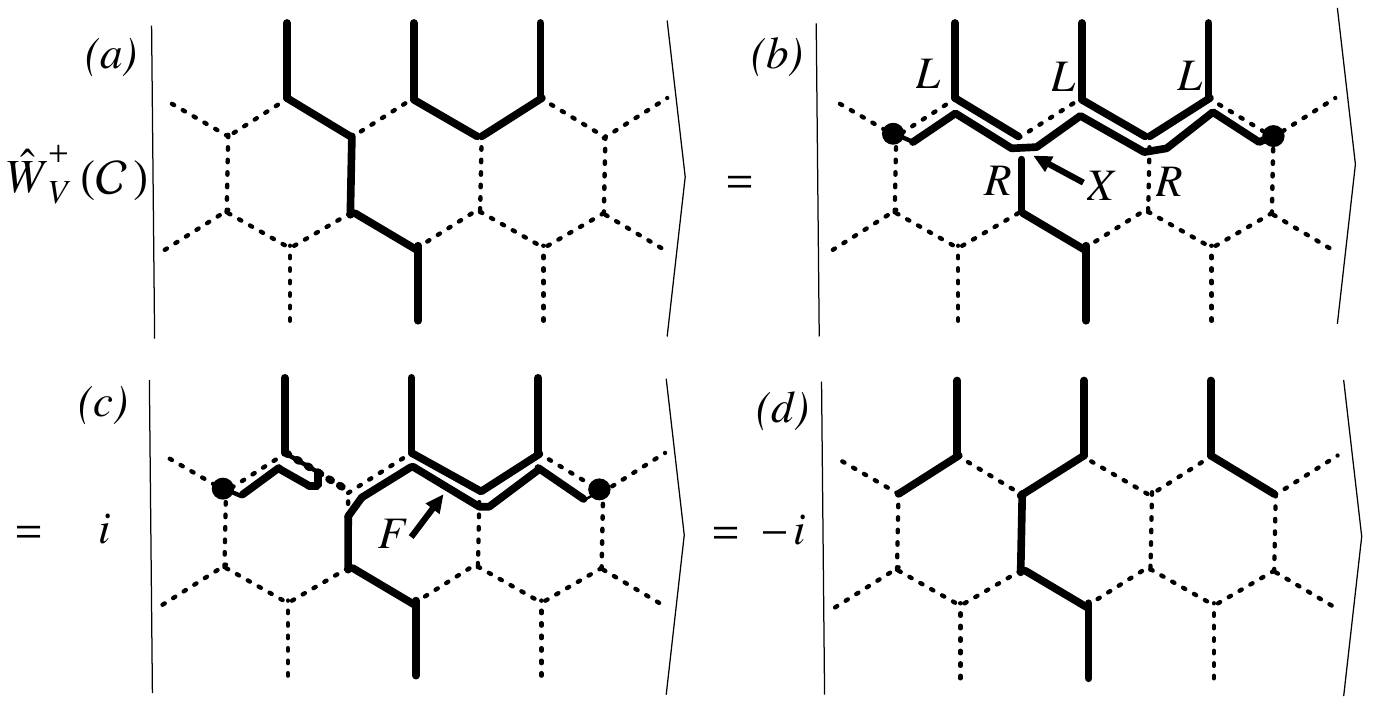}
\caption{This diagram illustrates how we can use an off-lattice string to define a positive chirality string-operator $\hat{W}^{+}\db{C}$ in the doubled semion model; positive chirality corresponds to taking $+i$ in the concrete Pauli matrix form of the operator in \eqnref{DSemstring}. Starting with some ket in (a) we define the string operator using the diagram in (b), where the black dots denote the end-points of the string operator. In order to evaluate this diagram we need to use the braiding rule \figref{tcdsem}(e) at the string over-crossing labelled $X$, yielding a factor of $i$. We then use and the fusion rule \figref{tcdsem}(d) at point $F$ to fuse the string into the edge, yielding an additional factor of $-1$.}
\label{stringDSem}
 \end{figure}

There is again a simple graphical mnemonic which reproduces the required spin flips and configuration dependent phases along the length of the string away from its endpoints. Pairs of `positive chirality' vertex defects can be created by laying a string {\it over} the lattice along a prescribed path and using the rules \figref{tcdsem}(b)-(e) to fuse the string into the edges; an example of this procedure is shown in \figref{stringDSem}. The complex conjugate operator corresponds to laying the string {\it under} the edges it crosses in the lattice, and produces a pair of `negative chirality' vertex defects.

As in the toric code, it is useful to consider string operators that do not create excitations (i.e. where the curve $\maC$ has no endpoints). If $\maC$ encircles a single plaquette $p$, $\hat{W}^{\pm}_V(\maC) = B_p$, and hence $\hat{W}^{\pm}_V(\maC)$ commutes with $H$. If $\maC$ traces the perimeter of a cluster of plaquettes, $\hat{W}^{\pm}_V(\maC) = \prod_{p \in cluster} B_p$ and hence such string operators act trivially on the ground states. The operators of main interest are again those vertex type string operators which follow paths $\maC_x,\maC_y$ which wind completely around the system in the $x$ and $y$ direction respectively. The resulting operators $\hat{W}^{\pm}_V(\maC_x)$, $\hat{W}^{\pm}_V(\maC_y)$ flip the parities $P_y$, $P_x$ respectively. These non-contractable string operators cannot be written as a product of $B_p$ operators, but still commute with the Hamiltonian and have the effect of toggling between the four different ground state sectors in \figref{explicitgs}.

We end this section by noting a technical point. The above expression \eqnref{DSemstring} gives the correct form of vertex type string operators away from their endpoints. Capturing the precise form of the string operator at its endpoints requires a careful treatment.  The details of this are not essential for our purposes, but are shown in \appref{ExplicitDSem}.

\section{Simple 3D topological lattice models on closed manifolds}\label{WW}
\begin{figure}
 \includegraphics[width=\linewidth]{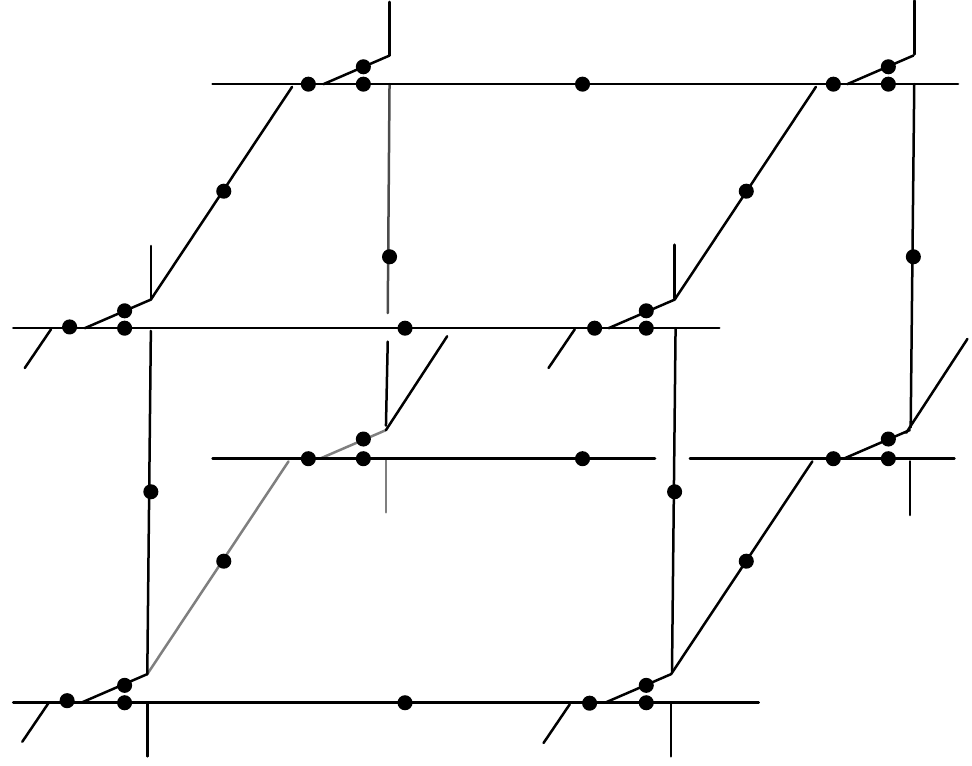}
 \caption{This figure shows the point splitting and fixed projection used to define the 3D lattice for the Walker-Wang models. The dot in the middle of each bond represents a spin variable.}
\label{fig:3Dlattice}
 \end{figure}

Levin and Wen\cite{Levin05} suggested a method generalizing a limited subset of their exactly solvable models to 3D. Their method can be used to generalize the 2D toric code to 3D but cannot be used to generalize the DSem model to 3D. In this section we follow the different prescription of Walker and Wang\cite{Walker11} which allows us to generalize both the 2D toric code and DSem models to 3D. The resulting 3D lattice models are described by a two-state system $\sigma^{z}=\pm 1$ on each edge of the trivalent lattice shown in \figref{fig:3Dlattice}. The generalization of the 2D toric code to 3D is well known\cite{Hamma05,Castelnovo09}. We nevertheless begin by reviewing the key signatures of topological order in this model (ground state degeneracy, possible defects and their statistics) for later contrast with the properties of the 3D cousin of the DSem model. In \secref{ss:TcT^3} we discuss the ground state degeneracy of the 3D Toric code on the 3-torus, and describe the low-lying excitations which take the form of deconfined point defects and linearly confined vortex rings.

In \secref{ss:SemT^3} we present the 3D analogue of the 2D DSem model (the `3D semion model'), which is our first example of a confined WW model. Again we will characterize the phase by studying its ground states and excitations. We find that the ground state is non-degenerate on the 3-torus, and in \secref{sss:Semconf} we relate this to the fact that its point defects are confined in the bulk. Combined with the study of these models on a manifold with boundary which we present in \secref{sec:DsemD^2xS^1}, this section provides a firm foundation for understanding the effective field theories for the lattice models in \secref{s:DSemFT}. Furthermore, the 3D toric code and semion models capture much of the qualitative behavior of more general Walker-Wang models which we study in \secref{s:MTCGS}.

	\subsection{The 3D toric code}\label{ss:TcT^3}
Viewed as a Walker-Wang model, the 3D toric code\cite{Hamma05,Castelnovo09} Hilbert space consists of a two state system $\sigma^{z}=\pm 1$ on each edge of the lattice shown in \figref{fig:3Dlattice}. The Hamiltonian takes the form
\be\label{HTC3D}
H = - \sum_v \underbrace{ \prod_{i\in s(v)} \sigma^z_i }_{B_v} - \sum_p \underbrace{\prod_{i\in \partial p} \sigma^x_i }_{B_p} \punc{,}
\ee
where $s(v)$ is the set of three edges attached to vertex $v$ and $\partial p$ is the set of ten edges of a plaquette $p$ (bold edges shown on the 3 types of plaquette in \figref{fig:3DSemplaquettes}(a)-(c)). As in 2D, the $B_v$ take the values $\pm 1$ depending on whether there are an even/odd number of down spins on the edges coming into vertex $v$, and $B_p$ flips the spins on each edge of $p$. The fact that $B_p$ flips a pair of spins at vertex $v$ implies that $\left[ B_p,B_{v}\right]=0$, and once again $\left[ B_v,B_{v'}\right]=\left[ B_p,B_{p'}\right]=0$, so the model is exactly solvable.

		\subsubsection{Ground states of toric code on $\mathbb{T}^3$}\label{sss:TCT^3}
The ground state space is defined by the conditions $B_p=B_v=1$ for all vertices and plaquettes. As in the 2D case in \secref{sss:DSemGS}, the condition $B_v=1$ forces the 3D ground state to be a superposition of closed loops. Recall that in 2D, the condition $B_p=+1$ implied that the rules in \figref{tcdsem} (b)-(d) relate the amplitudes of different spin configurations. Exactly the same type of calculation shows the rules in \figref{tcdsem} (b)-(d) relate the amplitudes of spin configurations in the ground state space of the 3D toric code.

\begin{figure}
	\includegraphics[width=1.\linewidth]{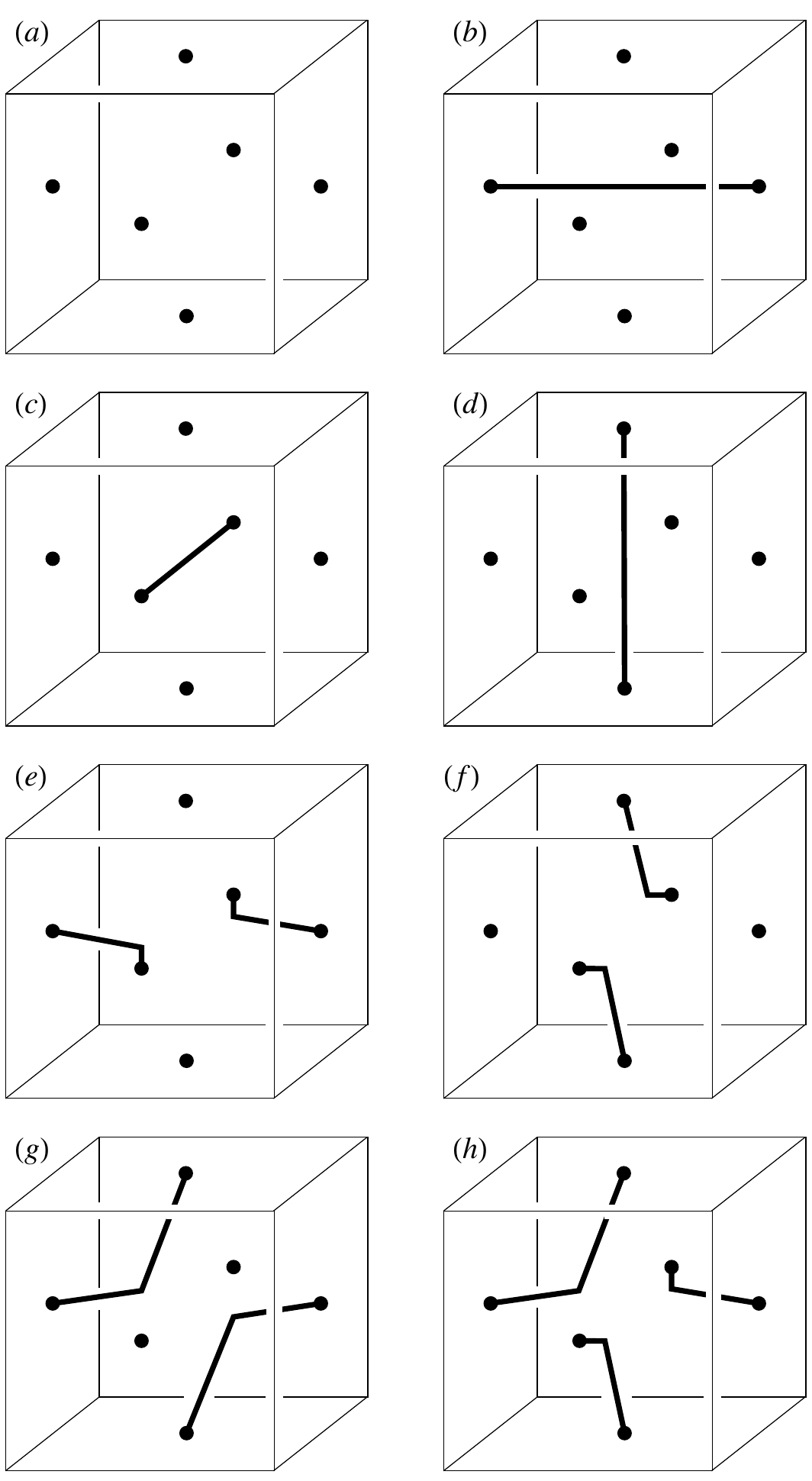}
	\caption{Shown are the eight canonical configurations with periodic boundary conditions in all three directions, where the thick black lines indicate edges with $\sigma^{z}=-1$. The underlying lattice is not drawn, for simplicity. Any basis ket without vertex violations can be related to one of these eight by using the graphical rules in \figref{tcdsem}.  In the case of the toric code, the ground state splits into eight orthogonal sectors labelled by these eight kets; two of these states are shown in \figref{tc3Dgs}.}
	\label{3torus}
\end{figure}

\begin{figure*}
	\includegraphics[width=1.\linewidth]{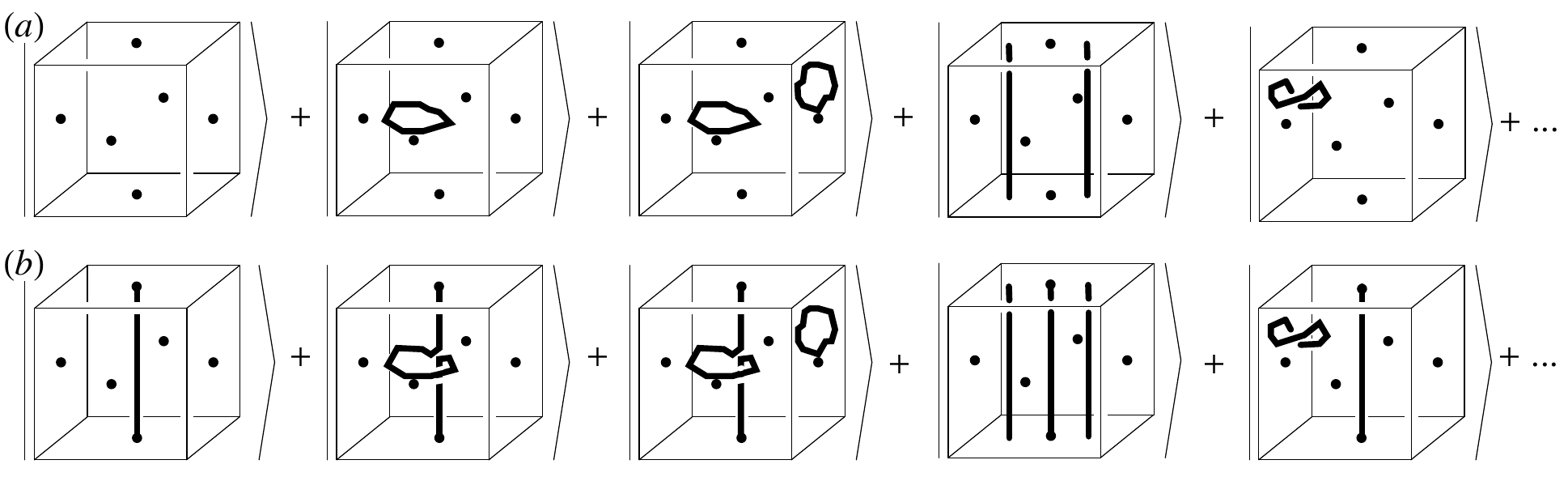}
	\caption{This figure shows two examples of ground states of the 3D toric code. (a) consists of an equal superposition of all kets related to \figref{3torus}a by \figref{tcdsem}(a)-(d), and has $P_{x_\perp}=P_{y_\perp}=P_{z_\perp}=1$. (b) consists of an equal superposition of all kets related to \figref{3torus}(d) by the same graphical rules, and has $P_{x_\perp}=P_{y_\perp}=1$ but $P_{z_\perp}=-1$. There are a further six ground states corresponding to the other kets in \figref{3torus}. The underlying lattice is not drawn, for simplicity.}
	\label{tc3Dgs}
\end{figure*}

In the 2D toric code on the 2-torus, the amplitude of any spin configuration in the ground state could be related to the amplitude of one of four canonical configurations (shown in \figref{explicitgs}) resulting in four degenerate ground states.   Analogously, on the 3-torus one can show that any configuration of closed loops appears in the ground state with the same amplitude as one of the eight canonical configurations shown in \figref{3torus}. These $2^3$ configurations can be labelled by three parities
\be\label{parities}
P_{n_\perp}=\prod_{i \in n_\perp} \sigma^z_i \ \ \ \ \ n=x,y,z
\ee
 which take values $\pm1$ depending on whether an even or odd number of loops wind around the $n$-cycle of the torus.   Here $z_\perp$ is the set of all edges emanating from the plane $(z=0)$ in the $+\hat{z}$-direction, and similarly for $x_\perp, y_\perp$.
 If we take an equal superposition of all configurations related to one of the eight canonical configurations by the equivalences in \figref{tcdsem}(b)-(d) (as shown in \figref{tc3Dgs}) then it is easy to verify that we get a ground state. This shows us that the ground state degeneracy is $2^3$, and each distinct ground state is labelled by the three eigenvalues $P_{x_\perp},P_{y_\perp},P_{z_\perp}=\pm 1$.
 On more general closed manifolds the ground state degeneracy is $2^{b_1}$, where $b_1$ is the number of independent non-contractible cycles on the manifold (also known as the first Betti number).


		\subsubsection{Excitations in the toric code}\label{sss:3Dtcexcitations}
The Hamiltonian \eqref{HTC3D} has two types of excitations: pairs of vertex defects where $B_v=-1$, and lines of plaquette defects where $B_p=-1$. We can create a pair of vertex defects with an operator
\be
\hat{W}_V({\mathcal{C}_{AB}})=\prod_{i\in \mathcal{C}_{AB}} \sigma^x_i\punc{,}
\ee
where $\mathcal{C}_{A B}$ is a path connecting the positions of the defects $A$ and $B$. Graphically we represent the string operator by laying a string along $\mathcal{C}_{A B}$, where it is understood that the operator acts on kets by fusing this string into the edges using the rules \figref{tcdsem}(b)-(e). The operator commutes with the Hamiltonian except at its endpoints, and so the defects are deconfined.

As in 2D, a closed vertex type string operator that wraps around the boundary of several plaquettes is just the product of $B_p$'s for the enclosed plaquettes and so trivially commutes with the Hamiltonian. Non-contractible string operators (i.e., string operators that wrap around the periodic boundary conditions) are more interesting because they commute with the Hamiltonian but cannot be expressed as a product of $B_p$'s. Operators of this form toggle between the different ground state sectors discussed in \secref{sss:TCT^3}; for example, a non-contractible string operator wrapping the z-direction of the torus toggles between the ground states in \figref{tc3Dgs}(a) and (b).

\begin{figure}
 \includegraphics[width=\linewidth]{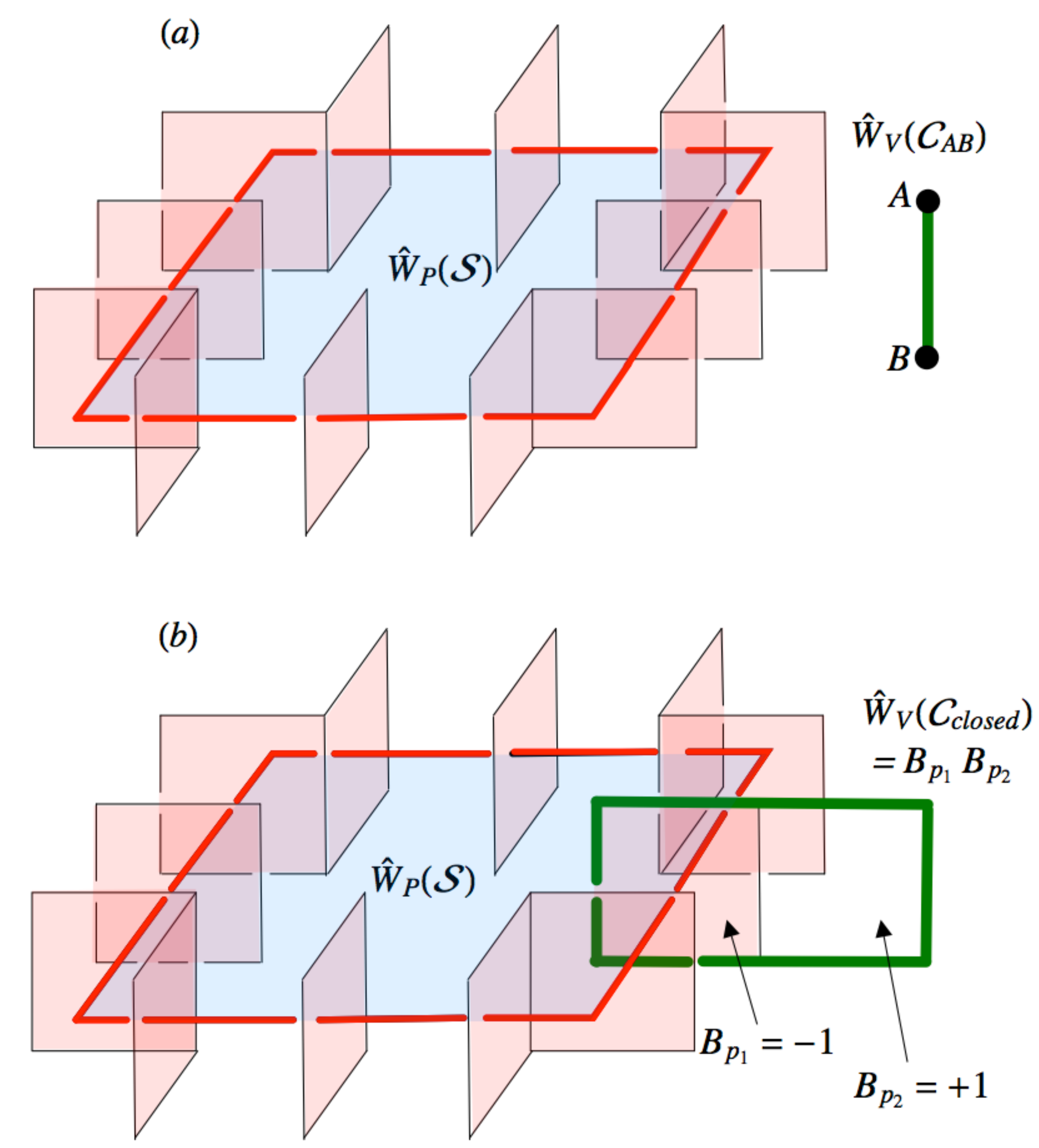}
 \caption{(Color online) This diagram illustrates the exchange between a line defect created by $\hat{W}_P({\mathcal{S}})$ and a pair of point defects created by $\hat{W}_V({\mathcal{C}_{AB}})$, both shown (a) for the 3D toric code. We show the process on a cubic lattice for simplicity, but the same reasoning applies on the trivalent Walker-Wang lattice. The defect $B$ is threaded through the surface $\mathcal{S}$ bounded by the line defect, and annihilated with $A$. The resulting closed loop can be written as a product of plaquette defects, yielding a result of $-1$ because one of the plaquettes lies on the line defect.}
\label{fig:3Dtoriclinedefect}
 \end{figure}

Thus, vertex defects in 3D are much the same as they were in 2D, being thought of as the end-points of a string operator. Plaquette defects, on the other hand, behave quite differently in $2$ and $3$ dimensions. They no longer appear at the end-points of string operators, but rather at the boundary of surface operators. To create a plaquette defect, one acts on an edge with a $\sigma^z$. This, however, creates defects in all four plaquettes associated with the edge. More generally, if we pick a surface on the dual lattice and act with $\sigma^z$ on each edge cutting the surface, then the resulting operator
\be\label{surfaceW}
\hat{W}_P({ \mathcal{S}})=\prod_{i\in \mathcal{S}} \sigma^z_i\punc{,}
\ee
creates plaquette defects along the boundary $\partial \mathcal{S}$ of the surface $\mathcal{S}$ (see \figref{fig:3Dtoriclinedefect}(a)). This is because $\sigma^z$ acts on an even number of edges of each plaquette cutting $\mathcal{S}$, except for those lying on the boundary $\partial \mathcal{S}$ which only have $\sigma^z$ acting on one of their edges. Thus this type of surface operator has an energy cost which scales linearly with the length of the boundary $\partial \mathcal{S}$ of the surface.

One of the defining features of the topological order of the 3D Toric code is the mutual statistics between the point-like vertex defects and the vortex lines. Moving a vertex defect all the way around a line defect gives a Berry phase of $-1$. The exchange process is defined as follows. First act on a ground state with the two operators to form $\hat{W}_V({\mathcal{C}_{AB}}) \hat{W}_P({ \mathcal{S}})\ket{GS}$ as shown in \figref{fig:3Dtoriclinedefect}(a). We now take the point defect $B$ and thread it through $\partial \mathcal{S}$, and then annihilate $A$ with $B$. This results in a string operator $\hat{W}_V({\mathcal{C}_{closed}})$ enclosing the plaquette defects on $\partial\mathcal{S}$ as shown in \figref{fig:3Dtoriclinedefect}(b). The string operator encircling the line defect can be written as a product of plaquette operators $\hat{W}_V({\mathcal{C}_{closed}})=\prod_{p \in \mathcal{R}} B_p$, where $\mathcal{R}$ is any surface bounded by $\mathcal{C}_{closed}$. However, one of the $B_p$ operators in $\mathcal{R}$ will lie on the line defect, and therefore take value $-1$. Thus a full exchange between the two types of defects leads to a sign of $-1$.

	\subsection{The 3D semion model }\label{ss:SemT^3}
\begin{figure}
 \includegraphics[width=\linewidth]{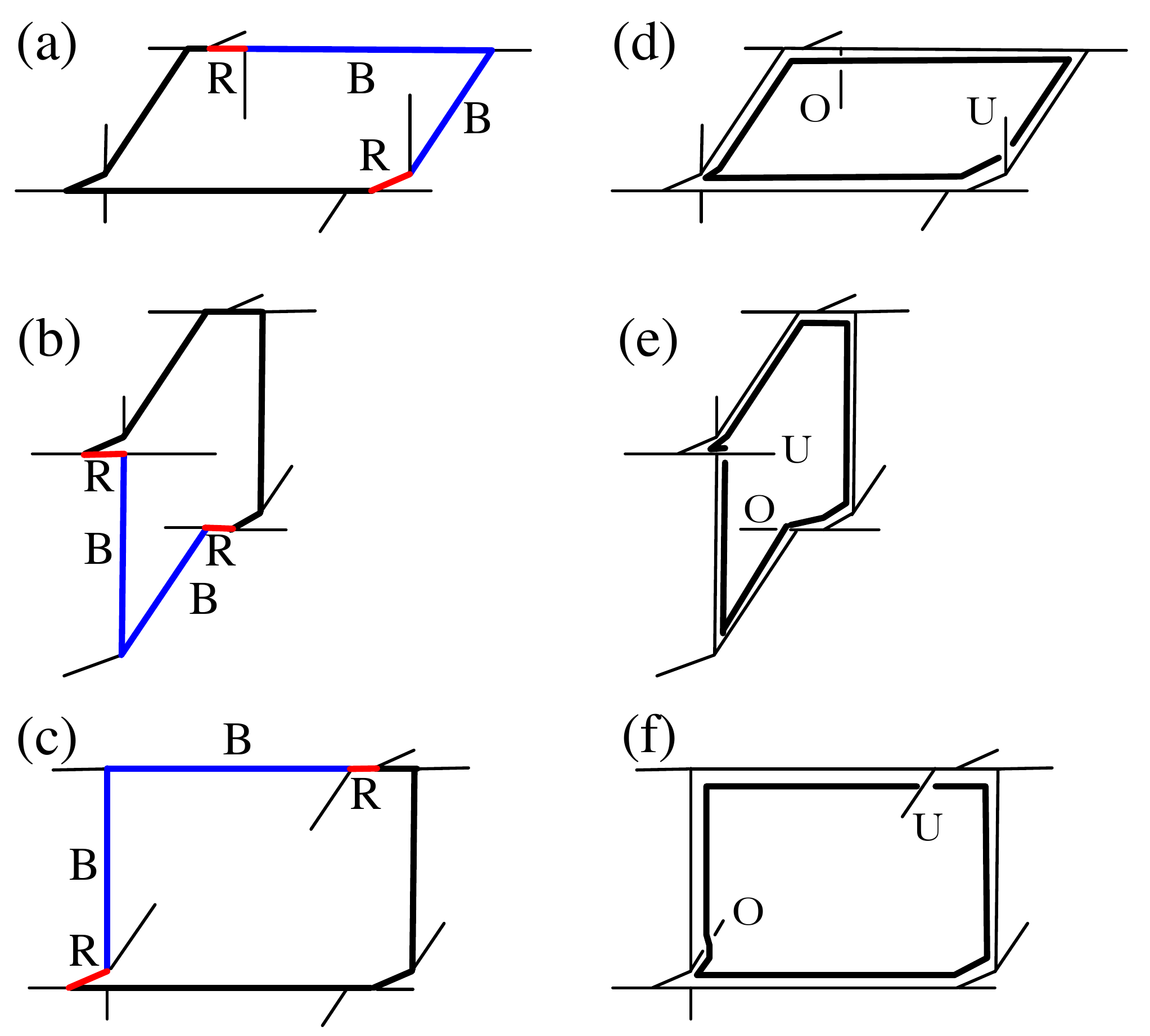}
 \caption{(Color online) (a)-(c) show the three different types of plaquette occurring on the lattice. The edges have been made either black, bold black, bold blue (labelled B) or bold red (labelled R) to aid in the definition of the 3D semion plaquette operator in \eqnref{DSemplaq}. The set $\partial p$ contains the ten bold edges, while the set $s(p)$ contains the ten black edges. (d)-(f) show how to define $B_p$ operators in terms of a string picture, where it is understood that the string is fused into the edges using \figref{tcdsem}(d) and (e); notice that the string under-crosses the edge labelled U, but over-crosses the edge labelled O.}
\label{fig:3DSemplaquettes}
 \end{figure}

We now define the `3D semion model', our first example of a Walker-Wang model which has not previously been examined in the literature. To define the operators in this model, we first fix a projection of the trivalent lattice onto 2D, as shown in \figref{fig:3Dlattice}.  (Note that we will always assume the lattice is defined on an orientable manifold.)  The Hamiltonian takes the form

\ba\label{3DSemplaq}
H &= &- \sum_v \underbrace{ \prod_{s\db{v}} \sigma^z_i }_{B_v}\\
&& + \sum_p \underbrace{  (\prod_{i\in \partial p} \sigma^x_i )   (\prod_{j\in s(p)} i^{n_j}  ) \,\, i^{\sum_{j \,\text{red }} n_{\text{j}}-\sum_{j\,blue} n_{\text{j}}} }_{B_p} \punc{,} \nonumber
\ea
where $n=\frac{1}{2}\db{1-\sigma^z}$ and, as for the 3D toric code, $s(v)$ is the set of three legs attached to vertex $v$, $\partial p$ is the set of ten edges bounding plaquette $p$, and it includes the two privileged blue and two privileged red edges used in the definition of $B_p$ above, while $s(p)$ is the set of ten edges radiating from plaquette $p$ (see \figref{fig:3DSemplaquettes}(a)-(c)).

As in 2D, for states $|\Psi_{loop} \rangle$ obeying $B_v|\Psi_{loop} \rangle=|\Psi_{loop} \rangle$ (so-called `loop gas states'), we can show that
\be
\left[B_p, B_v \right]|\Psi_{loop} \rangle =\left[B_p, B_{p'} \right]|\Psi_{loop} \rangle=0 \punc{,}
\ee
and furthermore that $B_p$ has eigenvalues $\pm1$. Thus, if we restrict our Hilbert space to the loop gas states, the Hamiltonian is a sum of commuting operators. Once again, this fact will allow us to solve exactly for the ground state of the Hamiltonian, because we will find that the ground state is a loop gas state. We can also define $B_p$ by using the graphical rules: Thread a string around the boundary of plaquette $p$ as shown in \figref{fig:3DSemplaquettes}(d)-(f), and fuse it into the edges using the graphical rules (\figref{tcdsem}(d) and (e)). The string will have a choice of under-crossing or over-crossing two edges in $s(p)$; these are the U and O edges marked in \figref{fig:3DSemplaquettes}(d)-(e). In the Walker-Wang prescription the plaquette string will under-cross the U edge, and over-cross the O edge, a choice which ensures adjacent plaquettes commute. Note that in 3D we need to use the crossing rule \figref{tcdsem}(e) in order to define the plaquette operators graphically, while in 2D we only needed this rule to define the string operators.

 Unlike the Toric code Hamiltonian \eqref{HTC3D}, the complex phase in \eqref{3DSemplaq} ensure that this Hamiltonian explicitly breaks time-reversal symmetry.   However, it is natural to define the parity operator such that it also acts by complex conjugation on $H$ (which corresponds to choosing an opposite-parity projection of the lattice).  With this definition $PT$ is a symmetry of the semion model.

 \begin{figure*}
	\includegraphics[width=1.\linewidth]{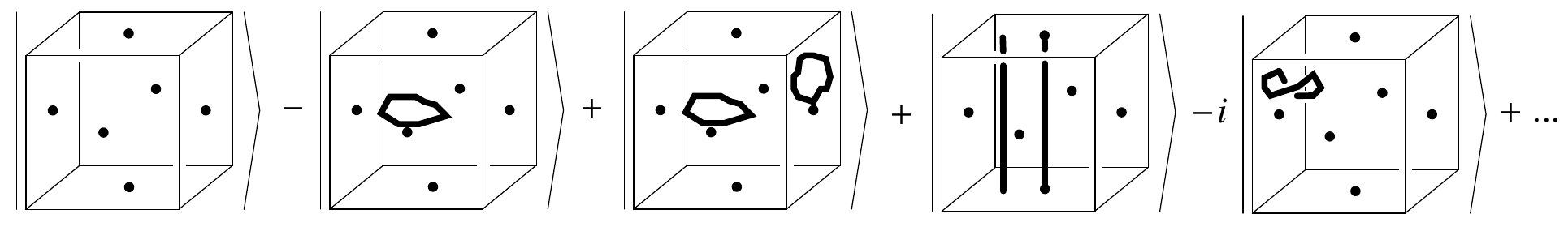}
	\caption{This figure shows the unique ground state of the 3D semion model on the 3-torus. It is a superposition of all configurations with all three parities $P_{x_\perp}=P_{y_\perp}=P_{z_\perp}=1$. To calculate the coefficient of any ket use \figref{tcdsem}. For example, a ket with a single loop has coefficient $-1$ using \figref{tcdsem}(c). One can show that the ket with the twisted loop has coefficient $-i$ by using \figref{tcdsem}(e) then (c).}
	\label{fig:3DSemgs}
\end{figure*}

\begin{figure}
	\includegraphics[width=.95\linewidth]{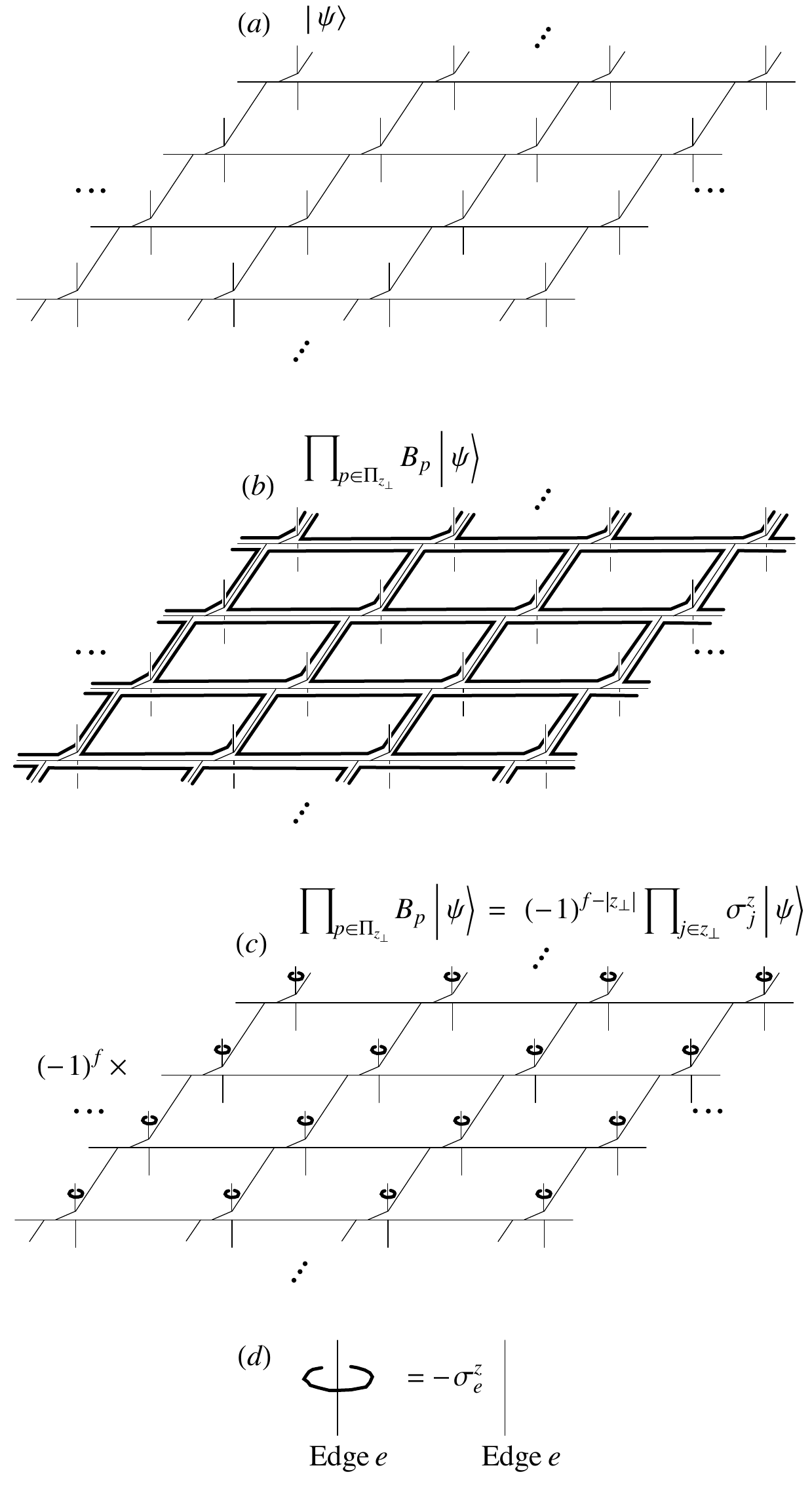}
\caption{This figure shows the graphical procedure for evaluating the operator $\prod_{p\in \Pi_{z_\perp}} B_p$ on the 3-torus. $\Pi_{z_\perp}$ denotes the set of plaquettes in the plane perpendicular to the $\hat{z}$ direction, while $z_{\perp}$ represents a set of edges encircled by loops in (c); $\left| \Pi_{z_\perp}\right|$ and $\left|z_{\perp}\right|$ represents the sizes of these sets. To begin, we evaluate (b) by fusing each neighboring string. This yields the picture in (c), where loops are left encircling edges coming up out of the plane as well as a factor of $(-1)^f$ for all the fusions. The loops encircling the vertical legs have the same action as $-\sigma^z$, as shown in (d). This gives $\prod_{p\in \Pi_{z_\perp}} B_p=(-1)^{f-\left| z_\perp \right|}\prod_{j\in z_\perp} \sigma^z_j$. Careful counting shows $(-1)^{f-\left|z_\perp \right|} = (-1)^{\left| \Pi_{z_\perp}\right|}$, which gives \eqnref{P_{xy}}.}
\label{3DSemsheet}
 \end{figure}

  		\subsubsection{Ground state of 3D semion model on $\mathbb{T}^3$}
We first contrast the topological order in 3DSem with that of the 3D toric code by comparing the ground states of the two models. The 3D toric code has a ground state degeneracy of $2^3$ on the 3-torus, where each ground state is labelled by three parity eigenvalues $P_{n_\perp} = \pm 1$ with $n=x,y,z$. We will find that 3DSem has precisely one ground state on the 3-torus, with all three parity values $P_{n_\perp} = + 1$.

In analogy with the 2D DSem model, the 3D semion ground state is defined by the conditions $B_v=1$ and $B_p=-1$ for all vertices and plaquettes. The vertex condition implies that the ground state is a loop gas. In the 2D DSem model, the condition $B_p=-1$ implied that the rules in \figref{tcdsem} (b)-(d) related the coefficients of spin configurations.
In the 3D semion model, something similar happens: the rules \figref{tcdsem} (b)-(d), as well as (e) relate the amplitudes of different configurations in the ground state superposition. To see the rule \figref{tcdsem}(e) in action, note the equation

\be\label{rulee}
\mathord{\includegraphics[height=14ex]{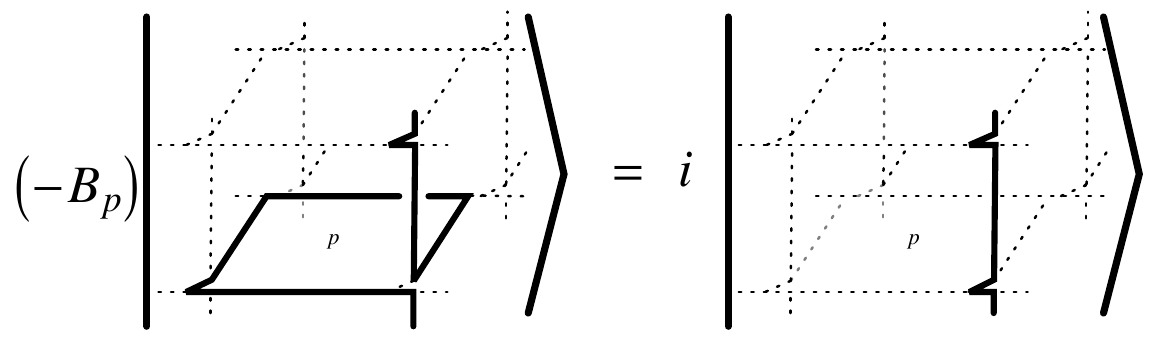}}\punc{,}
\ee
along with the condition $B_p=-1$, implies that the ground state amplitudes of these two configurations must be related\cite{LRamplitude} by a factor of $-i$. This is in accordance with the graphical rules, which tell us that

\be
\mathord{\includegraphics[height=14ex]{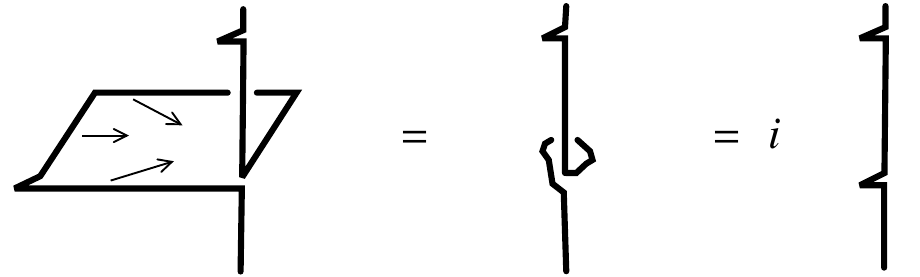}}\punc{.}
\ee

In other words, a twist in a loop segment (like that on the left hand side of \eqnref{rulee}) should be associated with a phase of $-i$ relative to a straight loop segment (such as the one on the right of \eqnref{rulee}). With the aid of these graphical rules, we show that the 3D semion model has a non-degenerate ground state on the 3-torus. In the case of the toric code, there were precisely $2^3$ ground states; each ground state was labelled by the three parity eigenvalues ($P_{n_\perp}$, where $n=x$, $y$ or $z$), and was formed by making an equal amplitude superposition of all configurations related to a canonical ket (with the corresponding parity) by the graphical rules. In the case of the 3D semion model, we can again use local rules to relate any loop configuration to one of the eight canonical configurations in \figref{3torus}, but this only proves that there is a ground state degeneracy of at most eight. In fact, there is only a single ground state, which has all $P_{n_\perp}=1$; it is a superposition of all loop configurations related to \figref{3torus}(a), with relative phases given by the rules \figref{tcdsem}(b)-(e) as shown in \figref{fig:3DSemgs}.

What happens to the other configurations, which have some $P_{n_\perp}=-1$? None of these configurations occur in a ground state superposition because they are all excited: $P_{n_\perp}=-1$ implies the existence of plaquette defects in the plane perpendicular to the $\hat{n}$-direction. This follows immediately from the identity

\be\label{P_{xy}}
P_{n_\perp}=\prod_{p \in \Pi_{n_\perp}}\left(-B_{p}\right)\punc{,}
\ee
which we prove graphically in \figref{3DSemsheet}, where $\Pi_{n_\perp}$ is the set of plaquettes lying in a plane perpendicular to the $\hat{n}$ direction. We see that any negative parity $P_{n_\perp}=-1$ is incompatible with the ground state conditions that $B_p=-1$ for all $p$.
%
Technically we still need to prove that the ket \figref{3torus}(a) has an overlap with the ground state. This can be shown by explicitly expanding the ground state projector $\proj$, as we do in \appref{trace}.

Although we have worked specifically with the 3-torus, the above method appears to generalize to any (orientable) manifold without boundary. Hence, as promised, we have shown that the ground state of 3DSem is unique on any (orientable) manifold without boundary -- unlike both the 3D Toric code, and the 2D semion model.

		\subsubsection{Bulk excitations in the 3D semion model}\label{sss:SemT^3Bulk}
A second important characterization of topological order is the identification of the low lying excitations and their statistics. In the 3D toric code, the low lying excitations are vertex string operators which produce deconfined point defects, and sheet operators which produce a line of plaquette defects at their boundary. The 3D semion model is very different because its point defects are confined: We will show that any attempt to separate a pair of point defects leads to the creation of a line of plaquette violations $B_p=1$ connecting them.

First we note that, unlike in the 3D toric code, flipping spins along a path using $\prod_{i\in \mathcal{C}_{AB}} \sigma^x$ does not produce deconfined point excitations. This operator fails to commute with the $B_p$ along its length due to the $\sigma^z$ -dependent phase terms in \eqnref{3DSemplaq}. The same kind of issue arose in 2D in the DSem model (\secref{sss:DSemExcitations}), where we found that the string operators producing deconfined point defects were not simply strings of $\sigma^x$ operators; in addition to flipping the spins along a path, the DSem string operators included compensating phases (depending on the states of edges touching the path) which ensured they commuted with the plaquettes along their length. However, in 3D there is no such assignment of phases, a fact we will prove in the next section \secref{sss:Semconf}. For now, we will demonstrate how a particularly important type of string operator fails to commute with plaquettes along its length.

\begin{figure}
 \includegraphics[width=.95\linewidth]{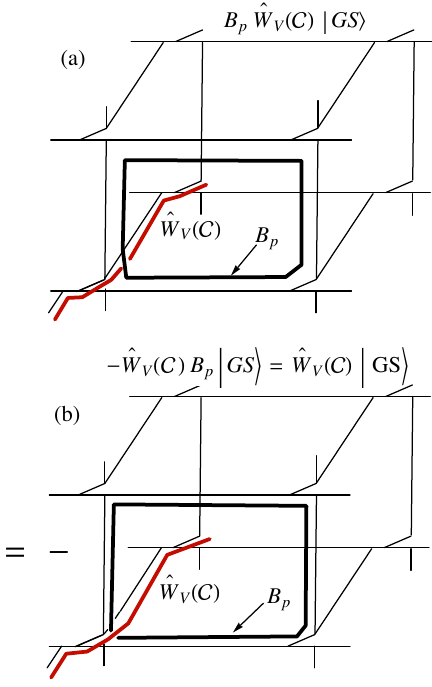}
 \caption{(Color online) This diagram shows how a vertex type string operator $\hat{W}_V(\maC)$ in the 3D semion model violates any plaquette $p$ it threads. Using the graphical rule in \figref{tcdsem}(e) shows that the state $B_p \hat{W}_V(\maC)\ket{\text{GS}}$ in (a) is equal to the state $- \hat{W}_V(\maC)B_p\ket{\text{GS}}=\hat{W}_V(\maC)\ket{\text{GS}}$ in (b). Hence $B_p=+1$ on the state $\hat{W}_V(\maC)\ket{\text{GS}}$, and so the string operator has excited the plaquette $p$ out of its low energy state $B_p=-1$ to an excited state $B_p=+1$.}
\label{fig:3DSemtcconf}
 \end{figure}

In the 2D DSem model there was a neat graphical representation for the action of string operators (\secref{sss:DSemExcitations}): lay a string along a path either over or under the lattice, and use the graphical rules \figref{tcdsem}(b)-(e) to fuse it into the edges. Now attempt to use this prescription in 3D: lay a string along a path $\maC$ connecting two defects, taking note of which edges the path over and under-crosses, and fuse it into edges using the rules \figref{tcdsem}(b)-(e) to form an operator $\hat{W}_V(\maC)$. In addition to flipping spins along the path, the resulting operator includes phases which ensure it commutes with many of the plaquettes touching $\maC$. However, if the path $\maC$ threads a plaquette $p$, as in \figref{fig:3DSemtcconf}(a), then $\hat{W}_V(\maC)$ flips the eigenvalue from the low energy state $B_p=-1$ to an excited state $B_p=+1$; we prove this statement in the caption to \figref{fig:3DSemtcconf}. Thus, attempting to create a pair of point defects with $\hat{W}_V(\maC)$ leads to a line of plaquette defects.

We might have expected from the outset that point particles are confined in the 3D semion model. In the 2D DSem model, point defects had relative semionic statistics: they were neither fermions nor bosons. The 3D semion model is built on the same semionic graphical rules, so we might have expected point defects in 3D to behave like semions. However, in 3D, {\it free} point-like particles are either fermions or bosons. The model complies with this requirement by confining the pairs of semionic point particles. Intriguingly, we will later find in \secref{sec:DsemD^2xS^1} that deconfined semionic defects do exist in these 3D lattice models, with the caveat that they are pinned to a boundary.

		\subsubsection{Connection between confinement and non-degeneracy}\label{sss:Semconf}

We saw above that the toric code has degeneracy $2^3$ on the 3-torus. This degeneracy is associated with the existence of de-confined point excitations in the following way. If the point defects are de-confined this means they are associated with the endpoints of vertex type string operators which commute with the Hamiltonian, except at their endpoints. Therefore closed string operators that wrap around the non-contractible cycles of the 3-torus have no energy cost. It is easy see how these operators toggle between the different sectors of the ground state. For example, acting on the ground state in \figref{tc3Dgs}(a) with the operator $\prod_{j\in\mathcal{C}_z}\sigma^x_j$ yields the ground state \figref{tc3Dgs}(b), where $\mathcal{C}_z$ is any path that goes once around the $z$-cycle of the torus, and none of the other cycles. This should be contrasted with the 3D semion model, which has a single ground state on the 3-torus. In this unique ground state (\figref{fig:3DSemgs}(a)), all parities are trivial ($P_{n_\perp}=1$). This observation proves our previous suggestion that point defects are confined in the bulk of the 3D semion model: if they were not confined, we would be able to form a closed string operator with no energy cost, and use it to thread a single string around a non-contractible cycle in the torus, forming another distinct ground state with non-trivial parity. This would contradict the fact that there is only one ground state.

Unlike the 3D Toric code, then, the 3D semion model has no topological order in the conventional sense.  Commensurate with this, its bulk entanglement entropy also vanishes, as we show in Appendix \ref{EntApp}.

Having established that the 3D semion model is topologically trivial on manifolds without boundary, we now turn to study the model on a manifold with boundary, where we will find that its physics is markedly different.

\section{3D topological lattice models on manifolds with boundary}\label{sec:DsemD^2xS^1}
In the previous section we investigated the 3D semion model on manifolds without boundary; all excitations were confined and we connected this with the uniqueness of the ground state. In this section we investigate the 3D semion model on manifolds with boundary, finding that the presence of the boundary leads to important qualitative changes in the topological order: There exist deconfined chiral anyonic excitations on the surface of the manifold, and these are associated with the existence of multiple ground states. In contrast, the presence of the surface will have little qualitative effect on the topological order in the 3D toric code; in particular, there are no new excitations. This section is organized as follows. In \secref{ss:3DtcD^2xS^1} we define the boundary conditions for the toric code on the solid donut, and then describe its ground states and excitations. Then in \secref{ss:3DSemD^2xS^1} we define the analogous boundary conditions for the 3D semion model, and show that there are two ground states which we tie to the novel deconfined anyonic surface excitations.

	\begin{figure}
\includegraphics[width=1\linewidth]{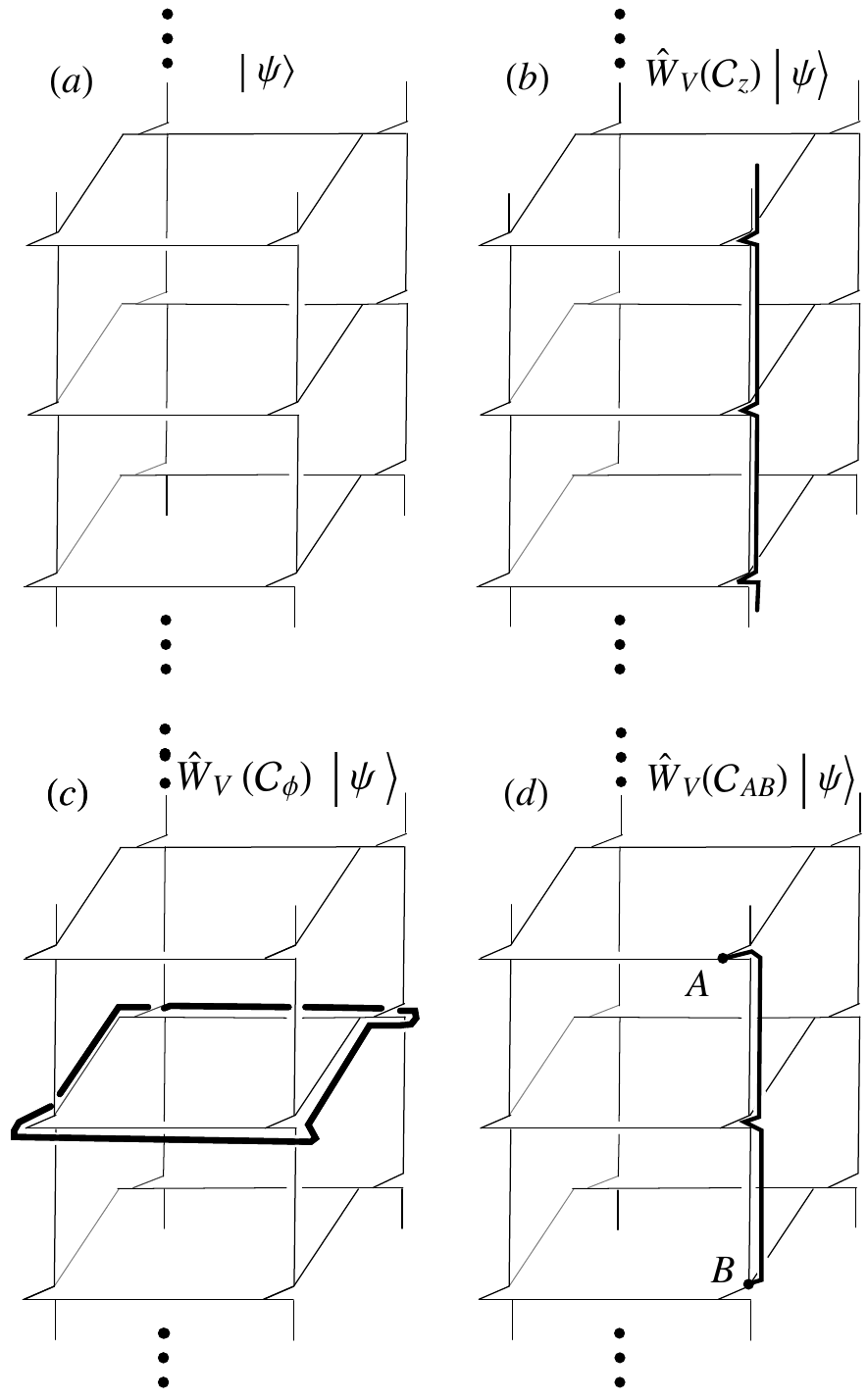}
\caption{(a) Shows a lattice representing a solid donut $D^2 \times S^1$ where periodic boundary conditions are enforced in the $z$-direction. Following that, (b)-(d) show how to define various string operators by laying a string over the lattice and fusing it into the edges using graphical rules. (b) Shows the action of a closed non-contractible string operator $\hat{W}(\maC_{z})$ that toggles between the two ground states, which are distinguished by whether an odd or even amount of flux goes around the $z$-cycle. (c) Defines the operator $\hat{W}(\maC_{\phi})$, which also commutes with the Hamiltonian but anti-commutes with $\hat{W}(\maC_{z})$. (d) Shows how to define deconfined surface excitations at $A$ and $B$ using an open string operator $\hat{W}(\maC_{AB})$. }
\label{soliddonutdsem}
 \end{figure}

\begin{figure}
\includegraphics[width=1\linewidth]{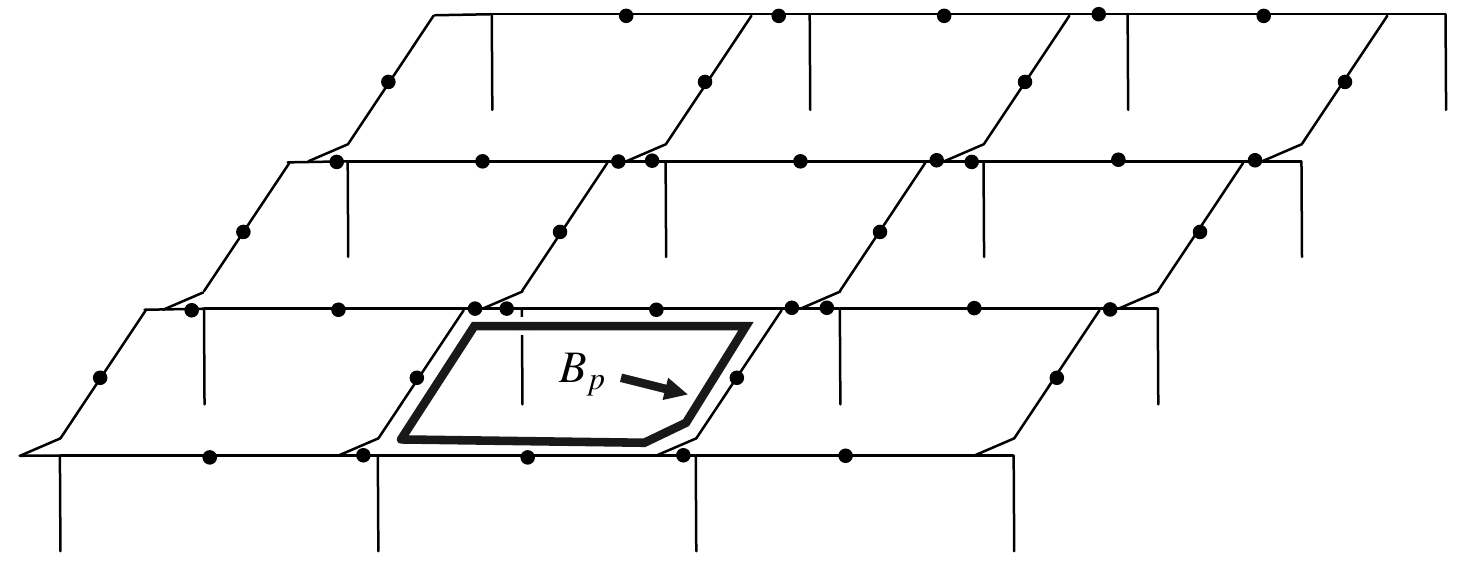}
\caption{This figure shows an example of the surface of a lattice that has been `smoothly' cut-off. The boundary Hamiltonians of both toric code and 3D semion models will include vertex and plaquette projectors for every (trivalent) vertex and plaquette in the surface.}
\label{boundaryconditions}
 \end{figure}

 The choices of boundary conditions in 3D are similar to those in 2D, which have been discussed at length in Refs.~\onlinecite{Bravyi98,Kitaev11}. In this section we focus on one of the 3D analogues of this, the `smooth' boundary shown in \figref{boundaryconditions}.  (We will also briefly discuss the rough boundary condition, which does not admit an exactly solvable surface Hamiltonian, in \appref{SurfaceOps}.)  We will  show that adding a chiral 2D anyon layer to this boundary can destroy the topological order associated with the surface.

\subsection{The toric code on the solid donut}\label{ss:3DtcD^2xS^1}\label{soliddonuttc}
In this section we examine the toric code on the solid donut manifold $D^2\times S^1$, where $D^2$ is a 2D disk and $S^1$ is a circle. The choices of boundary conditions in 3D are similar to those in 2D, which have been discussed at length in Refs.~[\onlinecite{Bravyi98,Kitaev11}]. Here we will consider one of the 3D analogues of this, the `smooth' boundary shown for a small (one plaquette thick) version of the solid donut in \figref{soliddonutdsem}, and more generally in \figref{boundaryconditions}. The toric code Hamiltonian is again defined by \eqnref{HTC3D}, where $v$ runs over all trivalent vertices, and $p$ runs over all plaquettes, including those plaquettes and vertices lying on the surface of the manifold. These boundary conditions gap the surface to vertex and plaquette defects, which will ensure that the ground state degeneracy is independent of the number of sites on the lattice.

The ground state is defined by $B_v=1$ and $B_p=1$. This again implies that the ground state is a loop gas state, and that we can relate different configurations using the graphical rules \figref{tcdsem}(a)-(d) as before. In the case of the solid donut, there are two `canonical' configurations to which all other loop gas states can be deformed, each distinguished by the winding parity $P_{z_{\perp}}$. There are $2$ ground states labelled by $P_{z_{\perp}}=\pm 1$, and each is an equal sum superposition of all loop configurations related by graphical rules to the corresponding canonical configuration. The ground state degeneracy is $2^{b_1}$ on more general manifolds with these boundary conditions, where $b_1$ is the number of independent non-contractible cycles.

		\subsubsection{Excitations}
The excitations of the toric code on the solid donut are much the same as they were for the 3-torus. There are deconfined point particles in the bulk and on the surface, created by vertex type string operators $\hat{W}_V(\maC)=\prod_{j\in \maC} \sigma^{x}_j$. As before, we are particularly interested in string operators which are closed and non-contractible like $\hat{W}_V\db{\maC_z}$, where $\mathcal{C}_z$ is any path that wraps once around the (non-contractible) $z$-cycle of the solid donut. This operator commutes with the Hamiltonian, but flips $P_{z_\perp}$, so toggles between the two possible ground states.

As on the 3-torus, we can also use operators of the form $\hat{W}_P(\mathcal{S})=\prod_{i\in \mathcal{S}} \sigma^z_i$ to create lines of plaquette defects along the boundary $\partial \mathcal{S}$ of a surface on the dual lattice $\mathcal{S}$. An interesting feature of our chosen boundary conditions is that $\hat{W}_P(\mathcal{S})$ does not have an energy cost for parts of $\partial \mathcal{S}$ which lie on the surface. We can use this to create open lines of plaquette defects in the bulk, provided they end on plaquette defects on the surface of the manifold. To create a line of plaquette defects connecting surface plaquettes $p_1, p_2$ along a line $\maC'$ in the bulk, find a surface $\mathcal{S}$ which has $\maC'$ as part of its boundary. Make the rest of the boundary of $\mathcal{S}$ lie on the surface of the manifold. The resulting operator $\hat{W}_P(\mathcal{S})$ produces the line $\maC'$ of plaquette defects ending at $p_1, p_2$, but no other defects.

%
%

In summary, the presence of the surface does not change the nature of defects in the 3D Toric code; they still consist of deconfined vertex defects, and linearly confined vortex lines (although these lines can now be open provided they end on the boundary). The excitations are either deconfined vertex defect, or lines of plaquette defects. This is very different to the behavior of 3DSem model on the solid donut, which we study next.

	\subsection{The 3D semion on the solid donut}\label{ss:3DSemD^2xS^1}
Evidently, the presence of a `smooth' surface has little effect on the toric code; the ground state degeneracy still only depends on the number of independent non-contractible cycles, and we can toggle between all sectors of the ground state by using vertex string operators that lie in the bulk or on the surface. We now contrast this with the 3D semion model, using the same `smooth' cut-off of the lattice and analogous boundary conditions. On the solid donut, the 3D semion model has Hamiltonian
\begin{align}\label{3DSemplaqD^2xS^1}
H =  - \sum_v B_v + \sum_{\text{$p$ in bulk}} B_p+ \sum_{\text{$p$ on surface}} B^{\text{surf}}_p \punc{,}
\end{align}		
where $v$ runs over all trivalent vertices, including those on the surface. For those plaquettes $p$ with both $\partial p$ and $s(p)$ in the bulk, the definition of plaquette operator $B_p$ is as it was in \eqnref{3DSemplaq}. A plaquette $p$ lying near or entirely in the surface will have some edges missing from $s(p)$; the corresponding operator $B^{\text{surf}}_p$ is defined simply by omitting these edges from the original definition of $B_p$ in \eqnref{3DSemplaq}. It is easy to show that the new plaquette operators still satisfy $(B^{\text{surf}}_p)^2=1$, and therefore have eigenvalues $\pm 1$. These boundary conditions are again chosen to gap the surface to vertex and plaquette defects.

Having defined the operators of the Hamiltonian, we note that the ground state is defined by $B_p=-1$ and $B_v=1$. These conditions again imply that the ground state is a loop gas state, and we can use the graphical rules of \figref{tcdsem}(a)-(e) to relate the coefficients of different configurations. In the present case there are two `canonical' states (labelled by $P_{z_{\perp}}=\pm 1$) to which all other loop configurations can be related using the graphical rules. This suggests that the ground state degeneracy is at most two. In fact it is precisely two, and one can readily prove this using the methods of \appref{trace}.

		\subsubsection{Surface excitations in the 3D semion model on a manifold with boundary}\label{sss:3DSemsurfaceexcitations}
So far we have shown (See also \appref{trace}) that the toric code and 3DSem have degeneracy $2$ on the solid donut. However, they exhibit very different kinds of topological order, a point we clarify by comparing their excitations. In the 3D toric code, point defects are deconfined on the surface and in the bulk. Being deconfined 3D particles, they must be bosonic or fermionic; it is easy to show that they are bosonic. In this section, we show that 3DSem has deconfined vertex defects on the surface of the manifold.
%
%
These defects are chiral semions. However, we cannot have semions in 3D, only bosons or fermions! The model complies with this requirement because the bulk confinement of vertex defects (\secref{sss:SemT^3Bulk}) pins the deconfined defects to the surface. Thus they are effectively $2$D particles, and therefore are allowed to have anyonic statistics.

In 2D we found that there were two chiralities of deconfined point defects; positive chirality and negative chirality corresponded to the choice of $+ i$ or $-i$ in \eqnref{DSemstring}. The string operators creating the point defects had a neat graphical mnemonic demonstrated in \figref{stringdefn}: To act with a string operator, lay a string along the prescribed path and fuse it into the edges using the graphical rules \figref{tcdsem}. Positive chirality strings lie above the lattice, while negative ones lie below.

However, on the boundary of 3-manifolds there is only one species of deconfined vertex defect. To create a pair of these vertex defects at $A$ and $B$, lay a string just outside the surface as shown in \figref{soliddonutdsem}(b) along a path $\maC_{AB}$ connecting $A$ to $B$, and fuse it into the edges using the graphical rules. (Recall that laying the string outside the surface, as opposed to inside, corresponds to a particular choice of phases $\pm i$ when this string crosses a spin-down edge).  Such a string operator ($\hat{W}(\maC_{AB})$) has phase choices which ensure it commutes with the plaquettes along its length; graphically, this corresponds to the fact that the strings do not thread any plaquettes. In contrast, strings lying just beneath the surface necessarily produce confined point excitations because they excite the plaquettes along their length in the same way that bulk string operators do (see \figref{fig:3DSemtcconf}). This preference for one chirality over the other is a symptom of the fact that the model breaks time reversal symmetry because of the $i$ phases appearing in the Hamiltonian \eqnref{3DSemplaq}. The chiral phase choice (needed in defining the surface string operators) is such that two vertex string operators on the surface will anti-commute if their paths cross (see \secref{sss:DSemExcitations}). This indicates a sensitivity to the order in which particle pairs are created and destroyed on the surface, and can be used to show that the deconfined surface vertex defects have semionic statistics\cite{Levin05}.

These chiral surface vertex defects are deconfined because the surface strings used to create a pair of defects commute with the plaquettes along their length. Therefore if we form a non-contractible surface string $\hat{W}_V(\maC_z)$ which winds once around the $z$-cycle of the torus as shown in \figref{soliddonutdsem}(b), it will commute with the Hamiltonian and flip the sign of the parity operator $P_{z_{\perp}}$. This operator toggles between the two ground states mentioned in the previous section.


For the 3D semion model on the 3-torus, we found there was a single ground state and that any state with (for example) $P_{z_{\perp}}=-1$ is excited. This latter fact followed from \eqnref{P_{xy}}, which showed that $P_{z_{\perp}}=-1$ implies the existence of plaquette defects in the $z_\perp$ plane. What is the analogue of \eqnref{P_{xy}} on the solid donut, and how does it allow a ground state with $P_{z_{\perp}}=-1$? Using the same methods used to derive \eqnref{P_{xy}}, we can show that
\be
P_{z_\perp}=- \hat{W}(\maC_{\phi})\!\!\! \prod_{\text{$p \in z_\perp$ plane}}\!\!\!\left(-B_{p}\right)\punc{,}
\ee
where $\hat{W}(\maC_{\phi})$ is another string operator on the surface which commutes with the Hamiltonian, defined in \figref{soliddonutdsem}(c). We see immediately that $P_{z_\perp}=-1$ is not incompatible with the ground state condition $B_p=-1$, provided $\hat{W}(\maC_{\phi})=1$. Note that, when there are no plaquette defects, $P_{z_\perp}=-\hat{W}(\maC_{\phi})$, and so $\hat{W}_V(\maC_\phi)$ anti-commutes with $\hat{W}_V(\maC_z)$.

\subsubsection{Other boundary conditions}
\label{subsub:other}

We have seen that in the presence of a smooth boundary, the confined Walker-Wang models are topologically ordered, with a topological ground state degeneracy and deconfined chiral anyons on their boundaries.  An exactly solvable zero correlation length (fixed point)\cite{Levin05} model capable of describing  chiral topological order is striking: as discussed in the introduction, there is good reason to believe that such a description is only possible at the boundary of a $3$D system.

However, it is certainly possible to generate chiral anyon theories in $2$D lattice models that are not fixed point Hamiltonians.   Given this, we might ask how robust the surface topological order is to our choice of boundary conditions?     One might imagine that the surface topological order might be altered or even destroyed by adding some 2D model to the boundary, i.e., by choosing an appropriate perturbation to the surface of the $3$D system.

To completely destroy the surface topological order, let us begin with a smooth boundary, and add a Laughlin $\nu = 1/2$ quantum Hall system of opposite chirality to the surface.  We take the quantum Hall system to have a finite density of quasi-particles, which are not strongly pinned in space.  Next, we perturb the surface Hamiltonian by adding a very strong attraction between the vertex defects on the surface of the 3D model and the quasi-particles of the quantum Hall system in such a way that each quasiparticle will bind exactly one surface vertex defect (we must make the binding sufficiently strong so as to overcome the energy gap to creating vertex defects, but much less strong than the quantum Hall gap).    The key to altering the surface topological order is that these composite particles have bosonic statistics with the semionic braiding phases of the vertex defects precisely canceling the phases from the quantum Hall quasiparticle braiding.

We now introduce a small kinetic term to the Hamiltonian so as to allow the vertex defects to hop from site to site. (Such a term can indeed be written down so as to commute with all of the bulk plaquettes.   Details of this are given in Appendix D).   We therefore expect that a bose condensate of these excitations will form at the surface.  Interestingly,  this bose condensate confines the chiral anyons present in the smooth boundary, as these braid non-trivially with the condensate\cite{SlingerlandBais}. This confinement ensures that the ground-state degeneracy is lifted, leaving only the even parity ground state ($P_{z_\perp}=1$).

In summary, unlike in the 3D toric code, the 3DSem model behaves very differently on systems with and without boundaries. In closed systems it appears topologically trivial (unique ground state, confined excitations).   With the an exactly solvable ``fixed point" boundary, however, we find deconfined chiral semions living at the boundary, and a corresponding ground-state degeneracy when the boundary is topologically non-trivial.  This topological order, which arises purely at the boundary, can nonetheless be destroyed by adding a sufficiently strong perturbation to the surface.

\section{Effective field theories for the 3D toric code and 3D semion model}\label{s:DSemFT}
Having described the ground states of the 3D toric code and 3D semion models, as well as the properties of their excitations, we are ready to investigate effective field theories for the two lattice models. In \secref{bFtoric} we review $(3+1)$D abelian U(1) $bF$ theory and its connections to the Toric code\cite{Hansson04}, and show that it captures two important aspects of topological order in the 3D toric code: the ground state degeneracy, as well as the types of excitations and their statistics. Then in \secref{ss:bFbbDsem}, we show that adding a $\lev b_{\mu \nu} b_{\rho \sigma}$ term to the $bF$ action similarly captures important aspects of the topological order of the 3D semion model. In particular, as outlined in [\onlinecite{Walker11}], this field theory gives rise to a chiral Chern-Simons anyon theory on the surface of the manifold, which accurately describes the deconfined semionic vertex defects found on the surface of the lattice model in \secref{sss:3DSemsurfaceexcitations}. We then go further to show that the field theory forbids pairs of free bulk point particles by insisting that they are bound to to the ends of line-like objects; this corresponds to the fact that vertex defects in the bulk of the confined Walker-Wang lattice model are found at the ends of lines of plaquettes defects (\secref{sss:SemT^3Bulk}).

	\subsection{The $bF$ description of the toric code}\label{bFtoric}
In this section we review the well-known fact that (3+1)D abelian $bF$ theory\cite{Horowitz89} captures the topological behavior of the 3D toric code\cite{Hansson04}. $bF$ theory has two fields: A vector field $A_\mu$, and an antisymmetric rank-2 tensor field $b_{\mu \nu}$. In what follows we will work, for sake of simplicity, in 4D Euclidean space with cartesian co-ordinates, where there is no distinction between upper and lower indices because the metric takes the form $g_{\mu\nu}=\delta_{\mu\nu}$. The field theory has an action
\be\label{eq:bFaction}
S_{bF}\left[J,\Sigma\right]=\!\int \!\!d^4 x \!\left(\frac{1}{2\pi} \lev b_{\mu \nu} \partial_{\rho } A_{\sigma} \!+\! \frac{1 } {2}  b_{\mu \nu} \Sigma^{\mu \nu}\!+\! A_{\mu}J^{\mu}\right)  \punc{,}
\ee
where $J^\mu$ and $\Sigma^{\mu \nu}$ source the world-lines of point particles, and the world-sheets of line-like objects respectively. More precisely, $J^\mu$ is defined as


\be\label{eq:explicitJ}
J^{\mu}(x) = \int d\sigma \, \delta^{(4)}(x - X(\sigma)) \frac{dX^{\mu}}{d\sigma}\punc{.}
\ee
where $\{X^{\mu}(\sigma)\}$ is a particle world-line parameterized by $\sigma$. Similarly, $\Sigma^{\mu \nu}$ is defined as
%
\be\label{eq:explicitSigma}
\Sigma^{\mu \nu}(x) = \! \int d^2\sigma\, \delta^{(4)}(x - X(\sigma)) \db{\frac{dX^{\mu}}{d\sigma^1} \frac{dX^{\nu}}{d\sigma^2}-\frac{dX^{ \mu}}{d\sigma^2} \frac{dX^{\nu}}{d\sigma^1}}\punc{,}
\ee
where $\{ X^{\mu} ({\sigma^1,\sigma^2}) \}$ is an embedding of a 2D world-sheet into the 4D space, parameterized by $(\sigma^1,\,\sigma^2)$. The point particles and line-like objects of the field theory correspond respectively to the vertex defects and lines of plaquette defects in the lattice model.

We expect the field theory to give us the ground state degeneracy of the 3D toric code, as ground state degeneracy is an important topological feature of the lattice model. More precisely, the dimension of the Hilbert space of the field theory  in the absence of sources should equal the ground state degeneracy of the lattice model. Indeed, it can be shown that the Hilbert space corresponding to the $bF$ action in \eqref{eq:bFaction} has dimension $2^3$ on the 3-torus\cite{Hansson04}, in accordance with the lattice calculations in \secref{sss:TCT^3}. On a more general manifold \cite{Szabo00} $\mathcal{M}$, it is $2^{b_1\db{\mathcal{M}}}$, where $b_1\db{\mathcal{M}}$ is the number of independent non-contractible cycles in the spatial manifold. This matches our calculation of the lattice model ground state degeneracy in \appref{tcDSem3D}. In addition to giving the correct ground state degeneracy, the (3+1)D $bF$ theory also yields the correct statistics between point particles (sourced by $J^\mu$) and line defects (sourced by $\Sigma^{\mu \nu}$)\cite{Hansson04}.

			\subsection{The $bF+bb$ description of the 3D semion model}\label{ss:bFbbDsem}
Having seen that (3+1)-D $bF$ theory gives an effective description of topological order in the 3D toric code, we now seek to motivate a field theory which has been claimed\cite{Walker11} to give a similar effective description of 3DSem. Concentrating on 3DSem on the solid donut (see \secref{sec:DsemD^2xS^1}), we expect the field theory to capture two important features of topological order: the ground state degeneracy (which was $2$), and the associated deconfined semionic excitations on the surface of the manifold (which is $\mathbb{T}^2$). These are precisely the topological properties of a $\nu=1/2$ bosonic Laughlin state on $\mathbb{T}^2$. Moreover, it is well known that the topological properties of $\nu=1/2$ Laughlin state are described by a $U(1)_2$ Chern-Simons theory. Therefore, in the absence of sources in the bulk, we expect our effective (3+1)D field theory on the solid donut to reduce to a (2+1)D $U(1)_2$ Chern-Simons theory on the boundary $\mathbb{T}^2$.

In fact, in the lattice model the surface string operators ($\hat{W}(\maC_z)$ and $\hat{W}(\maC_{\phi})$) used to toggle and differentiate between the different ground state sectors have exact analogues in the field theory. Because the semion graphical rules \figref{tcdsem} are based on a $\text{U(1)}_2$ abelian Chern-Simons theory, the non- contractible string-operators on the surface obey a similar algebra to the non-contractible Wilson-line operators that determine the dimension of the ground state Hilbert space in an abelian Chern-Simons theory living on $\mathbb{T}^2$.

This suggests a (3+1)D field theory with action

\be\label{eq:FFaction}
S_{FF}\left[A\right]=\int d^4 x \, \, \left(\frac{k}{16 \pi} \lev F_{\mu \nu}F_{\rho \sigma} + A_{\mu} J^{\mu}\right)\punc{,}
\ee
with $k=2$ and $F_{\mu \nu}=\partial_{\mu} A_{\nu}-\partial_{\nu} A_{\mu}$, because it reduces to a surface Chern-Simons term using $\lev F_{\mu \nu}F_{\rho \sigma}=4\,\partial_{\mu} \db{ \lev A_{\nu} \partial_{\rho} A_{\sigma} } $ to give action

\be \label{eq:FFaction2}
S_{FF}\left[A\right]=\frac{k}{4 \pi} \underbrace{\int dt \int_{\partial \mac{M}} dS \,n_\mu  \lev A_{\nu} \partial_{\rho} A_{\sigma}}_{S_{CS}[A]} + \int d^4 x A_{\mu} J^{\mu}\punc{,}
\ee
where $dS$ is the area element on $\partial \mac{M}$, and $n_\mu$ is the outer normal. The first part of this expression is just a Chern-Simons action of the gauge field on the surface of the manifold. Heuristically, the total partition function takes the form
\begin{align}
&Z_{FF}\left[J\right]\nonumber\\
&=\int DA_S \, e^{ \frac{i k}{4\pi} S_{CS}\left[A_S\right]+i\int J_S \cdot A_S} \int DA_B \, e^{\int J_B \cdot A_B}\nonumber\\
\label{eq:deltaconstrFF}
&= \int DA_S \, e^{\frac{i k}{4\pi}S_{CS}\left[A_S\right]+i\int J_{S}\cdot A_S} \delta\left[J^\mu=0 \, \text{in bulk}\right]\punc{,}
\end{align}
where we have smoothly split $A_{\mu}$ and $J^{\mu}$ into bulk ($B$) parts and surface ($S$) parts. $J^{\mu}$ can serve as a source term for semionic excitations on the surface. However, when one tries source a particle world-line in the bulk using a $J^{\mu}$ term the field integral disappears upon integrating out the bulk degrees of freedom of $A_\mu$, as represented by the delta function constraint.

This suggests that the $S_{FF}$ action gives an incomplete description of 3DSem, because we saw in \secref{sss:SemT^3Bulk} that point defects {\it do} appear in the bulk and moreover they are attached to the end-points of line defects. It would be reassuring to see these compound `point + line' objects emerge naturally from a field theory. To this end we introduce a source $\Sigma^{\mu \nu}$ and a corresponding rank-2 antisymmetric field $b_{\mu \nu}$ to give a `$bF+bb$' action

\begin{align}\label{bf+bb}
&S_{bF+bb}\left[A,b\right]= \int d^4 x\,  \lev\db{\frac{k}{4\pi} b_{\mu \nu} \partial_{\rho } A_{\sigma} -\frac{k}{16\pi}  b_{\mu \nu}b_{\rho \sigma}}\nonumber\\
&+ \int d^4 x\left(\frac{1 } {2}  b_{\mu \nu} \Sigma^{\mu \nu} + A_{\mu}J^{\mu}\right),
\end{align}
where $J^{\mu}$ sources point particle world-lines (\eqnref{eq:explicitJ}), and $\Sigma^{\mu \nu}$ sources the world-sheets of line defects (\eqnref{eq:explicitSigma}). Upon integrating out $b_{\mu \nu}$ we are left with an effective action

\begin{align}\label{effaction}
&S^{\text{eff}}_{bF+bb}\left[A\right]=\frac{k}{4\pi} S_{CS}\left[A\right]\nonumber \\
&+\int d^4 x \left[\frac{\pi}{4 k}\lev \Sigma_{\mu \nu}\Sigma_{\rho \sigma}  + A_{\mu}\db{J^\mu + \partial_{\nu}\Sigma^{\mu \nu}}\right]\punc{,}
\end{align}
%
where $S_{CS}$ is again the Chern-Simons action on the surface of the manifold and we have assumed for now that $\Sigma$ vanishes on this surface. In the absence of $\Sigma$ sources this reduces to the $S_{ FF }$ action in \eqnref{eq:FFaction2}, therefore the Hilbert space of the field theory will be isomorphic to the ground state subspace of the 3D semion model. Furthermore, $J$ can again be used to source semions on the surface of the manifold. Leaving the $\Sigma$ sources in, we can further deduce (heuristically) that
\begin{align}
&Z\left[J,\Sigma\right]\nonumber\\
&=e^{  \frac{i \pi}{4 k}I\db{\Sigma}}\!\! \int DA_S \, e^{ \frac{i k}{4\pi} S_{CS}\left[A_S\right]+i\int J_S\cdot A_S}\! \!\int DA_B \, e^{\int \db{J+K} \cdot A_B}\nonumber\\
\label{eq:deltaconstrBF}
&=e^{  \frac{i \pi}{4 k}I\db{\Sigma} }  \int DA_S \, e^{\frac{i k}{4\pi}S_{CS}\left[A_S\right]+i\int J_S\cdot A_S} \delta\left[J=-K \, \text{in bulk}\right]\punc{,}
\end{align}
where $K^{\mu} = \partial_{\nu}\Sigma^{\mu \nu}$, and $I\db{\Sigma}=\int d^4x\lev \Sigma_{\mu \nu}\Sigma_{\rho \sigma}$. In the equation above we have smoothly split $A$ into a bulk ($B$) part and a surface ($S$) part, the only restriction on this splitting is $A_B=0$ on the boundary. The delta function constraint in Eq. \eqref{eq:deltaconstrBF} ties bulk vertex defects ($J^{\mu}$) to $K^{\mu}$. It is a general feature (see example below) of line sources that $K^{\mu}=0$ everywhere but at the ends of the line source. With this observation, the delta function constraint forces bulk vertex defects ($J^{\mu}$) to lie at the end-points of line sources (where $K^{\mu}\neq0$).

As an example, let us look at a line defect in the $z$-direction which extends from $(0,0,-a)$ to $(0,0,a)$. Using the prescription \eqnref{eq:explicitSigma}, it can be shown that the $\Sigma^{\mu \nu}$ source takes the form
\begin{align}
&\Sigma^{\mu \nu}= I(z\in[-a,a]) \,\delta^{(2)}\db{x,y}(\delta_{\mu t}\delta_{\nu z}-\delta_{\mu z}\delta_{\nu t})\\
&K^{\mu}= \delta^{(2)}\db{x,y}(\delta(z+a)-\delta(z-a))\delta_{\mu t}
 \end{align}
where $I(z\in [-a,a])=1$ when $z\in [-a,a]$ and $I(z\in[-a,a])=0$ otherwise. Note that $K^{\mu}$ is zero everywhere but at the endpoints of the line source. Therefore, from the delta function constraint in \eqnref{eq:deltaconstrBF}, point charges are found precisely at the end-points of the line source i.e. $J^{\mu}= \delta^{(2)}\db{x,y}(\delta(z-a)-\delta(z+a))\delta_{\mu t}$.

Therefore bulk point particles are found precisely at the endpoints of line-like defects where $K^{\mu} \neq 0$; the only other place we can have non-vanishing $J^\mu$ is on the boundary of the manifold. This corresponds to what we found on the lattice in \secref{ss:3DSemD^2xS^1}: point defects in the bulk always lie at the ends of line defects of plaquette violations because point defects are confined, but point defects on the surface of the manifold need not lie at the ends of line defects. We have shown that the $bF+bb$ and $FF$ field theories capture important aspects of the topological order in the lattice model, namely the ground state degeneracy and the surface anyonic excitations. However, we also showed that, unlike the $FF$ theory, the $bF+bb$ field theory can be used to source the bulk `line+point' defects characteristic of the 3DSem lattice model.

While the $FF$ and $bF + bb$ field theories can describe the topological features of the lattice model, they also contain extra gapless degrees of freedom not present in the lattice model. These extra degrees of freedom are not apparent in either field theory as both have vanishing Hamiltonians; however, introducing a Maxwell (or Yang-Mills) term in the bulk will render these degrees of freedom dynamical, gapping many of these formerly gapless degrees of freedom but leaving a gapless photon in the bulk. In order to obtain a field theoretic description that faithfully captures the low-energy (gapped in the bulk physics of the lattice model, these gapless degrees of freedom must be eliminated from the field theory\cite{QiThanks}.

\section{General Walker-Wang models}\label{s:MTCGS}

So far we have studied two of the simplest Walker-Wang models, built from the graphical rules in \figref{tcdsem}. There are many more examples of Walker-Wang models, each based on a different set of self-consistent graphical rules, taking the general form shown in \figref{gencat}; we refer to such a collection of rules as a unitary braided fusion category (or just `category' for short). In this section we introduce the general Walker-Wang models in \secref{GenWW} and find that, of the many possible models, there is a family which behaves in a similar manner to 3DSem. We call these `confined Walker-Wang models' because, like 3DSem,
all of their bulk excitations are confined. In \secref{MTCexpert2} we show that a Walker-Wang model is `confined' if and only if it is based on a modular tensor category (MTC) (which we define in \secref{MTCexpert2}).

We then go on to demonstrate that confined Walker-Wang models share many other properties with 3DSem. First we show that the confined models have a unique ground state on the 3-torus in \secref{MTCT^3}. We then show in \secref{MTCD^2xS^1} that confined models can have multiple ground states on manifolds with boundary like the solid donut, and these degenerate ground states are associated with deconfined anyonic vertex excitations which are pinned to the boundary. These results support the suggestion of Walker and Wang\cite{Walker11} that the lattice models based on certain MTC's (specifically those MTC's coming from a Chern-Simons theory) have a description in terms of $bF+bb$ theory, analogous to the $bF+bb$ description of 3DSem in \secref{ss:bFbbDsem}.

	\subsection{The general Walker-Wang Hamiltonian, and excitations}\label{GenWW}
The Hilbert spaces of the 3D toric code and 3D semion model consisted of a $2$ state system $\sigma^z=\pm1$ on each edge of the lattice shown in \figref{fig:3Dlattice}, and we represented spin configurations graphically by coloring in only the $\sigma^z=-1$ edges. The ground states were determined by using the graphical rules: \figref{tcdsem}(a) told us that the ground state involved configurations with an even number of colored edges coming into a vertex (so the ground state was a `loop gas'), while the graphical rules \figref{tcdsem}(b)-(e) provided relationships between the amplitudes of different loop gas configurations.

\begin{figure}
  \includegraphics[width=.95\linewidth]{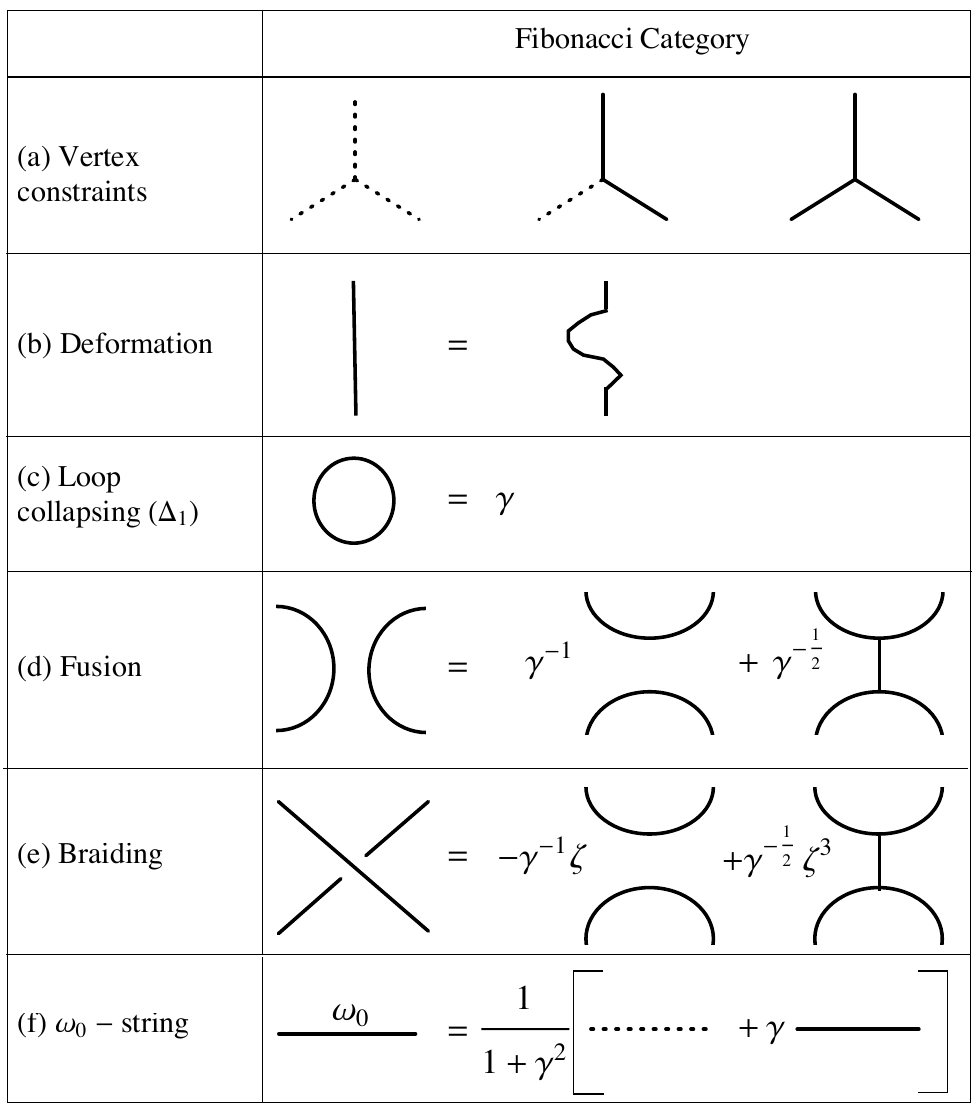}
 \caption{The figure above shows the graphical rules for the Fibonacci category, where $\gamma=(1+\sqrt{5})/2$ and $\zeta=e^{ i \pi/5}$. (a) represents the fact that the ground state of a Walker-Wang model based on this category is a superposition of configurations involving vertices with either zero, two or three edges colored in. The rules (b)-(e) give graphical relations between the amplitudes of different spin configurations on the ground state, as well as provide a graphical calculus for defining string operators. Note that there is a rule conjugate to (e) obtained by turning the over-crossing into an under-crossing on the left hand side, and sending $\zeta\rightarrow \zeta^{*}=\zeta^{-1}$ on the right hand side. The string type $\omega_0$ in (f) is used in the definitions of plaquette projection operators.}
\label{Fib}
 \end{figure}

More general Walker-Wang models are based, in an analogous way, on more complicated sets of graphical rules. For example, probably the next simplest Walker-Wang model is based on the Fibonacci category. The graphical rules for this model are given in \figref{Fib}(a)-(e). The Walker-Wang model based on this category (3DFib) still consist of a $2$ state system $\sigma^z=\pm1$ on every edge, and again we graphically represent different spin configurations by only coloring in edges with $\sigma^z=-1$. The Hamiltonian (which we define below) is set up so as to ensure the following: The ground state is a superposition of configurations obeying the modified vertex constraints \figref{Fib}(a), and the graphical rules \figref{Fib}(b)-(e) give new linear relations between the amplitudes of different kets in the ground state superposition. As for the semion rules, the Fibonacci rules \figref{Fib}(a) permit vertices with none or two of their incoming edges colored, and forbid vertices with a single colored edge. However, unlike the semion rules, the Fibonacci rules also permit vertices with all three incoming edges colored. This means that the ground state does not quite look like a loop gas -- we say that the ground state is a `string-net'\cite{Levin05} obeying \figref{Fib}(a).

Further generalizations of Walker Wang models can include Hilbert spaces where each edge is in one of $k$ possible states, as compared to the above cases where each edge is in only one of two possible states. More generally, a Walker Wang model can be defined by any `unitary braided fusion categories'\cite{Walker11}. A `unitary braided fusion category' is a set of possible edge labels along with a set of graphical rules (shown in \figref{gencat}) that satisfy certain consistency conditions\cite{Wang10,Bonderson07} which we will not enumerate here. For a category with $k$ possible labels, the Hilbert space of the corresponding WW model consists of giving a label $i \in \left\{0,\ldots,k-1\right\}$ to each edge in the system\cite{fusionmultiplicity}.

The Hamiltonian, which we introduce in the next section, is constructed to ensure two things. Firstly, it ensures that the ground state is a string-net with $k-1$ different `colored' strings with labels $\left\{1,\ldots,k-1\right\}$ obeying a more general set of vertex constraints represented by \figref{gencat}(a). Secondly, it ensures that the rules \figref{gencat}(b)-(e) give linear relations between the amplitudes of different string-net configurations in the ground state superposition.

%

		\subsubsection{Hamiltonians}
Having explained that all Walker-Wang Hamiltonians are contrived so as to ensure the ground state is described by a set of graphical rules, we now define the Hamiltonians. The details in this section are not essential to understanding the qualitative behavior of confined Walker-Wang models, which we resume studying in \secref{ss:MTCexamples}. The casual reader may therefore wish to skip directly to \secref{ss:MTCexamples}.

\begin{figure}
 \includegraphics[width=1\linewidth]{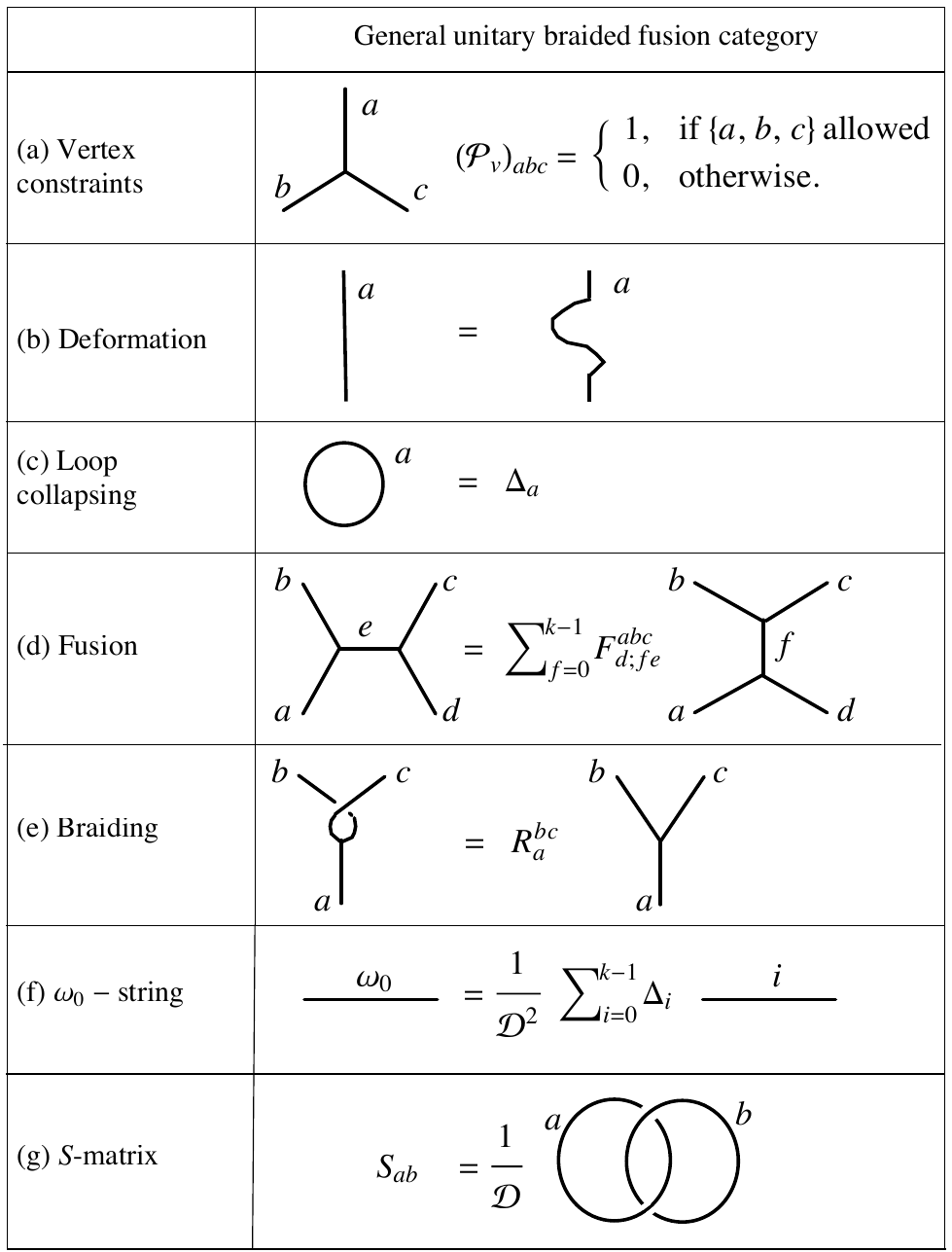}
 \caption{(a) Represents the vertices allowed by the category; the ground state of a Walker-Wang model will involve only these types of vertices. The diagrams in (b)-(d) serve two purposes. Firstly, they tell us the relative amplitudes of `string-net' configurations in the ground state e.g. row (c) tells us that configurations related by removing a closed loop carrying label $a$ occur with a relative factor of $\Delta_a$ in the ground state. Second, these diagrams provide a neat graphical mnemonic for the definitions of string operators. Note that there is a rule conjugate to (e) obtained by turning the over-crossing into an under-crossing on the left hand side, and sending $R\rightarrow R^{*}$ on the right hand side. }
\label{gencat}
 \end{figure}

The general Walker-Wang Hamiltonian takes the form:

\be\label{eq:genHam}
H_{WW} = -\sum_{v}\proj_v -\sum_{p}\proj_p
\ee
where $\proj_v,\proj_p$ are a commuting set of projectors. The ground state(s) can be solved for exactly, and has the defining property that $\proj_v = \proj_p=1$ for all vertices and plaquettes. The operator $\proj_v$ projects onto configurations where only certain combinations of edge colors come into a vertex (see \figref{gencat}(a)), and so the condition $\proj_v=1$ for all $v$ ensures that the ground state is a string-net involving only these types of configurations. The plaquette operators will be defined below, but we reiterate that they are constructed to ensure that \figref{gencat}(b)-(e) relate the amplitudes of different string-net configurations.

%

All that remains is to define the plaquette projectors generally. For explicit expressions for the $\proj_p$, we refer readers to [\onlinecite{Walker11}]; in what follows, we will define the $\proj_p$ using a graphical calculus. We saw in the 3D toric code and 3DSem models that there exists a graphical mnemonic for acting with string-operators. Exactly the same mnemonic works here. $\hat{W}^{i}_V({\maC})$ denotes an operator formed by laying down a string of type $i\in\{0, 1,\ldots,k-1\}$ along a path $\maC$ and fusing it into the edges using \figref{gencat}(b)-(e). While there are only $k-1$ colored strings in the category, it is notationally convenient to introduce an `empty' string labelled $i=0$ with the convention that it braids and fuses trivially with everything, and has $\Delta_{i=0}=1$ (see \figref{gencat}(c)). This implies that empty string operators always acts trivially on the Hilbert space i.e. $\hat{W}^{i=0}_V(\partial p)=\mathbb{I}$.

\begin{figure}
  \includegraphics[width=\linewidth]{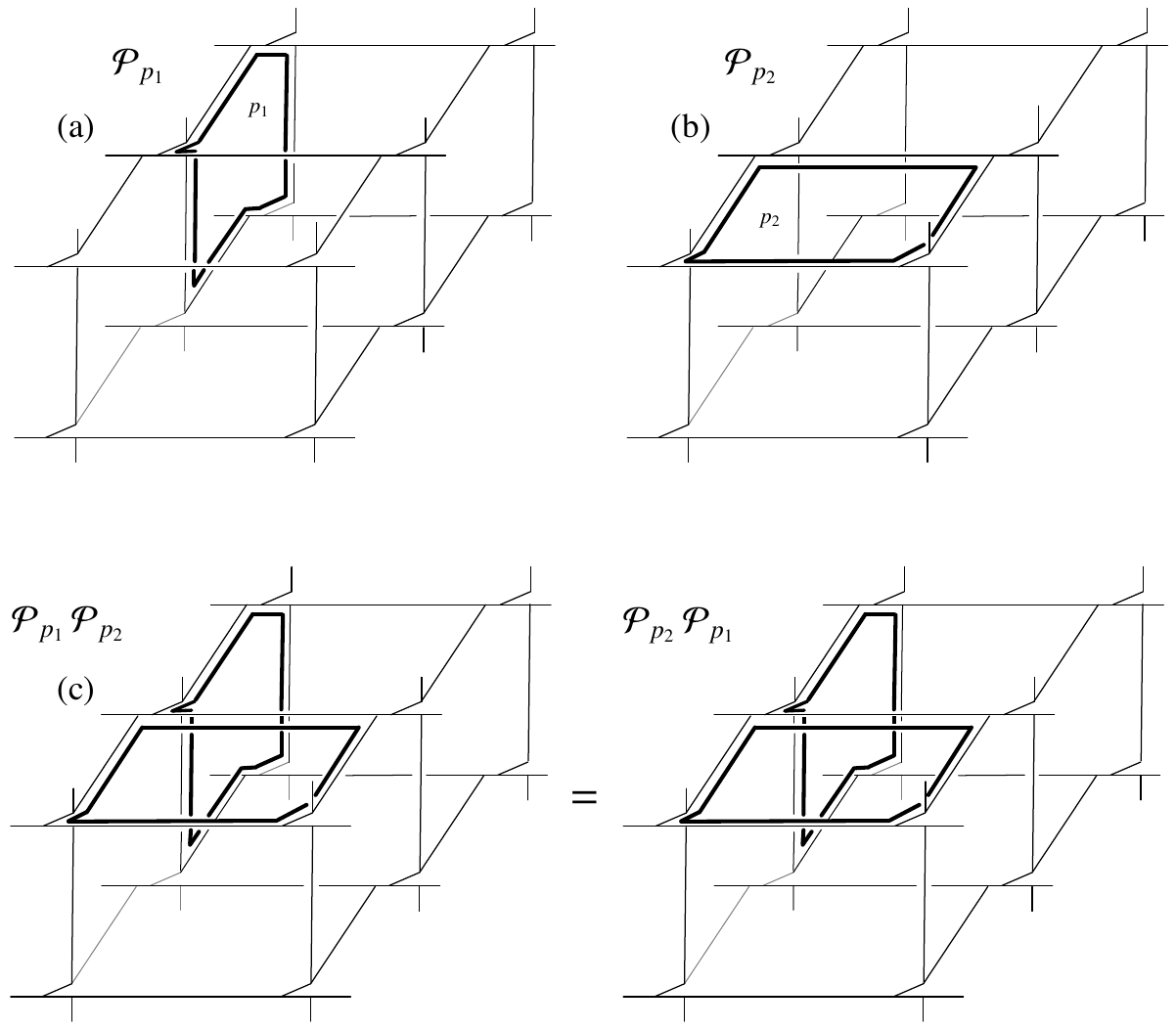}
 \caption{This figure shows the graphical definition of plaquette operators in general Walker Wang models. Having fixed a projection we define a plaquette projector by threading a $\omega_0$ string just inside the perimeter of the plaquette. Two examples are shown in (a) and (b). Then in (c) we show that these adjacent operators commute with each other because the diagrams representing $\proj_{p_1} \proj_{p_2}$ and $\proj_{p_2}\proj_{p_1}$ are identical.}
\label{WWlattice}
 \end{figure}

We now apply this formalism to define the plaquette operators. First we define plaquette operators $\hat{W}^{i}_V(\partial p)$ for each string type $i$: Using the fixed projection of the lattice onto 2D, thread a string carrying label $i$ around $\partial p$ using the under and over-crossing prescription shown in \figref{fig:3DSemplaquettes}(d)-(f) and fuse it into the edges of $p$  using the graphical rules (\figref{gencat}(b)-(e)); this prescription ensures that adjacent plaquette operators commute, as shown in \figref{WWlattice}. Now, the plaquette projector is defined by
\be\label{genplaq}
\proj_p = \frac{1}{\mathcal{D}^2}\sum^{k-1}_{i=0} \Delta_i \hat{W}^{i}_V(\partial p)
\ee
where the $\Delta_i$ 
is known as the quantum dimension of $i$ (see \figref{gencat}(c)), and $\mathcal{D}^2=\sum^{k-1}_{i=0}\Delta^2_i$. This operator will sometimes be referred to as $\hat{W}^{\omega_0}_V(\partial p)$, where $\omega_0$
 is the superposition of string types
\be
\omega_0 = \sum^{k-1}_{i=0} \Delta_i \hat{i} /\mathcal{D}^2 \ \ \ .
\ee
where $\hat i$ indicates a string of type $i$.   While the form of this $\omega_0$ operator may seem complicated, it is constructed to have one crucial property, namely the `handle-slide' property\cite{Wang10} shown and explained in \figref{Omegaloop}. It is precisely this property which ensures that \figref{gencat}(b)-(e) relate the amplitudes of different configurations in the ground state.

		\subsubsection{DSem and 3DSem in the more general language}\label{sss:gen3DSem}
Let us now try to understand 3DSem in this new language. The 3DSem category has two string types. There is the `empty' string carrying label $i=0$, which for simplicity we did not mention in \figref{tcdsem} because it braids and fuses trivially with everything. Then there is the single colored string type which has label $i=1$ and has semionic self-braiding \figref{tcdsem}(e).

The allowed vertices in 3DSem had only an even number of down spins attached to each vertex, which corresponded with $B_v=+1$, where we defined $B_v$ in \eqnref{3DSemplaq}. Therefore we should define $\proj_v=(1+B_v)/2$ so that the condition $\proj_v=1$ is equivalent to $B_v=1$, a condition which ensured that the ground state was a loop gas (see \secref{ss:SemT^3}).

According to \eqnref{genplaq} the definition of $\proj_p$ in DSem is $\proj_p= (1-\hat{W}^{1}_V(\partial p))/2$, because $\Delta_0=1,\Delta_1=-1$. But we noted in \secref{ss:SemT^3} that $B_p$ has a graphical definition which is precisely the definition of the string operator $\hat{W}^{1}_V(\partial p)$. Hence $\proj_p= (1-B_p)/2$. Plugging these relations, as well as $\proj_v=(1+B_v)/2$, into \eqnref{eq:genHam} gives back \eqnref{3DSemplaq} up to an unimportant re-scaling of the Hamiltonian (and an overall shift of the Hamiltonian by a constant).

	\subsection{Modular tensor categories and excitations}\label{MTCexpert2}

In \secref{sss:SemT^3Bulk} we found that the bulk vertex excitations of the 3DSem model were confined, and this was related to the statement that string operators have an energy cost for every plaquette they thread. For these reasons, we called 3DSem a `confined Walker-Wang model'. We now turn our attention to Walker-Wang models based on modular tensor categories (MTC's). We will show that these are precisely the Walker-Wang models which are confined, in the sense that their string operators have an energy cost for every plaquette they thread.

\begin{figure}
  \includegraphics[width=.9\linewidth]{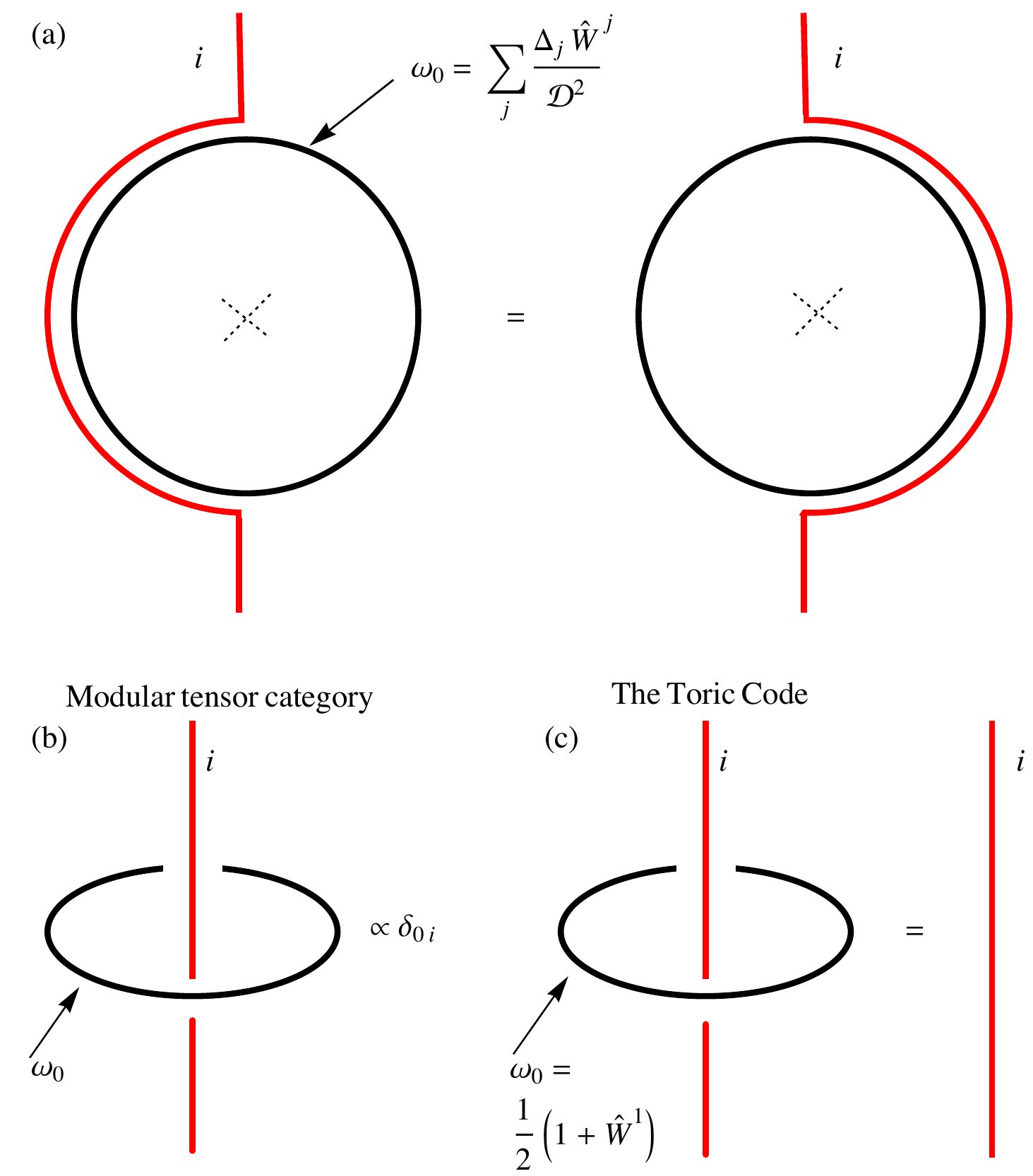}
 \caption{(Color online) (a) demonstrates the handle-slide property of the $\omega_0$ loop which appears in the definition of plaquette operator. A special property of $\omega_0$ allows us to deform other string operators (or indeed edge strings) $i$ over $\omega_0$ regardless of whether other strings thread $\omega_0$.   The X in the middle of the $\omega_0$ loop indicates that other strings may be passing through the middle of the loop. (b)  In a modular category no string (except the empty string) can pass through an $\omega_0$ loop.   However, in a  non-modular category, such as in the toric code, this is not the case. }
\label{Omegaloop}
 \end{figure}

MTC's are categories with a unitary `S-matrix'\cite{Bonderson07}. They have the special property that a loop of the string type used to define $\proj_p$ (i.e. $\omega_0=\sum_i \Delta_i  \hat{i}/\mathcal{D}^2$ in \figref{gencat}(f)) kills all flux going through it i.e using the fusion and braiding rules in \figref{gencat} to evaluate \figref{Omegaloop}(b) yields an answer of zero if the flux $i\neq0$. Examples of MTC's include the semion, Fibonacci, and Ising categories as well as $\text{SU(2)}_n$ for all $n$\cite{Bonderson07}. An important non-example is the toric code; we see in \figref{Omegaloop}(c) that a string operator $(1+\hat{W}^1_p)/2$ does not kill flux going through it.

We can define a string operator $\hat{W}^i_V(\mathcal{C})$ by threading an $i$ string along $\mathcal{C}$, and fusing it into the edges along the path. This produces vertex defects at the end-points of $\mathcal{C}$. For the 3D toric code, such string operators only produced vertex defects at their endpoints. But for 3DSem, the operator also produced plaquette violations along $\mathcal{C}$, which represented the fact that bulk vertex excitations are confined in 3DSem. We will now show that MTC based models behave like 3DSem: Acting on the ground state of any MTC based model with string operator $\hat{W}^i_V(\mathcal{C})$ will excite plaquettes $p$ threaded by the string in addition to creating vertex defects at the ends of $\maC$. Put another way, point defects are confined because they are tied to a line of plaquette defects.

The previous paragraph establishes that vertex string operators that act in accordance with the graphical rules produce confined excitations.  In \appref{App:3DGStrconf}, we show that this holds for any string operator in a confined Walker-Wang model, completing the proof that all excitations are confined in the bulk.

\begin{figure}
 \includegraphics[width=1.\linewidth]{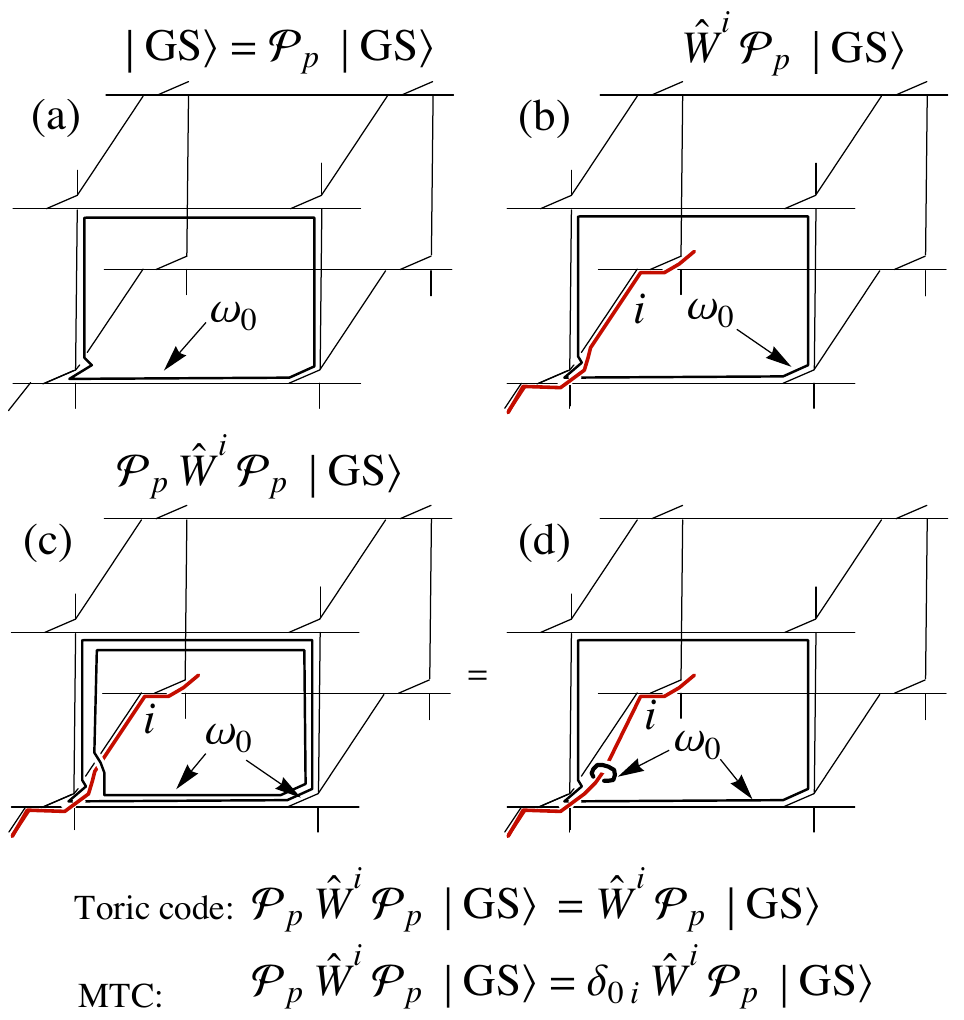}
 \caption{(Color online) This diagram shows that acting on the ground state of a lattice model with a string operator carrying label $i\neq0$ and threading plaquette $p$ will excite the plaquette to $\proj_p=0$ if the underlying category is modular, but will not excite the plaquette if the underlying category is $\mathbb{Z}_2$ (the toric code case). (a) Represents the ground state, which we have acted on with $\proj_p$ which equals $1$ on the ground state. In (b) we act with a string operator. In (b) is we act with plaquette projector $\proj_p$ to ascertain the new eigenvalue of $\proj_p$ and this forms (c). Using the handle-slide property of the plaquette projectors strings, we handle-slide the inner projector over the outer projector. The result is (d) in which a $\omega_0$-string is left encircling the $i$-string giving $\delta_{0i}$ for a MTC, which is zero because of our choice of $i$. }
\label{bulkconfinement}
 \end{figure}

Algebraically, confinement means that $\proj_p \hat{W}_V^i(\maC) \ket{\text{GS}}=0$ for $i\neq 0$,  where $p$ is any plaquette threaded by $\mac{C}$. We prove this relation graphically in \figref{bulkconfinement}(a)-(d). The crucial feature of MTC's is that loops of $\omega_0$ string kill the flux going through them (see \figref{Omegaloop}), and this key property is used to evaluate \figref{bulkconfinement}(d). This behavior should be contrasted with that seen for the toric code, which is not a modular tensor category. For the toric code we can follow the same reasoning as for a MTC in \figref{bulkconfinement}(a)-(d). The difference comes in evaluating diagram (d); in the toric code the $\omega_0$ string does not kill the $i$ flux, it gives back the same answer \figref{bulkconfinement}(a) because of the identity shown in \figref{Omegaloop}(c). Algebraically this can be expressed as $\proj_p \hat{W}^i \ket{\text{GS}}=\hat{W}^i \ket{\text{GS}}$, and this implies that the toric code string operator does not produce any plaquette violations along its length, a point we have already discussed in a simpler setting in \secref{sss:SemT^3Bulk}.

Having seen that all MTC based models are confined, we ask the reverse question: If a category is bulk confined, is it necessarily modular? The answer is yes. If a category is {\it not} modular, then it can be shown\cite{Bonderson07} that there must exist a non-vacuum particle which braids trivially with all other particles. If we write down a string operator carrying the label of such a particle, we will find that it commutes with every plaquette projector, and therefore gives rise to deconfined bulk point excitations. Therefore `not-MTC' implies `not bulk confined', or equivalently all bulk confined models must be based on MTC's. With the reasoning in the previous paragraph, this implies that a model is bulk confined if and only if it is based on a MTC.

	\subsection{Confined Walker-Wang models on various manifolds}\label{ss:MTCexamples}
We have shown that string operators in confined Walker-Wang models violate every plaquette they thread. This suggests that confined Walker-Wang models on manifolds without boundary have a unique ground state, an assertion we verify in \secref{MTCT^3} (this result is known to the mathematical community\cite{Walker11,Sikora}). We then go on to show in \secref{MTCD^2xS^1} that, like 3DSem, confined Walker-Wang models on manifolds with boundary have deconfined anyonic excitations pinned to the boundaries, and that these deconfined excitations are associated with a ground state degeneracy.
		
	\subsubsection{The 3-torus}\label{MTCT^3}
One consequence of bulk confinement is that there are no non-contractible string operators in the bulk that commute with the Hamiltonian of an MTC based model. This naively suggests that the ground state is non-degenerate on manifolds without boundary because there are no operators which commute with the Hamiltonian and toggle between different ground states. We have already written out an argument specific to the 3D semion model, proving that it has non-degenerate ground states on the 3-torus. In \appref{App:3DGSnondeg} we present a general proof that the ground state is non-degenerate on $\mathbb{T}^3$ for {\it any} modular tensor category. For MTC Walker-Wang models it appears as though the interesting physics occurs on the surface, and not in the bulk. This motivates the study of a simple manifold with boundary.

		\subsubsection{The solid donut $D^2 \times S^1$}\label{MTCD^2xS^1}
		 \begin{figure}
 \includegraphics[width=1\linewidth]{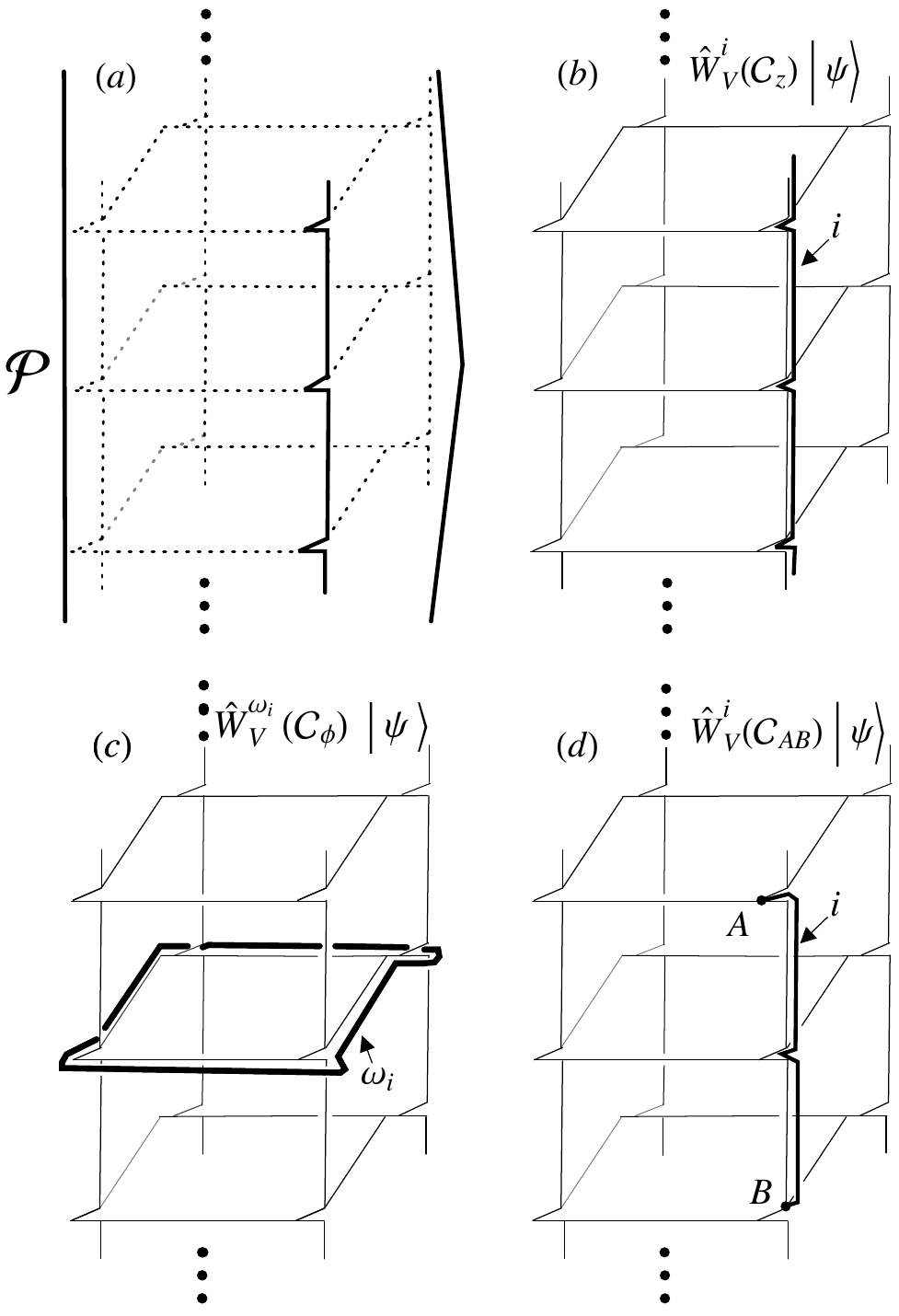}
 \caption{(a) Shows the $k$ independent ground states $\ket{GS_i}$ where $i\in\left\{0,1,2,\ldots,k-1 \right\}$ and $\proj$ is the ground state projector. The dotted lines are edges in state $0$ while the bold black edges are in states corresponding to the $i$-label. (b) Defines a string operator $W^i_V\db{\maC_z}$ winding around the non-contractible $z$-cycle of the solid donut. This operator toggles can be used to toggle between ground states e.g. $W^i_V\db{\maC_z}\ket{GS_0}=\ket{GS_i}$. (c) Defines yet another operator $W^{\omega_i}_V\db{\maC_\phi}$ which can be used to distinguish between the different ground states. This operator involves a string type $\omega_i$ with a special superposition of labels $\omega_i = \frac{\Delta_i}{\mathcal{D}} \sum_h S^*_{h i} \hat{h}$ where $S$ is the S-matrix of the underlying category. In a MTC, this operator has the property $W^{\omega_i}_V\db{\maC_\phi}\ket{GS_j}=\delta_{i j }\ket{GS_j}$. (d) Defines an open string operator, creating defects at $A$ and $B$ but otherwise commuting with the Hamiltonian along its length.}
 \label{soliddonut}
 \end{figure}
We now put a MTC based model on a solid donut choosing the `smooth' boundary conditions shown in \figref{soliddonut}. These are completely analogous to the smooth boundary conditions described for the 3D toric code and 3DSem model in \secref{sec:DsemD^2xS^1}, and are similarly constructed to gap the surface to vertex and plaquette excitations.

As in the case of 3DSem on the donut, we can create pairs of deconfined anyons on the surface of the manifold by using graphically defined operators of the form $\hat{W}^{i}_V \db{C_{AB}}$, where $\mathcal{C}_{AB}$ is a path lying just above the surface of the manifold connecting points $A$ to $B$ and $i\in\left\{1,\ldots,k-1\right\}$ is any label in the category. The anyons are deconfined because the path of the string does not thread any plaquettes (see \figref{soliddonut}(d)).

For such an MTC with $k-1$ non-trivial string types, the ground state degeneracy for the corresponding Walker-Wang model on the solid donut is $k$  -- exactly that of the chiral anyon theory living on its surface. To see this, note that any ket without vertex violations can be related, using graphical rules \figref{gencat}, to a combination of the $k$ canonical kets shown in \figref{soliddonut}(b). The ground states labelled by these canonical kets are mutually orthogonal; we can verify this by showing that they have distinct eigenvalues under flux measurement, which involved acting with a closed string operator of the form $\hat{W}^{\omega_i}_V \db{C_{\phi}}$ defined in \figref{soliddonut}(d). Hence there are precisely $k$ distinct ground states, and we can toggle between them using closed string operators of the form $\hat{W}^{i}_V \db{C_{z}}$ (see \figref{soliddonut}(c)). This is consistent with the degeneracy of $2$ found for 3DSem in \secref{sec:DsemD^2xS^1}. As a further example, an MTC based on $\text{SU(2)}_n$ has a degeneracy of $n+1$. We note again that the ground state degeneracy is tied to the existence of surface anyon excitations, a point which motivates the field theoretic description in the next section.

%

			\subsection{The $bF+bb$ description of MTC's}\label{ss:bFbbMTC}	 
We saw in \secref{ss:bFbbDsem} that a level $k=2$ U(1) $bF+bb$ theory captured the ground state degeneracy and types of defects of the 3DSem model. We now turn our attention to the more general statement\cite{Walker11}: If a Walker-Wang model is based on a MTC which looks like a Chern-Simons anyon theory, for example $\text{SU}(2)_n$\cite{Bonderson07}, then the resulting lattice model has an effective description as a $bF+bb$ field theory with gauge group $\text{SU}(2)$ and level $n$.

For concreteness, let us fix our attention on the modular Walker-Wang models based on $\text{SU(2)}_n$. From our work in \secref{ss:MTCexamples}, we see that these models have a ground state degeneracy of $n+1$ on the solid donut, which is precisely the dimension of the Hilbert space of a non-abelian $\text{SU(2)}_n$ Chern-Simons theory. Furthermore, these models have deconfined anyons on their surface, which obey the statistics found in the $\text{SU(2)}_n$ Chern-Simons theory. This information so far suggests the non-abelian effective field theory

\be\label{eq:FFactionNonab}
S_{FF}\left[A\right]=\int d^4 x \, \, \tr \left(\frac{n}{16 \pi} \lev F_{\mu \nu}F_{\rho \sigma}\right)\punc{,}
\ee
which reduces to the correct surface Chern-Simons theory. Above we have assumed that $A$ lives in the fundamental representation of the gauge group, and $F$ is the corresponding non-abelian field strength. Creating surface nonabelions corresponds to introducing path-ordered\cite{Peskin95} Wilson lines of the form
\be
\tr_{\mathcal{R}} \db{e^{i \int d^4 x A_\mu J^{\mu}}}
\ee
where $\mathcal{R}$ is a representation of the gauge group, which labels the particle type. However, using the same heuristic reasoning that led to the $\delta$-function constraint in \eqnref{eq:deltaconstrFF}, this field theory seems to forbid the presence of bulk point particles. As we saw in \secref{MTCexpert2}, bulk point particles do exist in the lattice model, but are tied to the ends of lines of plaquette defects. This again motivates the introduction of a two form $b_{\mu\nu}$ transforming in the adjoint representation to give
\begin{align}\label{bf+bbnonab}
&S_{bF+bb}\left[A,b\right]= \!\!\int d^4 x\, n \lev  \tr\db{\frac{1}{8\pi} b_{\mu \nu} F_{\rho\sigma} -\frac{1}{16\pi}  b_{\mu \nu}b_{\rho \sigma}}\nonumber\\
\end{align}
which reduces to \eqnref{eq:FFactionNonab} upon integration over $b$, so reproduces the correct ground state degeneracy and surface anyonic statistics. To source a compound `line+point' defect, one introduces Wilson line and surface operators of the form
\be
\tr_{\mathcal{R'}} \db{e^{i \int d^4 x b_{\mu \nu}\Sigma^{\mu \nu}}}\tr_{\mathcal{R}} \db{e^{i \int d^4 x A_\mu J^{\mu}}}
\ee

There are many subtleties in defining surface operators\cite{Cattaneo05} which we do not discuss here. That aside, we expect that for such a `line+point' defect to exist in the bulk, the following analogue of the delta function in \eqnref{eq:deltaconstrBF} arises: $\delta_{\mathcal{R'} \mathcal{R}}\delta\left[J=-K= -\partial_{\nu}\Sigma^{\mu \nu}\, \text{in bulk}\right]$. In other words, the point defects are bound to the ends of line defects, and both carry the same quantum label.

\section{Connection to models with surface fractional quantum Hall states} \label{QHSect}
We have seen that the 3DSem model has a surface that is topologically a bosonic Laughlin state at $1/2$-filling (i.e. described by a $k=2$ abelian Chern-Simons theory\cite{Wenbook}, as discussed in \secref{ss:bFbbDsem}). In this section we present lattice models with the surface topological order of  fermionic Laughlin states. In particular we concentrate on a model which, on its surface, has topological order resembling a $\nu=1/3$ fermionic fractional quantum Hall effect.

Though the microscopic degrees of freedom of Walker-Wang models are $k$-state spins (i.e., $k$ possible quantum numbers on each edge), their low-energy properties can be made to resemble those of fermionic systems. The resemblance that we will describe is reminiscent of that between the 2D toric code and a superconductor in two dimensions\cite{Hansson04}. Though the Toric code does not describe superconductivity per se, both systems have a ground state degeneracy of $4$ on the torus, and two types of low-lying excitations (vertex and plaquette for the Toric code, or BdG quasi-particles and $\pi$ vortices for the superconductor) which acquire a Berry phase of $-1$ when braided around one another. In other words, the 2D thin-film superconductor and the 2D toric code exhibit the same kind of topological order\cite{Hansson04}.

In this section we give examples of lattice models whose surfaces exhibit the same kind of $T$-symmetry breaking topological order as a fermionic fractional quantum Hall state. The existence of these phases is suggestive: In the same way that the toric code has the topological order of a superconductor, the lattice model in \secref{FQHE1/3} may well have the same topological order as some system of electrons. We will not discuss such electron systems in detail, but we note that the lattice model in \secref{FQHE1/3} shares some of the topological properties one might expect of a fractional topological insulator undergoing confinement\cite{Maciejko12}: namely, time-reversal symmetry is explicitly broken, leading to surface states with the topological order of a fermionic Laughlin state, and the only bulk deconfined excitation is the fermion.


This section is structured as follows. In \secref{FQHE1/2} we revisit the 3DSem model, which has a confined bulk and the topological order of a $\nu= 1/2$ bosonic fractional Hall effect on its surface. Then in \secref{FQHE1/3} we attempt to find a lattice model which has surface topological order resembling a fermionic $\nu= 1/3$ fractional Hall effect. We succeed in part: we find a lattice model with the expected deconfined anyons pinned to its surface, but we also find that there exists a deconfined vertex defect in the bulk with fermionic statistics. In \secref{genhall} we give a dictionary for going between more general fermionic fractional Hall effect, and a Walker-Wang model with the corresponding topological order on its surface. Once again we find that lattice models with a fermionic surface fractional Hall effect have a deconfined fermionic vertex excitation in the bulk.

	\subsection{Bosonic $\nu=1/2$ surface Hall state}\label{FQHE1/2}
We saw in \secref{sss:3DSemsurfaceexcitations} that the 3D semion model has surface excitations with semionic statistics -- in other words, with exactly the statistics of the charge-$1/2$ excitations of the bosonic $\nu = 1/2$ Laughlin state (which are described by a $\text{U}(1)$ Chern-Simons theory at level $k = 2$)\cite{Wenbook}. Because excitations in the bulk are confined and the ground-state degeneracy is determined only by the surface, the topological order of 3DSem on the ($3D$) solid donut is exactly that of the bosonic $\nu = 1/2$ Laughlin state on the torus.  	
	
We begin by making more precise the analogy between the surface of the lattice model and a system of bosons of charge $q_B$ in a magnetic field. The charge $q_B/2$ excitations have the same semionic statistics as surface vertex defects (corresponding to vertices at the ends of the string operators $\hat{W}(\maC_{AB})$ discussed in \secref{sss:3DSemsurfaceexcitations}). The bosonic charge $q_B$ excitations, however, do not exist in the lattice model: excitations that carry charge but no interesting statistics are, in the string-net framework, indistinguishable from the identity. This represents an important difference between the Walker-Wang lattice models of this section and actual quantum Hall fluids: The lattice models have $\mathbb{Z}_{n}$ rather than $U(1)$ charge conservation. In the present example fusing together two of the $q_B/2$ excitations in the bosonic $\nu=1/2$ state gives a charge $q_B$ excitation, which is physically measurable but topologically trivial, while in 3DSem fusing the corresponding vertex defects gives the identity -- so the lattice model has only $\mathbb{Z}_2$ charge conservation.

	\subsection{Fermionic $\nu=1/3$ surface Hall state}\label{FQHE1/3}
	\begin{figure}
 \includegraphics[width=1.\linewidth]{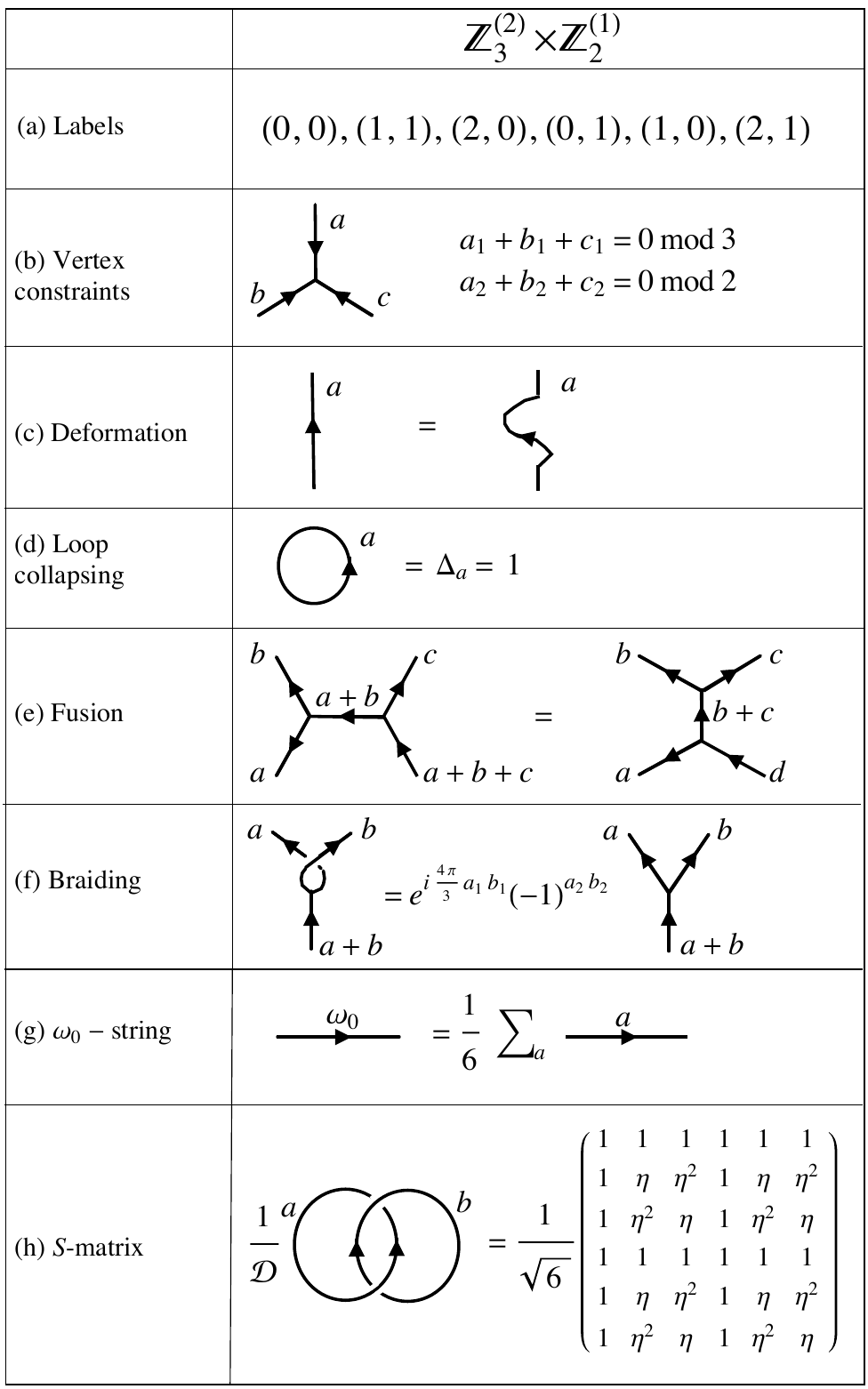}
 \caption{(a) Shows the six possible edge labels of the category. All of the string labels should be viewed as coming from $\mathbb{Z}^{(2)}_{3}\times\mathbb{Z}^{(1)}_{2}$. (b) to (f) show the category rules, which we have explained in \secref{s:MTCGS}. Note that labels can be added component by component e.g. $a+b=(a_1+b_1\mod 3,a_2+b_2 \mod 2)$. (g) Shows the $\omega_0$-string in the category (which, as always, has the handle-slide property), while (h) shows the $S$-matrix of the theory (where $\eta=e^{i 2 \pi/3 }$). The columns and rows of the $S$-matrix are ordered according to the list in (a). }
\label{FQHEcategory}
 \end{figure}

We have seen that the 3D semion model has the same topological order as a $\nu=1/2$ bosonic fractional Hall effect on the boundary of a confining bulk. Can we construct a Walker-Wang model with surface topological order resembling a $\nu=1/3$ fermionic Hall effect? We will find that there does exist such model but, unlike 3DSem, it has a deconfined vertex excitation (with fermionic self-braiding) in the bulk which is associated with an additional ground state degeneracy. Therefore, the lattice model does not quite have the topological order of a surface $\nu=1/3$ Laughlin state on a confining bulk.  This difference can, however, be eliminated by adding a term to the bulk Hamiltonian.

We choose a category with labels which correspond with the quasi-particle content of the $\nu=1/3$ quantum Hall effect (summarized in \tabref{tab:nuthird}). The category is known as $\mathbb{Z}^{(2)}_{3}\times\mathbb{Z}^{(1)}_{2}$, and it can be thought of as arising from a compact $k=3$ $\text{U}(1)$ Chern-Simons theory\cite{Bonderson07}.

\begin{table}
 \begin{center}
    \begin{tabular}{ | c | c |c | }
    \hline
    Charge &  Exchange Phase   & Category Label 	\\ \hline
    $0$       	& $1$               		& $\db{0,0}$		\\ \hline
    $e/3$   		& $e^{i\pi/3}$   		& $\db{1,1}$ 		 \\ \hline
    $2e/3$ 	& $e^{i4\pi/3}$ 		& $\db{2,0}$		\\ \hline
    $e$      		& $e^{i\pi}=-1$ 		& $\db{0,1}$		 \\ \hline
    $4e/3$ 		& $e^{i4\pi/3}$		& $\db{1,0}$		\\ \hline
    $5e/3$ 	 	& $e^{i\pi/3}$		& $\db{2,1}$	 	\\ \hline
        \end{tabular}
\end{center}
 \caption{This table shows the edge labels in the category $\mathbb{Z}^{(2)}_{3}\times\mathbb{Z}^{(1)}_{2}$ used to construct the state with surface resembling a $\nu=1/3$ fermionic  Hall effect. Each of the labels in the category correspond with a quasi-particle in the Laughlin $\nu=1/3$ state.  The exchange phase is the phase accumulated by exchanging two particles of the same type.}
\label{tab:nuthird}
\end{table}


The category in question has labels represented by a doublet $i=(i_1,i_2)$ where $i_1$ is an integer modulo $3$ and $i_2$ is an integer modulo $2$. The Hilbert space thus consists of a $6$-state system on each edge of a lattice\cite{orientation subtlety}. The label $(i_1=1,i_2=1)$ represents a Laughlin quasi-particle of charge $e/3$, while $(i_1=0,i_2 =1)$ represents a charge $e$ quasi-particle (which, unlike the boson of the $\nu = 1/2$ Laughlin state, we must keep track of explicitly because of its fermionic statistics). Excitations such as $(i_1 =1, i_2=0)$ can be viewed as the combination of Laughlin charge $e/3$ and charge $e$ quasi-particles. The correspondence between the doublets and the various Laughlin quasi-particles is summarized in \tabref{tab:nuthird}.

		\subsubsection{Excitations of the $\mathbb{Z}^{(2)}_{3}\times\mathbb{Z}^{(1)}_{2}$ model}
In this section we summarize the forms of excitations present in the Walker-Wang model based on $\mathbb{Z}^{(2)}_{3}\times\mathbb{Z}^{(1)}_{2}$. First we discuss the vertex defects, which carry a $\mathbb{Z}_6=\mathbb{Z}^{(2)}_{3}\times\mathbb{Z}^{(1)}_{2}$ charge (see \secref{FQHE1/2}). Unlike the confined Walker-Wang models such as 3DSem, the Walker-Wang model based on $\mathbb{Z}^{(2)}_{3}\times\mathbb{Z}^{(1)}_{2}$ has deconfined point-like excitations in the bulk and on its surface. The deconfined bulk excitations are fermionic vertex defects that occur at the end-points of strings labelled $(0,1)$. The other four possible vertex defects, which on the surface correspond to Laughlin quasi-particles, or Laughlin quasi-particles bound to electrons, are deconfined only on the surface. This is consistent with the fact that point particles in 3D must be bosons or fermions.

It is enlightening to see how this arises in practice. From the $S$-matrix (\figref{FQHEcategory}(h)) we see that strings carrying label $f=(0,1)$ ($f$ for fermion) braid trivially with all other particles (represented by the fact that the column $S_{i f}$ is identical to $S_{i 0}$). This in turn implies that string operators of the form $\hat{W}^f_V(\maC)$ commute with all the plaquettes along their length because they commute with all the string operators used to define the plaquette operator. On the other hand, all other string types braid non-trivially, so there is only one deconfined vertex defect. Thus the other four types of excitation listed in \tabref{tab:nuthird} are confined in the bulk, but as in the case of MTC based models, they are deconfined on the surface of the manifold.

In the bulk, the four kinds of confined vertex defect are associated with lines of plaquette defects (like those in \secref{sss:SemT^3Bulk}) which we can think of as `chiral vortex' lines. However there is another type of confined vortex line which has the same relative statistic with the deconfined $\db{0,1}$ vertex defects as we found in \secref{sss:3Dtcexcitations} between vortex lines and vertex defects in the 3D toric code. We can define this vortex line using a surface operator (see \eqnref{surfaceW}) of the form
\be\label{surfaceWFQHE}
\hat{W}_P({ \mathcal{S}})=\prod_{j\in \mathcal{S}} \wt{\sigma}^z_j\punc{,}
\ee
where $\wt{\sigma}^z_j$ acts on edge $j$ in the following way: if edge $j$ is in state $i=(i_1,i_2)$ then $\wt{\sigma}^z_j=(-1)^{i_2}$. Here $\mathcal{S}$ is a surface on the dual lattice, and the product is over edges cutting the surface transversally. The operator $\hat{W}_P({ \mathcal{S}})$ creates a line of plaquette violations on its boundary in the manner shown for the 3D toric code in \figref{fig:3Dtoriclinedefect}. In the bulk, the $\mathbb{Z}^{(2)}_{3}\times\mathbb{Z}^{(1)}_{2}$ model is almost identical to the 3D toric code: the deconfined $\db{0,1}$ vertex defects behave like the vertex defects in the toric code (except $\db{0,1}$ have fermionic self-braiding), and the vortex lines created by \eqnref{surfaceWFQHE} are like the vortex lines in the toric code. Having understood the spectrum of this Walker-Wang model, we now investigate its properties on some simple manifolds.

			\subsubsection{The $\mathbb{Z}^{(2)}_{3}\times\mathbb{Z}^{(1)}_{2}$ model on the 3-torus}\label{laughlint^3}
The Walker-Wang model based on the $\mathbb{Z}^{(2)}_{3}\times\mathbb{Z}^{(1)}_{2}$ category has $2^3$ degenerate ground states on the 3-torus, just like the 3D toric code. We will not go through the arguments for this result thoroughly, as it strongly resembles those presented for the toric code in \secref{sss:TCT^3}, but we emphasize that the ground state degeneracy is associated with the presence of deconfined excitations.  The string operator creating a pair of fermions at either end of a curve $\mac{C}$ is:
\be
W^f_V(\maC)=\prod_{j\in \mathcal{C}}\widetilde{\sigma}^{x}_j \prod_{ \text{crossed edges}}\widetilde{\sigma}^{z}_i
\ee
where the second product is over edges which are over- or under-crossed by $\maC$. Here $\widetilde{\sigma}^{x}_j$ flips the $i_2$ label between $0$ and $1$ on an edge $j$ in state $(i_1,i_2)$, while $\widetilde{\sigma}^{z}_j=(-1)^{i_2}$. Since $W_V^f$ commutes with the Hamiltonian, as before, one can toggle between the different ground states on the 3-torus using the $\mathbb{Z}_2$ operators $W^f_V(\maC)$, where $\maC$ is a closed path winding around one of the non-contractible cycles of the torus. Furthermore, one can distinguish between the $2^3$ different ground states using the new parity operators:

\be\label{paritiesFQHE}
\wt{P}_{n_\perp}=\prod_{i \in n_\perp}\wt{\sigma}^z_i \ \ \ \ \ n=x,y,z
\ee
which take values $\pm1$ depending on whether an even or odd number of $f$ loops wind around the $n$-cycle of the torus. Here $n_\perp$ is a set of edges parallel to the $\hat{n}$-direction, whose centers all lie on a plane perpendicular to the $\hat{n}$-direction. This generalizes the construction in \eqnref{parities}. Having understood the Walker-Wang model on a manifold without boundary, we now investigate its properties on a manifold with boundary.

		\subsubsection{The $\mathbb{Z}^{(2)}_{3}\times\mathbb{Z}^{(1)}_{2}$ model on the solid donut}
In the case of the solid donut, the ground state degeneracy is $2\times3=6$. Heuristically, this result arises for the following reasons. The deconfined fermion can have a parity $\pm1$ around the single non-contractible direction in the solid donut (this gives the factor of $2$). To see where the factor of $3$ comes from, let us fix the fermion parity. This is like considering a Walker-Wang model based on the category $\mathbb{Z}^{(2)}_{3}$. As this category is modular\cite{Bonderson07}, and has $3$ types of string, we can use the reasoning in \secref{MTCD^2xS^1} to show that it has a degeneracy of $3$ on the solid donut associated with the surface deconfined anyons. Therefore, together with the two values of fermion parity, the ground state degeneracy is $6$.

Unlike 3DSem, then, the topological order we have found on the solid donut does not correspond exactly to that of a quantum Hall state on the boundary of a confining bulk. The discrepancy is due to the fact that we have had to explicitly keep track of the charge $e$ quasi-particle in our category, due to its fermionic statistics -- which effectively doubled the number of ground states. The appearance of this deconfined fermionic vertex excitation is not unexpected: if this state does have the topological order of some system of fermions, then the fermions should be able to exist in the bulk where they would be deconfined in general.


The extra ground-state degeneracy due to these deconfined fermions is, however, somewhat unexpected.  It arises because the fermions in the Walker-Wang model are non-dynamical:
The ground state sectors of our model can be split into two `halves', distinguished from each other by the number of fermionic strings (mod $2$) running through the donut.
In order to lift this degeneracy we add a term resembling a bulk fermion kinetic term $t \sum_{\left< ij \right>} (c^\dag_i c_j + c^\dag_i c^\dag_j + h.c. )$ (which also serves to create and destroy pairs of charge $e$ fermions). In the lattice model, the analogue of this term takes the form
\be \label{kinetic}
t \sum_{j} \widetilde{\sigma}^x_j\punc{,}
\ee
where $\widetilde{\sigma}^{x}_j$ flips the $i_2$ label between $0$ and $1$ on an edge $j$. This strongly mixes states $(i_1, 0)$ and $(i_1, 1)$, such that at low energies there remain only three string types, which can be labelled by their value of $i_1$. Similarly, pairs of ground states differing only by a fermionic flux are strongly split, reducing the ground state degeneracy from $6$ to $3$.

		\subsubsection{The effective field theoretic description}\label{bfbbnu1/3}
		
Given that the surface theory of the $\mathbb{Z}^{(2)}_{3}\times\mathbb{Z}^{(1)}_{2}$ lattice model looks like a $\nu=1/3$ quantum Hall effect, one might expect that its topological features are described by a $bF+bb$ theory with $k=3$ (\eqnref{bf+bb}). In fact, the field theory appears to capture some, but not all, of the properties of the lattice model. In the field theoretic language, the world-lines of anyons on the surface of the sample correspond to Wilson lines carrying ($\mathbb{Z}_6$) labels in $\left\{1,2,4,5\right\}$, while the fermions carry label $3$; the field theory reproduces the correct statistics for these particles.

When we try to source point particles in the bulk, we again encounter the constraint in \eqnref{eq:deltaconstrBF}: this constraint implies that the particles necessarily lie at the ends of line-like objects. This is again consistent with the fact that particles labelled $\left\{1,2,4,5\right\}$ are at the ends of lines of plaquette defects defects when in the bulk. In the case of the fermionic vertex defect, the line-like object is not energetic in the lattice model, but is measurable insofar as it keeps track of any self-twisting in the trajectory of the vertex defect.

The $k=3$ Chern-Simons theory certainly captures surface topological order, but apparently fails to account for the additional ground state degeneracy due to the deconfined fermion in the bulk. Furthermore, it is not obvious how to use the field theory to create the vortex line defined in \eqnref{surfaceWFQHE}, or to reproduce the relative statistic between the vortex line and the fermionic quasi-particle. To capture these additional properties, one might introduce another two-component quantum field $c_{\mu \nu}$ and add it to the $bF+bb$ action using a term $\frac{1}{2\pi} \lev c_{\mu\nu} \partial_{\rho }A_{\sigma}$. The world-sheets of the vortex lines (\eqnref{surfaceWFQHE})  correspond to sources coupled to the new field $c$ via $\frac{1}{2}c_{\mu\nu}\Gamma_{\mu\nu}$.

While we have seen that $bF+bb$ theory at $k=3$ does not describe the lattice model, there is a deep relation between the field theory and the {\it category} $\mathbb{Z}^{(2)}_{3}\times\mathbb{Z}^{(1)}_{2}$. Chern-Simons theories have a well known relation to chiral conformal algebras\cite{Witten89}. Many chiral conformal algebras, on the other hand, admit a description in terms of categories. In the case of $k=3$ (more generally odd $k$) Abelian Chern-Simons theory, the corresponding conformal algebra is in fact a superalgebra\cite{Dijkgraaf90,Bonderson07}, and the  category corresponding to the algebra is $\mathbb{Z}^{(2)}_{3}\times\mathbb{Z}^{(1)}_{2}$. The $\mathbb{Z}_2$-grading corresponds to the fact that Wilson lines carrying label $3$ are fermionic under self-twisting; it is precisely this $\mathbb{Z}_2$-grading which leads to the deconfined bulk fermion in our lattice models.

%

		\subsubsection{More general surface quantum hall states}\label{genhall}
Having seen how two specific examples give rise to surfaces with topological order resembling fractional Hall states, we briefly summarize how these results generalize. The Walker-Wang model based on category $\mathbb{Z}^{(1/2)}_{k}$ for even $k$ (listed in [\onlinecite{Bonderson07}]) gives a fully confined bulk, and a surface with the topological order of a $\nu=1/k$ bosonic Laughlin state. This topological order is captured by an abelian $bF+bb$ theory at even level $k$, written down in \eqnref{bf+bb}.

On the other hand, to obtain the surface topological order resembling the hierarchical fermionic Laughlin states at $\nu=n/m$ ($m$ odd, $n<m$ and $n$ coprime to $m$), we can construct a Walker-Wang model based on category $\mathbb{Z}^{(2p)}_{m}\times\mathbb{Z}^{(1)}_{2}$ (where $p$ is odd and $n p\equiv 1\mod m$, see [\onlinecite{Bonderson07}]). As in the case of $\nu=1/3$, all of these models have their excitations confined in the bulk except for a single deconfined fermionic vertex defect. The lattice models have a description in terms of an abelian $bF+bb$ similar to that in \secref{bfbbnu1/3}, but generalized using the $K$-matrix formalism\cite{Wen90}.

It is natural to ask whether this correspondence generalizes to more exotic Hall states. Using the 12-particle category in [\onlinecite{Bonderson07}] called `Moore-Read'\cite{Moore91} we can produce a Walker-Wang model with surface Moore-Read topological order, and a deconfined fermionic vertex in the bulk (labelled $\psi_{(2,4)}$ in the reference). On the solid donut, this model has a degeneracy of $12$, but adding the Moore-Read analogue of the kinetic term in \eqnref{kinetic} lifts the degeneracy to $6$.


\section{Conclusion}

In this work, we have discussed two distinct families of topological phases in 3D, exemplified by the 3D Toric code and 3DSem respectively. The first family (which has been studied in a variety of contexts\cite{DiscreteGauge}) has a ground-state degeneracy that depends on the topology of 3D space: for example, the ground state is non-degenerate on $\mathbb{T}^3$. The excitations of these models can be grouped into point-like `charges', which are deconfined everywhere, and vortex loops; the interesting mutual statistics occurs between point charges and vortex loops. These features (topological ground state degeneracy, and mutual anyonic statistics between point defects and vortex lines) are the most natural extension of the notion of topological order to 3 dimensions.

The second family, however, suggests a different notion of 3D topological order: it has a unique ground state on any closed (orientable) 3-manifold, and all excitations in the bulk are confined. Nonetheless on a manifold with boundary these models do have degenerate ground states, whose number depends on the topology of the boundary. More than this, their surfaces admit deconfined anyonic excitations. For 3DSem these have (abelian) semionic statistics, but other models exist for which these surface excitations are non-abelian anyons. Hence these are 3D models whose topological characteristics are largely restricted to their boundaries -- but at these boundaries, we recover the full richness of possible topological orders of 2D systems.

The interesting low-energy physics of the confined WW models thus occurs almost entirely at their boundaries. It is well-established that in some systems, such as topological insulators\cite{Schnyder08}, the low-energy physics at the boundary can be used to classify distinct bulk phases of matter (in that case, due to the presence of symmetry-protected gapless surface states).  We have established that the interesting surface physics of the Walker-Wang models does not play quite so strong a role: since  $2$D systems with the same topological order exist, these surface states do not uniquely identify a bulk phase of matter, and indeed can be eliminated by adding a 2D quantum Hall layer to the surface.

  However, within the realm of ``fixed-point" lattice models (whose correlation length is smaller than the lattice constant, and thus effectively $0$) the deconfined chiral surface anyon states cannot be eliminated.  Further, the existence of chiral anyons at all in an exactly solvable model is notable: we believe that no $2$D fixed-point Hamiltonian can emulate these\cite{WenPersonal}.  Thus within the sphere of fixed-point Hamiltonians, one can make a strong case that the topological order of the surface does uniquely identify the bulk.  A physical understanding of the meaning of this bulk-boundary correspondence, which here is a feature of the fixed-point Hamiltonian rather than the phase, remains an interesting open question.


Finally, we might ask what kinds of real physical systems could be expected to share the long-wavelength characteristics of the Walker-Wang models.  One possibility is that a confined phase of fractional topological insulators, where fractionalized excitations are confined in the bulk, might in some circumstances be expected to support deconfined fractionalized excitations on its surface.
 (One of us\cite{BurnellCardy} will discuss this possibility in more detail in a future work.)   However, much remains to be understood on the subject of realising such phases in more physically motivated systems.

In summary, we have shown that the Walker-Wang Hamiltonians present an interesting playground in which to examine possible 3D topological states of matter, in a context where concrete calculations can be carried out.  This raises the interesting possibility that some $3$D systems have a topological order that is purely $2$ dimensional, while others (such as the fermionic models discussed here in \secref{QHSect}) are topologically ordered in the bulk, but admit much richer anyon models at their surfaces.

\acknowledgments

We thank Victor Gurarie, Xiao-Gang Wen, Xiao-Liang Qi, and Zhenghan Wang for useful discussions. We are particularly grateful to Kevin Walker for
his very careful reading of this manuscript. CVK acknowledges the financial support of the EPSRC. SHS acknowledges funding from EPSRC grants EP/I032487/1 and EP/I031014/1.
%

\appendix

\section{MTC's are non-degenerate on manifolds without boundary} \label{App:3DGSnondeg}
In this section we show that all confined Walker-Wang models have a single ground state on the 3-torus, although this method of proof generalizes to other simple manifolds without boundary. This can also be viewed as a proof that there are no deconfined excitations in the three dimensional bulk. In \secref{GenWW} we reasoned that all confined Walker-Wang models are based on modular tensor categories. One of the key features of modular tensor categories is the `Flux killing property', described in \figref{Omegaloop}(c): A loop of the string-type $\omega_0$ (used to define plaquette operators) projects anything it encloses onto zero flux. In the following proof we will make heavy use this fact, as well as the handle-slide property \figref{Omegaloop}(a) of $\omega_0$ strings (which holds in all unitary braided fusion categories).

\begin{figure*}
 \includegraphics[height=.7\pdfpageheight,width=1\linewidth]{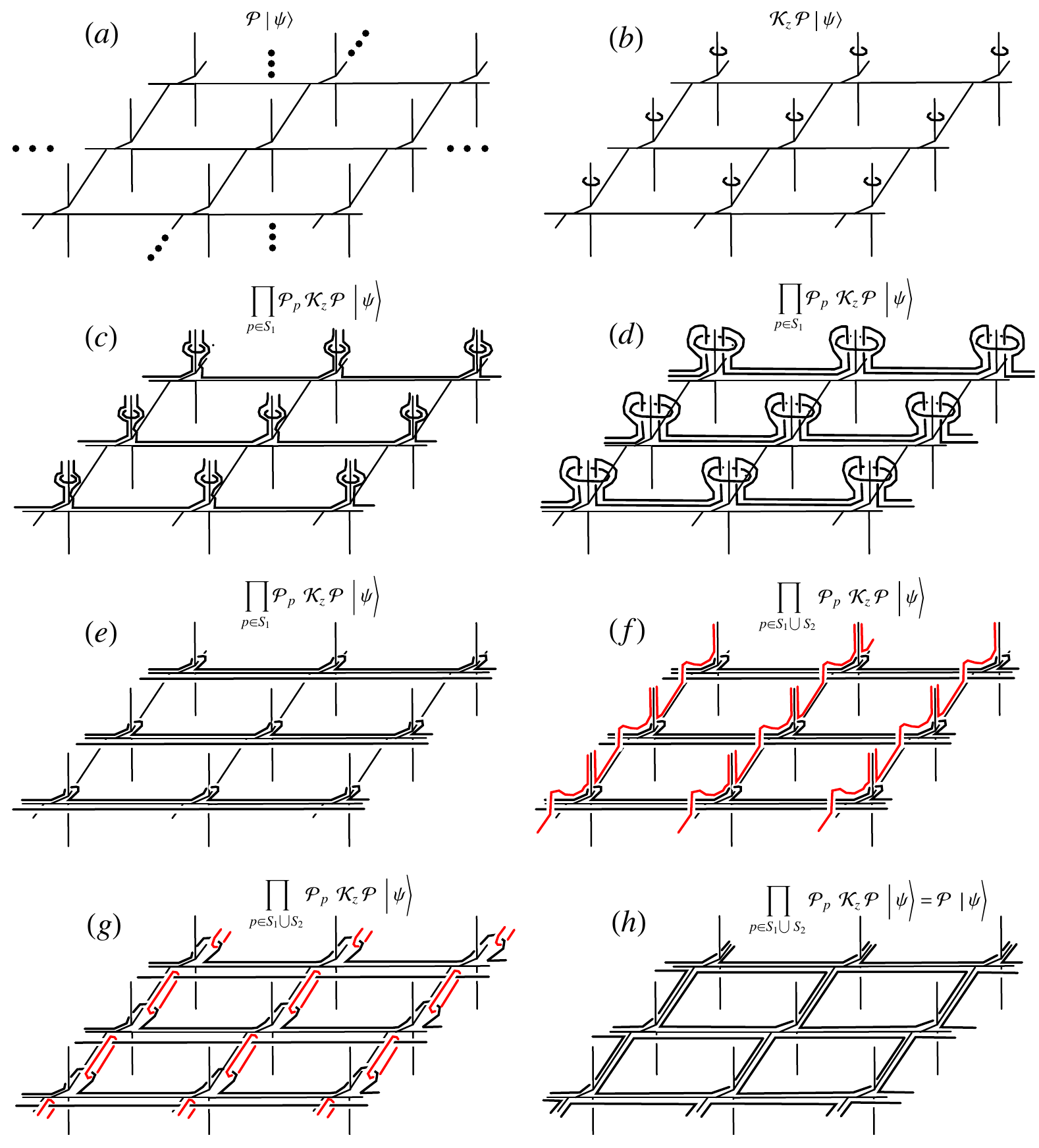}
 \caption{(Color online) This figure shows the sequence of steps used in proving that Walker-Wang models based on MTC's are non-degenerate on the 3-torus.}
 \label{fig:MTCproof}
 \end{figure*}

This proof is structured as follows. In order to show there is just one ground state, we need only show that $\proj\ket{\psi}\propto\proj \ket{0}$ holds for any string-net configuration, where $\ket{0}$ is the configuration with no edges colored in (meaning all edge bonds have label $i=0$)  and $\proj$ is the ground state projector. We do this in three parts. First, we show that
\be\label{eq:xy}
\proj \ket{\psi}= \sum_i a_i \proj\ket{\psi_{xy}}_i\punc{,}
\ee
where $\ket{\psi_{xy}}_i$ are configurations with edges colored only in the $xy$ plane. We will then apply the same procedure to show that, for any of the configurations lying in the $xy$ plane,

\be\label{eq:yz}
\proj \ket{\psi_{xy}} =  \sum_i b_i \proj\ket{\psi_{y\text{-axis}}}_i
\ee

 where $\ket{\psi_{y\text{-axis}}}$ are configurations with edges colored only along the y-axis. Applying the procedure one last time we will show that, for any of the configurations along the y-direction
 \be\label{eq:zx}
 \proj \ket{\text{y-axis}} =  c \proj\ket{0}\punc{.}
 \ee
Hence, feeding this all into \eqref{eq:xy}, we get

\be\label{eq:xyfinal}
\proj \ket{\psi}\propto \proj\ket{0}
\ee

for all $\ket{\psi}$, as required. Let us first prove \eqref{eq:xy}. Consider the sequence of diagrams \figref{fig:MTCproof}, which show a section of the $xy$ plane on the 3-torus. \figref{fig:MTCproof}(a) represents some string-net configuration acted on by the ground state projector $\proj \ket{\psi}$. We remind the reader that the ground state projector $\proj=\prod_v\proj_v \prod_p\proj_p$ is a just a product of all vertex and plaquette projectors, and that a plaquette projector $\proj_p$ can be represented by drawing a $\omega_0$ string in plaquette $p$. In (b) we act on this state with $\omega_0$ loops on the vertical edges sticking up and out of the $xy$ plane. We call the operator $\mathcal{K}_z$, and so we have formed a state $\mathcal{K}_z \ket{\psi}$. We then act with a set $S_1$ of plaquette projectors just above the $xy$ plane, each lying in a $xz$ plane, to form the state $\prod_{p\in S_1} \proj_p \mathcal{K}_z \proj\ket{\psi}$ in (c). Using the plaquette projectors in the original expression $\proj \ket{\psi}$, we can handle-slide the plaquette projectors in $\prod_{p\in S_1} \proj_p$ to form the figure in (d), which is another graphical representation of $\prod_{p\in S_1} \proj_p \mathcal{K}_z \proj\ket{\psi}$. We can then use the flux killing property of $\omega_0$ strings (in MTC's) to show that this state is diagrammatically represented by (e).

We are now near the end of the proof. We act on (e) with a set $S_2$ of plaquette projectors just above the $xy$ plane to form the state $\prod_{p\in S_1\cup S_2} \proj_p \mathcal{K}_z \ket{\psi}$ shown in (f); we color the new plaquette strings red for clarity. We can again use the handle slide properties of the plaquette projectors in $\proj \ket{\psi}$ to manipulate the red strings into the equivalent form shown in (g). Using the flux killing properties of the red strings, we can close up the black plaquette strings to form (h). But (h) simply represents the state $\prod_{p\in xy} \proj_p \proj\ket{\psi}$ which is equal to $\proj\ket{\psi}$. Hence, we have proved the following equation:

\be
\prod_{p\in S_1\cup S_2}\proj_p   \mathcal{K}_z \proj \ket{\psi} = \proj\ket{\psi}
\ee

Acting on this equation with $\proj$, we get the following equation:

\be\label{MTCfinal}
 \proj \mathcal{K}_z \proj \ket{\psi} =  \proj\ket{\psi}\punc{.}
\ee

This equation is the main result of this section because the state $\proj\mathcal{K}_z \proj \ket{\psi}$ has an interesting property: The operator $\mathcal{K}_z$ kills all the flux on the vertical legs just above the $xy$ plane. This implies that the string-net configurations in $\mathcal{K}_z \proj \ket{\psi}$ are contractible, at least around the z-cycle of the torus. So the string-net configurations in $\mathcal{K}_z \proj\ket{\psi}$ can be deformed to lie in the $xy$ plane. Hence

\be
\proj \mathcal{K}_z \proj \ket{\psi}= \sum_i a_i \proj\ket{\psi_{xy}}_i \punc{,}
\ee and using \eqnref{MTCfinal} we get \eqnref{eq:xy}. We now repeat the process. Define an operator $\mathcal{K}_x$ by putting $\omega_0$ loops around edges leaving the $yz$ plane and note the result \eqnref{MTCfinal} also holds for $\mathcal{K}_x$. As the operator $\mathcal{K}_x$ kills all the flux on the legs just to the right of the $yz$ plane, the string-net configurations in $\mathcal{K}_x \proj \ket{\psi_{xy}}$ are contractible around the $x$-cycle of the torus (as well as the z-cycle). So the string-net configurations in $\mathcal{K}_x \proj \ket{\psi_{xy}}$ can be deformed to lie along the y-axis. Hence

\be
\proj \mathcal{K}_x \proj \ket{\psi_{xy}}= \sum_i b_i \proj\ket{\psi_{y\text{-axis}}}_i \punc{,}
\ee and using \eqnref{MTCfinal} we get the result \eqnref{eq:yz}. Now to the final stage of the argument. Define an operator $\mathcal{K}_y$ by putting $\omega_0$ loops around edges leaving the $zx$ plane and note the result \eqnref{MTCfinal} also holds for $\mathcal{K}_y$. As the operator $\mathcal{K}_y$ kills all the flux on the legs leaving the $zx$ plane, the string-net configurations in $\mathcal{K}_y \proj \ket{y\text{-axis}}$ are contractible around the $y$-cycle of the torus (as well as the $z$-cycle and $x$-cycle). So the string-net configurations in $\mathcal{K}_x \proj\ket{\psi_{y\text{-axis}}}$ can be deformed to $\ket{0}$. Hence

\be
\proj \mathcal{K}_x \proj \ket{\psi_{y\text{-axis}}}= c \proj\ket{0} \punc{,}
\ee and using \eqnref{MTCfinal} we get the result \eqnref{eq:zx}. The statement \eqnref{eq:xyfinal} follows from this. Thus we have shown that the flux killing property of MTC's implies that Walker-Wang models based on MTC's are non-degenerate on the 3-torus.

\section{All string operators are confined in MTC's}   \label{App:3DGStrconf}
In this section we prove that for all MTC's, any vertex types string operators (obeying mild constraints) produce plaquette violations along their length. This result supplements the discussion in the text, where we considered a special type of string operator ($\hat{W}_V(\maC)$) and showed that it fails to commute with the plaquettes along its length.

Suppose we construct a string operator $\hat{\mac{V}}(\maC)$ with the following qualities. It can change the color of edges along a path $\maC$, and assign a configuration dependent phase to edges touching the path $\maC$. Furthermore, we assume that if the edges on $\maC$ have label $i=0$, $\hat{\mac{V}}(\maC)$ will flip the edges to some superposition of states, some of which have all the edges on the path $\maC$ in a state $i\neq0$. In this section we will show that such an operator necessarily fails to commute with plaquettes along its length.

\begin{figure}
 \includegraphics[width=1\linewidth]{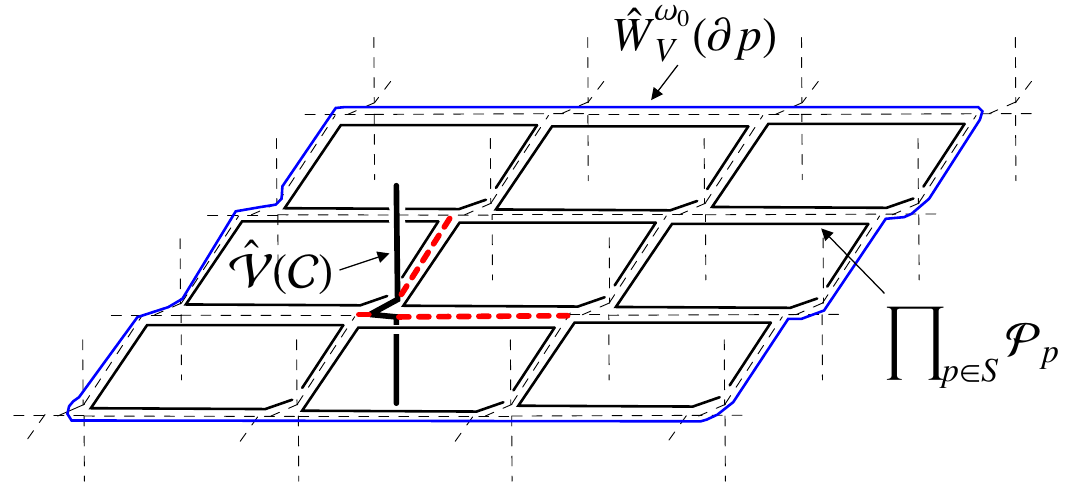}
 \caption{(Color online) This figure represents a string operator $\mathcal{V}(\maC)$ piercing the $xy$ plane. The black loops are the plaquette operators in a set $S$, pierced by the string operator, while $\hat{W}^{\omega_0}_V(\partial p)$ (blue line) is another string operator (with label $\omega_0$) which encircles the perimeter of the set $S$). In the text we show that $\hat{\mac{V}}(\maC)$ fails to commute with some of the plaquettes in $S$, with fairly mild assumption on the form of  $\hat{\mac{V}}(\maC)$.}
 \label{pierce}
 \end{figure}

Suppose that $\maC$ pierces the $xy$ plane once. Take a large set $S$ of plaquettes in the $xy$ plane with the condition that $\maC$ pierces the set, but the plaquettes near the boundary of $S$ are not acted on by $\hat{\mac{V}}(\maC)$ (see \figref{pierce}). We will show that, with the above conditions, $\hat{\mac{V}}(\maC)$ cannot possibly commute with some plaquettes in the $xy$ plane. This shows that the defects created by $\hat{\mac{V}}(\maC)$ are necessarily linearly confined.
	
With these above assumptions about $\hat{\mac{V}}(\maC)$, we can say that upon acting on the empty ket $\ket{0}$ (all edges carry label $i=0$), the string operator flips all the edges along $\maC$ to some superposition of labels $s =\sum_i s_i \hat{i}\neq0$. Without loss of generality (by subtracting a multiple of the identity operator from $\hat{\mac{V}}(\maC)$) we can assume that $s_0=0$. This implies that

\be\label{genstring1}
\hat{W}^{\omega_0}_{V}(\partial S) \prod_{p\in S}\proj_p \hat{\mac{V}}(\maC)\ket{0}=0
\ee where $\hat{W}^{\omega_0}_{V}(\partial S)$ is a string operator with the special handle-slide property, and $\partial S$ is some path on the boundary of the set of plaquettes in the $xy$ plane (see \figref{pierce}). This equation follows simply from the fact we can handle-slide the perimeter string $\partial S$ inwards so that it encloses the edges on path $\maC$ which carry label $s$. As we are in a MTC based model and $s_0=0$, we can use the flux killing property of the \figref{Omegaloop}(b) to get zero in \eqnref{genstring1}.

If we assume that $\prod_{p\in S}\proj_p$ and $ \hat{\mac{V}}(\maC)$ commute then we are led to a contradiction. Continuing from the previous equation:

\begin{align}
0&=\hat{W}^{\omega_0}_{V}(\partial S) \prod_{p\in S}\proj_p \hat{\mac{V}}(\maC)\ket{0} \nonumber \\
&=\hat{W}^{\omega_0}_{V}(\partial S) \hat{\mac{V}}(\maC) \prod_{p\in S}\proj_p \ket{0} \nonumber \\
&=\hat{\mac{V}}(\maC)\hat{W}^{\omega_0}_{V}(\partial S)  \prod_{p\in S}\proj_p \ket{0} \nonumber \\
\end{align}

To get to the second line, we assumed that the plaquettes commute with $\hat{\mac{V}}(\maC)$. In the third line, we used the fact that $\left[\hat{\mac{V}}(\maC),\hat{W}^{\omega_0}_{V}(\partial S)\right]=0$ because they do not act on the same edges. Now, $\hat{W}^{\omega_0}_{V}(\partial S) \prod_{p\in S}\proj_p \ket{0} = \proj_p \ket{0} $ because we can handle-slide the string over the plaquette operators to get a $\omega_0$ loop enclosing no flux. Therefore,
\begin{align}
0&=\hat{\mac{V}}(\maC) \prod_{p\in S} \proj_p \ket{0} \nonumber \\
&= \prod_{p\in S} \proj_p \hat{\mac{V}}(\maC) \ket{0}
\end{align}

But this state is just a cluster of plaquette projectors acting on a ket with some edges in superposition $s$. It can be shown that such a state is not zero by expanding out the plaquette projectors in terms of plaquette string operators $W^{j}_V(\partial p)$. Consider a string configuration where all edges have label $0$ except for those on $\maC$ (which are in superposition $s$). This configuration can be shown to occur once in the expansion of $\prod_{p\in S} \proj_p \hat{\mac{V}}(\maC) \ket{0}$, and with a non-zero coefficient. Hence the expression  $\prod_{p\in S} \proj_p \hat{\mac{V}}(\maC) \ket{0}$ cannot be zero. Thus, the assumption that $\prod_{p\in S}\proj_p$ and $ \hat{\mac{V}}(\maC)$ commute leads to a contradiction, implying $ \hat{\mac{V}}(\maC)$ is linearly confined.


\section{Explicit forms for operators in DSem model}\label{ExplicitDSem}
 \begin{figure}
 \includegraphics[width=.6\linewidth]{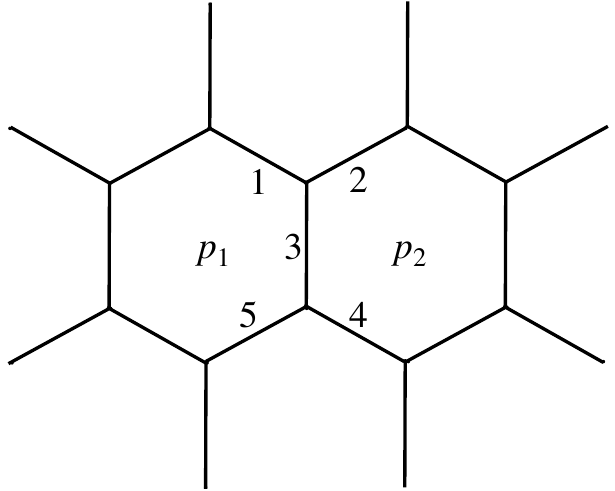}
 \caption{The figure shows an adjacent pair of plaquettes ($p_1$ and $p_2$), used to show that adjacent plaquette operators in the DSem model (as defined in \eqnref{DSemplaq}) do not commute if vertex defects are present.}
 \label{noncommplaq}
 \end{figure}
In our main treatment of the DSem model, we defined the plaquette operators only in the absence of vertex defects. Furthermore, we never explicitly showed the form of string operators near their end-points. In this appendix, we address some of these issues. In the main text we defined the plaquette operators as

\be \label{Psem1}
B_p= (\prod_{i\in \partial p} \sigma^x_i ) \prod_{j\in s(p)} i^{(1 - \sigma^z_{j})/2} \punc{.}
\ee
The difficulty with \eqnref{Psem1} is that as operators, $B_{p_1}$ and $B_{p_2}$ do not commute when acting on neighboring plaquettes. Specifically, on the $5$ edges on which both operators act (see \figref{noncommplaq}) we have

\ba
B_{p_1} B_{p_2} &=&  \left ( \sigma^x_1 \sigma^x_3 \sigma^x_5 e^{ i \frac{ \pi}{4}(1 - \sigma^z_{2} ) } e^{ i \frac{ \pi}{4}(1 - \sigma^z_{4} ) } \right )  \n
&&  \left ( \sigma^x_2 \sigma^x_3 \sigma^x_4 e^{ i \frac{ \pi}{4}(1 - \sigma^z_{1} ) } e^{ i \frac{ \pi}{4}(1 - \sigma^z_{5} ) } \right ) \n
&=&  -\sigma^x_1  \sigma^x_2  \sigma^x_4 \sigma^x_5  e^{ i \theta_{1245}}  \n
B_{p_2} B_{p_1}
&=&  -\sigma^x_1  \sigma^x_2  \sigma^x_4 \sigma^x_5  e^{ - i \theta_{1245} }
\ea
or
\be
\frac{1}{2} [B_{p_1}, B_{p_2} ] = -  i \sigma^x_1  \sigma^x_2  \sigma^x_4 \sigma^x_5 \sin  \theta_{1245}
\ee
where
\be
\theta_{1245} = \frac{ \pi}{4}(\sigma^z_2+ \sigma^z_4 - \sigma^z_{1}  - \sigma^z_{5} )
\ee
In the ground state, where
\be \label{BvS}
B_v =
\prod_{i \in \partial V} \sigma^z_i =1 \ \ \ ,
\ee
 $\theta_{1245}$ is a multiple of $\pi$, and the two products are equal.
Thus the plaquette operators (\ref{Psem1}) commute only when acting on states without vertex violations.
In states with vertex violations, $\theta_{1245}$ is an even or odd multiple of $\pi/2$, depending on whether there is an even or odd number of vertex violations on the two vertices $(123)$, $(345)$.

\begin{figure}
 \includegraphics[width=1\linewidth]{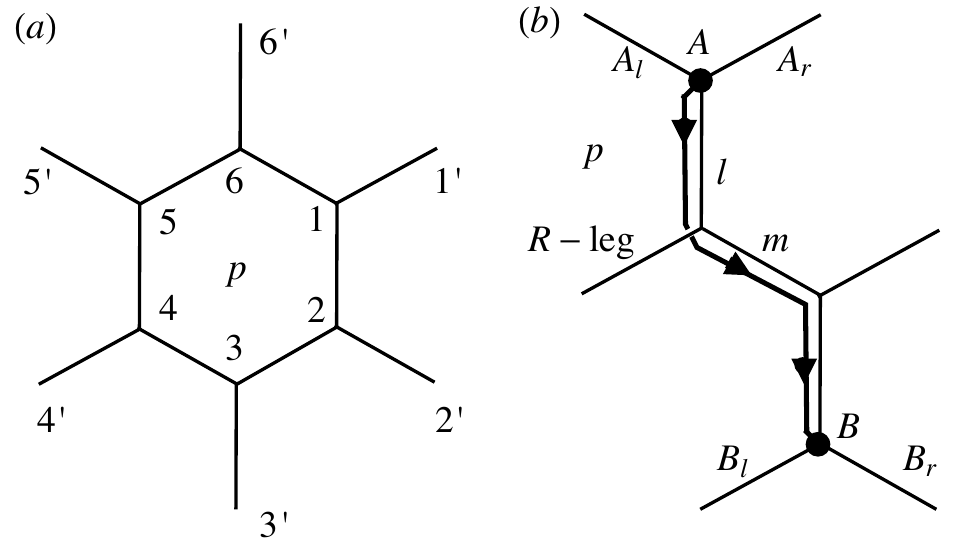}
 \caption{(a) Shows a labelled plaquette, used to define the complete plaquette operator. (b) Shows a graphical representation of the string operator defined in the text.}
 \label{fullplaquette}
 \end{figure}

For the following definition of $B_p$ it is more convenient to label the vertices of $p$ rather than edges (\figref{fullplaquette}(a)), and denote edges by the pair of vertices connecting them. Edges are now labelled by the pair of vertices they connect:

\begin{align}\label{eq:DSemfull}
B_p = &\prod^{6}_{j=1} \sigma^{x}_{j \, j+1} (-1)^{n_{j-1 \, j}\db{1-n_{j \, j+1}}}\\
&i^{-(1-Q_1) n_{11'}\Lambda_1}i^{(1-Q_2)n_{22'}\Lambda_2}\nonumber\\
&i^{(1 - Q_ 3) (1 - n_ {33'}) (n_ {23} -n_ {34})} i^{-(1 - Q_ 4) n_ {44'}\Lambda_ 4}\nonumber\\
&i^{(1 - Q_ 5) n_ {55'}\Lambda_ 5} i^{(1 - Q_ 6) n_ {66'} (n_ {56} - n_ {61})}\nonumber
\end{align}

We have defined $Q_v= (I- B_v)/2$,  and $\Lambda_v=\sum_{e\in s(v)} n_e$ where $e$ runs over the edges connected to vertex $v$ e.g. $\Lambda_1= n_{61}+n_{11'}+n_{12}$. This plaquette term represents a rather complicated twelve spin interaction term, although we remind the reader that this reduces to the somewhat simpler form (the first line of \eqnref{eq:DSemfull}) in the absence of vertex defects.

	\subsection{String operators, and their endpoints.}
In this section we present a formalism for introducing pairs of vertex defects into the DSem model. We will write down the form of an open vertex-type string operator, near and away from its endpoints. We use the convention that positive chirality particles are associated with an over-crossing type string, and negative chirality under-crossing type string. In the formalism used, positive chirality particles will also be associated with a plaquette defect; in fact the most natural formalism requires that we introduce an additional label at vertices, but we opt to use another convention where one chirality of string involves a plaquette and vertex defect at the end-point.

For \figref{fullplaquette}(b), we make the negative chirality string operator take the following form if acting on a state without vertex violations:
\be
\prod_{i\in\mathcal{C}}\sigma^x_i \prod_{\expecval{j k }\in\mathcal{C}}\ (-1)^{n_j(1-n_{k})} \prod_{\text{$\expecval{\vec{l m}}$ at R vertex}} i^{n_m -n_l} \phi_A \phi_B\punc{.}
\ee

Here the $\phi_A, \phi_B$ factors are associated with the precise state of the string end-points. The phases are defined $\phi_A = i^{(1-n_{A_r})n_{A_l}}$ while $\phi_B = (-i)^{(1-n_{B_r})n_{B_l}}$, where $A_l,A_r,B_l,B_r$ are the edges marked in \figref{fullplaquette}(b). The reader should be warned that the exact forms of these phases will depend on which direction the string approaches the vertex, as well as the convention adopted for incorporating vertex defects.

To produce a defect of opposite chirality, we conjugate the operator, except $\phi_A, \phi_B$ parts. The resulting operator is of positive chirality, and will anti-commute with the plaquette labelled $p$ at its endpoint, but no other plaquettes.

\section{Topological entanglement entropy in 3DSem} \label{EntApp}

Here we show that unlike its $2$D counterpart, the $3$D semion model has no topological entanglement entropy.

\begin{figure}
 \includegraphics[width=\linewidth]{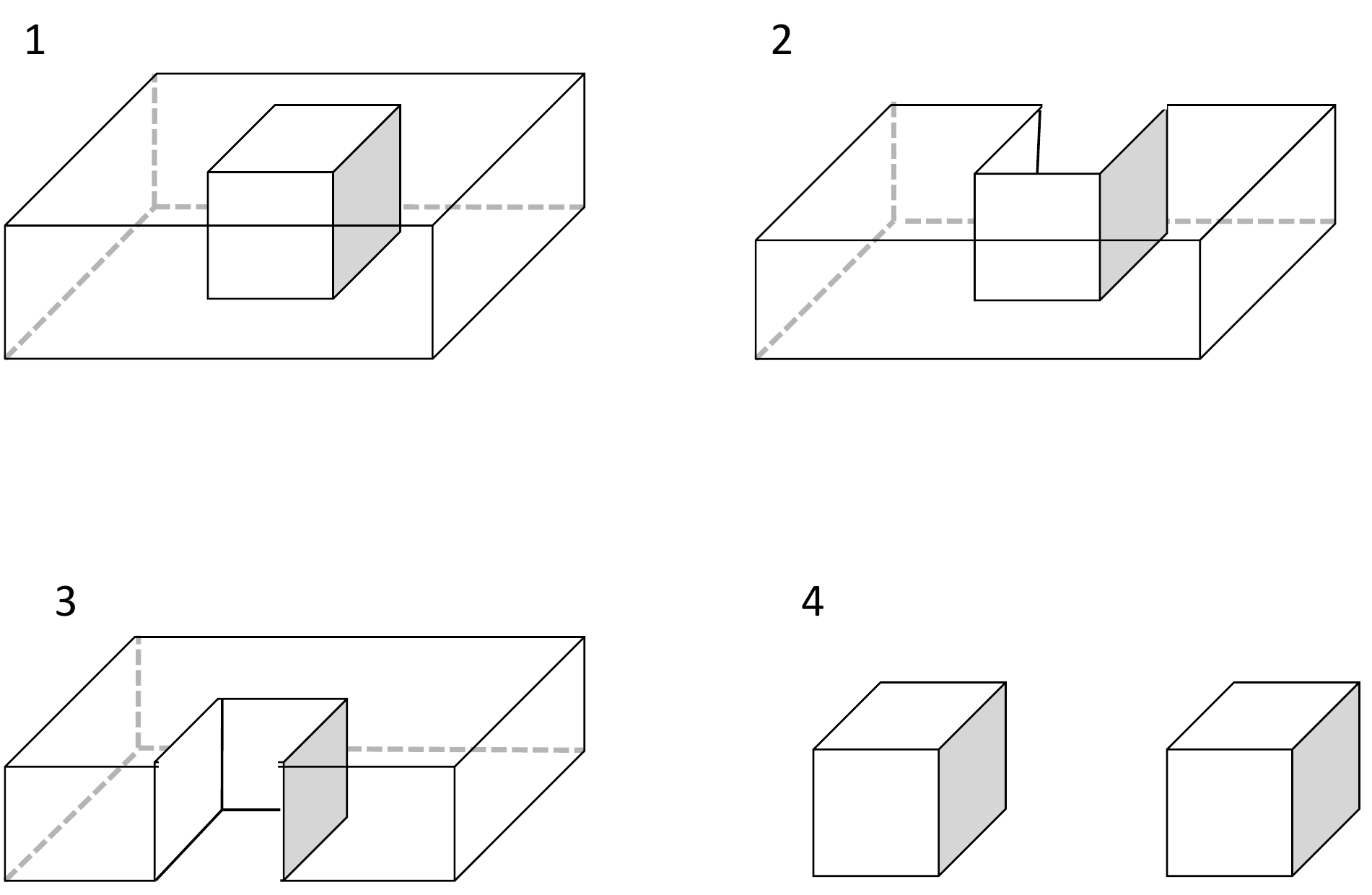}
 \caption{The combination of partitions used to calculate the entanglement entropy.  The figure shows the boundary of the regions $A$ and $B$.}
\label{EntroFig}
 \end{figure}

As described in Ref. \onlinecite{CastelnovoChamon}, there are two sensible prescriptions for defining topological entanglement entropy in $3$D.  We will use one of these for our calculation, but the result is independent of the choice of prescription.  Both prescriptions operate on the same principle, generalizing the approach of Levin and Wen\cite{LWEntanglement} for $2$D systems: we add together the entanglement entropies of several possible partitions of a system into subsystems $A$ and $B$, with coefficients such that the net boundary and corner terms all cancel.  Fig. \ref{EntroFig} shows the combination of partitions that we will use.  We will always cut the two subsystems along the middle of a set of edges, so that the two subsystems must share a set of edge labels along the boundary.

It is useful to briefly recall the source of the topological entanglement entropy in the Toric code.  The Toric code ground state is a superposition of all configurations of closed loops, with a relative amplitude of $1$.  Thus, for any partition of the system into two subsystems $A$ and $B$, we may write
\be \label{FactorEq}
|\Psi_0 \rangle = \sum_{i_c}  \alpha_{i_c}| \Psi_A^{(i_c)} \rangle   | \Psi_B^{(i_c)} \rangle
\ee
where $i_c$ denotes a particular choice of edge labels on the boundary between $A$ and $B$, for which the total number of occupied edges crossing the boundary is even.  $ | \Psi_A^{(i_c)} \rangle $ and $ | \Psi_B^{(i_c)} \rangle $ are themselves superpositions over many different loop configurations, with the configuration $i_c$ of spins on the edge.  $|\Psi_0 \rangle$ factorizes according to Eq. (\ref{FactorEq}) because the relative coefficient of all elements in these superposition is always $1$.

Because configurations in which the edge labels are different are orthogonal, it is easy to compute the reduced density matrix:
\begin{eqnarray*}
\rho_A &=& \text{Tr}_B  \sum_{i_c} |\alpha_{i_c}|^2  | \Psi_A^{(i_c)} \rangle  | \Psi_B^{(i_c)} \rangle \langle   \Psi_A^{(i_c)} |  \langle   \Psi_B^{(i_c)} |  \\ &=&
 \sum_{i_c}  N_B(i_c)  |\alpha_{i_c}|^2 | \Psi_A^{(i_c)} \rangle \langle   \Psi_A^{(i_c)} |
 \end{eqnarray*}
where $N_B(i_c)$ is the number of configurations in $B$ with these boundary conditions.
Since $\langle   \Psi_A^{(i_c')} | \Psi_A^{(i_c)} \rangle \propto \delta_{c c'}$, $\rho_A$ is diagonal, and we may read off the entanglement entropy:
\be
S_{AB} = \sum_{i_c} N_B(i_c) N_A(i_c)  |\alpha_{i_c}|^2 \log \left [ N_B(i_c) N_A(i_c) |\alpha_{i_c}|^2 \right ]
\ee
subject to
\be
\sum_{i_c} N_B(i_c) N_A(i_c)  |\alpha_{i_c}|^2 = 1
\ee

In practise $N_A(i_c), N_B(i_c)$ will be independent of the particular boundary configuration, as will $|\alpha_{i_c}|^2$, so that $ N_A(i_c)N_B (i_c) |\alpha_{i_c}|^2 = 1/ \mac{N}_i$, where $\mac{N}_i$ is the number of possible boundary conditions.   In a loop gas, if the total number of edges on the boundary between $A$ and $B$ is $n_i$, then
\be
 \mac{N}_i = 2^{n_i -N^{(0)}_A}
 \ee
 where $N^{(0)}_A$ is the number of connected components of the boundary of region $A$.  The entagnlement entropy is thus :
\be \label{Sent}
S =   \sum_{i=1}^{\mac{N}_i } \frac{1}{\mac{N}_i} \log \mac{N}_i  = (n_i -N^{(0)}_A) \log 2
\ee

The topological entanglement entropy is given by the combination of subdivisions shown in Fig. \ref{EntroFig}.  In the first three, $A$ consists of a single component and $N_A^{(0)} =1$; in the last term $N_A^{(0)} =2$.  Since the regions are chosen such that $n_{i}^{(1)} - n_{i}^{(2)}- n_{i}^{(3)} + n_{i}^{(4)} =0$, we obtain:
\begin{eqnarray*}
S_{\text{Top} } &=& - \left [  (n_{i}^{(1)} - 1) -  (n_{i}^{(2)} - 1)   \right. \\ &-& \left. (n_{i}^{(3)} - 1) +  (n_{i}^{(4)} - 2)  \right] \log 2  = \log 2
\end{eqnarray*}

For the doubled semion model, if $A$ is simply connected then we may use exactly the same reasoning as for the Toric code to calculate the entanglement entropy (whether or not $A$ has multiple boundary components).
That is, in this case the ground state wave function can be decomposed according to Eq. (\ref{FactorEq}), with $| \Psi_A^{(i_c)} \rangle$ a superposition of all loop configurations in $A$ with the boundary spin configuration $i_c$, and similarly for $ | \Psi_B^{(i_c)} \rangle $.   Explicitly, when $A$ is simply connected every loop configuration in $A$ with a fixed boundary $i_c$ can be obtained from every other such configuration by acting with some number of plaquette projectors on plaquettes that are entirely inside $A$.  Unlike in the Toric code, there are non-trivial relative phases between these different loop configurations; however, these phases are entirely dictated by the action of the plaquette projectors within $A$, and are independent of the loop configuration in $B$, such that the wave function still factorizes.
The rest of the calculation is identical to that for the Toric code; 
hence if $A$ is simply connected, the entropy is again given by Eq. (\ref{Sent}).

\begin{figure}
 \includegraphics[width=\linewidth]{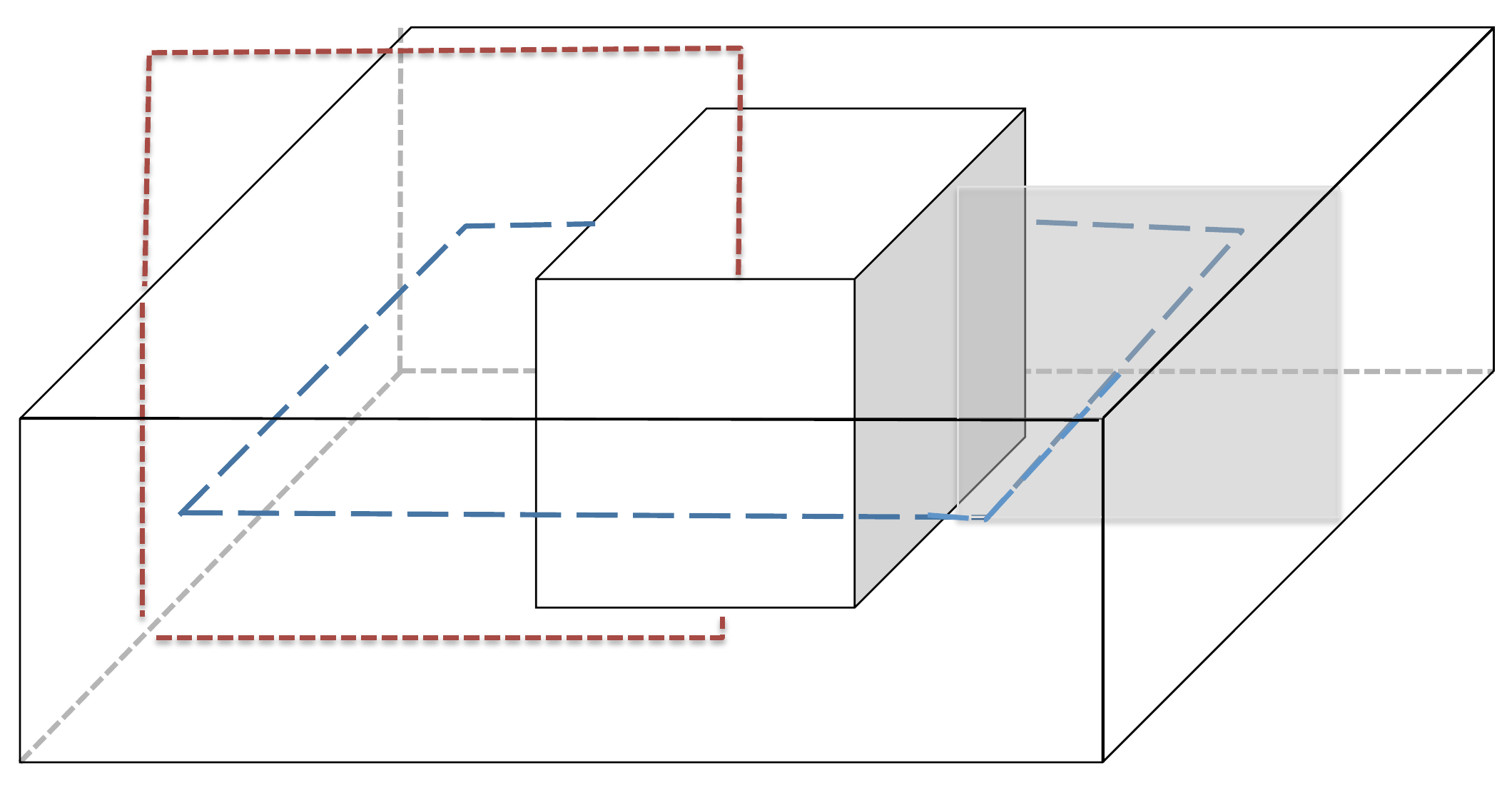}
 \caption{The scenario if $A$ is not simply connected.  If the total number of down spins crossing the cut through the donut is odd, then there is a loop that is not contractible within region $A$.  Similarly, if the number of down spins through the center of the donut is odd, then there is a loop that is not contractible within region $B$.  When both of these loops are present, the wave-function has an extra phase $-1$ due to their linking.  }
\label{TorConf1}
 \end{figure}

If $A$ is not simply connected,  in general the factorization (\ref{FactorEq}) fails because the doubled semion ground state contains phases that are sensitive to the linking number of loops in $A$ with loops in $B$.  The case of interest is where $A$ is topologically a donut.  Let us separate the wave function $|\Psi_A^{(i_c)} \rangle$ according to: 
\be
|\Psi_A^{(i_c)} \rangle = \frac{1}{\sqrt{2} } \left[  |\Psi_A^{(i_c)} \rangle_{\text{o} } +  |\Psi_A^{(i_c)} \rangle_{\text{e} }  \right ]
\ee
Here e and o refer to whether the total number of down spins measured along a slice through the donut is even or odd (Fig. \ref{TorConf1}).      $|\Psi_A^{(i_c)} \rangle_{\text{e} }$ (and similarly $|\Psi_A^{(i_c)} \rangle_{\text{o} }$) is a superposition of many different loop configurations with different relative phases, which are related by the action of some number of plaquette projectors acting entirely inside $A$.   To get from a configuration in the even set to one in the odd set, however, one must act with plaquette projectors in region $B$ as well.  Thus $|\Psi_A^{(i_c)} \rangle_{\text{o} } $ and $|\Psi_A^{(i_c)} \rangle_{\text{e} }$ need not appear in the ground state with the same phases, for a given loop configuration in $B$.

To see the implications of this, let us separate the loop configurations in region $B$ according to whether the number of down spins through the hole in the middle of the donut is even or odd (i.e., according to whether or not there is a loop in $B$ that is not contractible within $B$):
\be
|\Psi_B^{(i_c)} \rangle =  \frac{1}{\sqrt{2} } \left[  |\Psi_B^{(i_c)} \rangle_{\text{o} } +  |\Psi_B^{(i_c)} \rangle_{\text{e} }  \right ]
\ee
The ground state of 3DSem is given by
\begin{eqnarray} \nonumber
|\Psi_0 \rangle &=& \sum_{i_c} \frac{\alpha_{i_c} }{2}  \left [ |\Psi_A^{(i_c)} \rangle_{\text{e} } |\Psi_B^{(i_c)} \rangle_{\text{e} } + |\Psi_A^{(i_c)} \rangle_{\text{e} } |\Psi_B^{(i_c)} \rangle_{\text{o} } \right. \\ &+& \left. |\Psi_A^{(i_c)} \rangle_{\text{o} } |\Psi_B^{(i_c)} \rangle_{\text{e} } - |\Psi_A^{(i_c)} \rangle_{\text{o} } |\Psi_B^{(i_c)} \rangle_{\text{o} }
\right ]  \label{NotFactorEq}
\end{eqnarray}
The extra $-$ sign in front of the last term (in addition to any relative phases present in the definitions of $|\Psi_A^{(i_c)} \rangle_{\text{e,o} }, |\Psi_B^{(i_c)} \rangle_{\text{e,o} }$) arises because when both odd components are taken, the non-contractible loops in $A$ and $B$ are linked, which produces an extra $-$ sign in the wave function.

Because of this extra sign, Eq. (\ref{NotFactorEq}) cannot be expressed in the form Eq. (\ref{FactorEq}); rather, we have:
\begin{eqnarray} \nonumber
|\Psi_0 \rangle &=& \sum_{i_c} \frac{\alpha_{i_c}}{2} \left \{  \left ( |\Psi_A^{(i_c)} \rangle_{\text{e} }  + |\Psi_A^{(i_c)} \rangle_{\text{o} }  \right ) |\Psi_B^{(i_c)} \rangle_{\text{e} } \right.\\ &+& \left. \left( |\Psi_A^{(i_c)} \rangle_{\text{e} }  - |\Psi_A^{(i_c)} \rangle_{\text{o} } \right )  |\Psi_B^{(i_c)} \rangle_{\text{o} } \right \} \label{NotFactorEq2}
\end{eqnarray}
so that
\ba  
\rho_A& =& \sum_{i_c} \frac{ |\alpha_{i_c}|^2}{4}   N_{B}(i_c) \\ & & \left \{   \left ( |\Psi_A^{(i_c)} \rangle_{\text{e} } + |\Psi_A^{(i_c)} \rangle_{\text{o} }  \right )  \left ( \langle \Psi_A^{(i_c)} |_{\text{e} } + \langle \Psi_A^{(i_c)} |_{\text{o} }  \right )  \right .  \n
&& \left . +  \left( |\Psi_A^{(i_c)} \rangle_{\text{e} }  - |\Psi_A^{(i_c)} \rangle_{\text{o} } \right )  \left ( \langle \Psi_A^{(i_c)} |_{\text{e} } - \langle \Psi_A^{(i_c)} |_{\text{o} }  \right )
\right \} \nonumber
\ea
Here we have used the fact that the number of loop configurations in the even and odd sectors of $|\Psi_B^{(i_c)} \rangle$ are equal.  As this also holds for $|\Psi_A^{(i_c)} \rangle$, we have:
\be
 \left ( \langle \Psi_A^{(i_c)} |_{\text{e} } + \langle \Psi_A^{(i_c)} |_{\text{o} }  \right )   \left( |\Psi_A^{(i_c)} \rangle_{\text{e} }  - |\Psi_A^{(i_c)} \rangle_{\text{o} } \right )
 =0
 \ee
 so that
 \be
 \text{Tr} \rho_A = \sum_{i_c} | \alpha_{i_c}|^2 N_A(i_c) N_B(i_c)  = 1
 \ee
Since the coefficients $| \alpha_{i_c}|^2 N_A(i_c) N_B(i_c)$ are independent of the choice of boundary spin configuration, this fixes
\be
 \alpha_{i_c}|^2 N_A(i_c) N_B(i_c) = \frac{1}{ \mac{N}_i} = 2^{ 1 - n_i}
 \ee
 where $n_i$ is the total number of edges crossing the boundary between $A$ and $B$.  This gives the entanglement entropy:
 \begin{eqnarray*}
 S_{AB}  &=&  -   \sum_{i_c} |\alpha_{i_c}|^2 N_A(i_c) N_{B}(i_c)  \log \left[  \frac{1}{2} |\alpha_{i_c}|^2 N_A(i_c) N_{B}(i_c)  \right ]
 \\  &=& \mac{N}_{i}  \frac{1}{\mac{N}_i} \log 2 \mac{N}_i =  n_i  \log 2
 \end{eqnarray*}
for the solid donut.

Since the other three regions in Fig.  \ref{EntroFig} are simply connected, their entanglement entropy is identical to that of the toric code.  The topological entanglement entropy of the $3$D semion model is thus:
\begin{eqnarray*}
S_{\text{Top} } &=& - \left [  (n_{i}^{(1)} ) -  (n_{i}^{(2)} - 1)  - (n_{i}^{(3)} - 1) +  (n_{i}^{(4)} - 2)  \right ] \log 2 \\ &=& 0
\end{eqnarray*}
exactly as one might have expected for a system that has no bulk topological order.

\section{Surface plaquette operators of the 3D semion model with rough boundary} \label{SurfaceOps}

The choices of boundary conditions in 3D are similar to those in 2D, which have been discussed at length in Refs.~\onlinecite{Bravyi98,Kitaev11}.  In the main text we always assumed a ``smooth" boundary.  Here we consider a ``rough" boundary where  ``dangling" edges are sticking out of the surface.   Including these these  dangling edges on the boundary allows a source of a colored loops at the boundary with no energy cost.

For the toric code\cite{Bravyi98,Kitaev11} one can create ``partial-plaquette" operators which flip the value of two neighboring dangling edges as well as the bulk edge connecting them.  These partial-plaquette operators commute with the Hamiltonian and have the effect of allowing the sources to move freely along the surface.

The situation with the 3DSem model is a bit different.   We will similarly be able to construct such partial-plaquette operators which commute with the bulk Hamiltonian, however, they will not commute with each other.  This is to be expected since the sources should have semionic statistics with respect to each other.

\begin{figure}
 \includegraphics[width=1\linewidth]{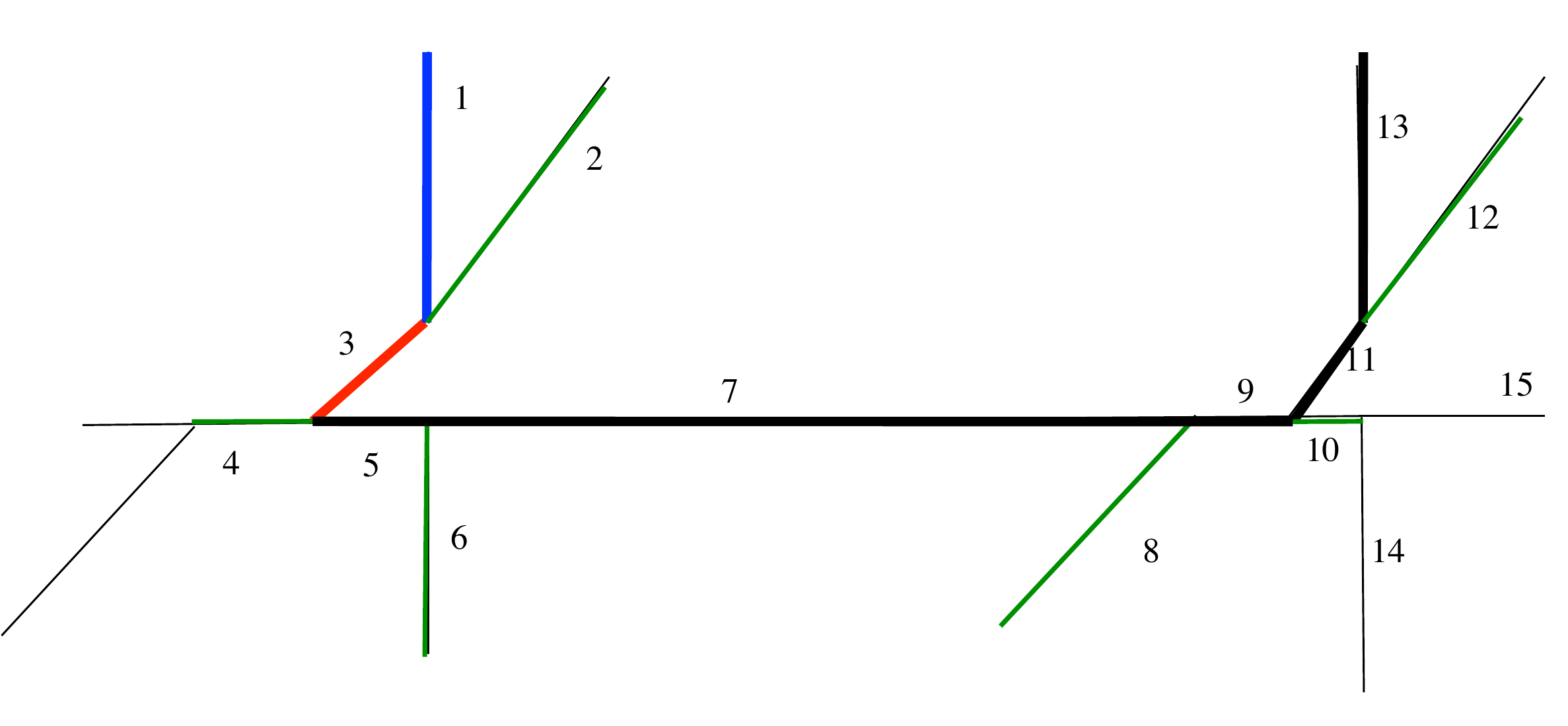}
 \caption{(Color online) A surface plaquette.  The surface plaquette operator described in the text flips the thickened edges $(1,3,5,7,9,$ and 13).  It has a phase that depends on $n_i$ on he edges bordering those that are flipped ($2,4,6,8,10,$ and $12$, shown here in green), as well as an extra phase depending on the occupancy of link $3$ (red) and link $1$ (blue).  In this figure the dangling edges are 1 and 13.}
 \label{PartPlaq}
\end{figure}

The form of such a partial plaquette is show in \figref{PartPlaq}.  The plaquette operator acts by flipping a pair of neighboring  dangling edges (labeled $1$ and $13$ in the figure), together with the bulk edges connecting them ($3,5,7,9,$ and $11$).  In order that this operator commute with the bulk plaquettes that also act on these bulk edges, we must include the usual phases depending on the edges bordering those that are flipped ($2,4,6,8,10,$ and $12$ in the figure), together with an extra phase depending on the value of some of the flipped edges before flipping.  (In the bulk plaquette operator \eqref{3DSemplaq}, these are the red and blue edges.  For the surface plaquette drawn here, only one red edge ($3$) and one blue edge ($1$) are included).  This gives
\be
B_{p 0}^{\text{surf} } = \sigma^x_1 \sigma^x_3  \sigma^x_5 \sigma^x_7  \sigma^x_9 \sigma^x_{13} i^{ n_2 + n_4 + n_6 + n_8 + n_{10} } i^{n_3 -n_1 }
\ee
where
\be
n_i = \frac{1}{2} \left( 1 - \sigma^z_i \right)
\ee

However, there is a problem here: $B_{p0}$ is not Hermitian if $n_1 - n_{13}$ is odd.  To compensate for this, we must add an extra phase factor $(-1)^{n_1 n_{13} }$.  The final form of the surface plaquette operator shown is:
\be
B_{p 0}^{\text{surf} } = \sigma^x_1 \sigma^x_3  \sigma^x_5 \sigma^x_7  \sigma^x_9 \sigma^x_{13} i^{ n_2 + n_4 + n_6 + n_8 + n_{10} } i^{n_3 -n_1 }(-1)^{n_1 n_{13} }
\ee
More generally, we have
\be
B_{p 0}^{\text{surf} } = ( \prod_{i \in \partial p}  \sigma^x_i ) \prod_{j \in s(p) } i^{n_i} i^{ \sum_{j \text{ red} } n_j - \sum _{j \text{ blue} } n_j } (-1)^{\prod_{j \text{ dangling} n_j } }
\ee
with red and blue edges defined differently for the different plaquette orientations, as in the bulk.

In \secref{subsub:other} we claimed that we could add a kinetic term to the vertex defects on the surface of the manifold.  This is done by adding dangling edges and including these partial plaquette operators.

\section{Trace methods}\label{trace}
In this section we give a more standard derivation of the ground state degeneracies of the toric code and 3DSem models: We will prove that the 3D toric code has ground state degeneracy $2^{b_1}$ on a manifold without boundary (and $b_1$ independent non-contractible cycles), whist the 3D semion model has a single ground state. For example the 3D toric code and semion model on the 3-torus have ground state degeneracies $2^{3}$ and $1$ respectively. In \secref{sec:2DAlg} we introduce a standard method used to calculate ground state degeneracy and apply it to the 2D models. Then, in \secref{tcDSem3D}, we generalize the method to the 3DSem model.

\subsection{2D ground state degeneracies via trace method}\label{sec:2DAlg}
In this section we algebraically deduce the ground state degeneracy of the toric-code and DSem model embedded on some 2D surface without boundary; we have already discussed the ground state degeneracy using a loop gas picture in \secref{s:2DLM}, but we opt here for a more concrete method. This section should be viewed as preparation for the work in 3D in the next section, where we use a similar method. We define a projector onto the ground state

\be\label{eq:proj_2D}
\mathcal{P} =\prod_{v\in V}\mathcal{P}_{v}\prod_{p\in P}\mathcal{P}_{p}\punc{,}
\ee

where $V$ is the set of all vertices, and $P$ is the set of all
plaquettes. In the toric code $\proj_p=(1+B_p)/2$ where $B_p$ is defined in \eqnref{HTC}. In the DSem model $\proj_p=(1-B_p)/2$ where $B_p$ is defined in \eqnref{DSemplaq}. In both models $\proj_v=(1+B_v)$ is the same. To calculate the ground state degeneracy $\tr \db{\proj}$ we write the trace as

\begin{align}\label{eq:tr_2D_proj1}
\text{Tr}\left(\mathcal{P}\right) & =2^{-V-F}\tr\left[\prod_{v\in V}\left(1+\mathcal{K}_{v}\right)\prod_{p\in P}\left(1+\mathcal{K}_{p}\right)\right]\\
&=2^{-V-F} \sum_{S_{V}\subseteq V} \sum_{S_{P}\subseteq P} \tr \left[ \prod_{v\in V}\mathcal{K}_{v}\prod_{p\in P}\mathcal{K}_{p}   \right]
\end{align}

where $\mathcal{K}_{v}=2\mathcal{P}_{v}-1$ and $\mathcal{K}_{p}=2\mathcal{P}_{p}-1$. Here $S_V$ and $S_p$ are subsets of the set of all vertices and plaquettes respectively. For the toric code and DSem model, all of the terms in the sum are zero except for the four corresponding to $\left(S_{V},S_{P}\right) \in \left\{  \left( \emptyset , \emptyset \right),\left(V,\emptyset  \right),\left( \emptyset ,P\right), \left(V,P\right) \right\}$. For each of these $2^2=4$ choices,
$\prod_{v\in S_{V}}\mathcal{K}_{v}\prod_{p\in S_{P}}\mathcal{K}_{p}=\mathbf{1}$,
and so

\begin{align}
\text{Tr}\left(\mathcal{P}\right) & =2^{2-V-F}\text{Tr}\left[\mathbf{1}\right]\nonumber \\
 & =2^{2-\chi}\label{eq:tr_2D_proj2}\end{align}

where $\chi=V-E+F$ is the Euler characteristic of the surface tiled
by the lattice. Therefore the ground state degeneracy of the 2D toric code, and DSem is $2^{2-\chi}=2^g$ on a surface of genus $g$.

To see how the above argument works in detail for the toric code note that $\mathcal{K}_{p}=B_p$
acts on each edge of plaquette $p$ with a $\sigma^{x}$ i.e. it flips
the spin on each edge of $p$. Therefore the product
$\prod_{p\in S_{P}}\mathcal{K}_{p}$ acts in the following way. Edges
that do not belong to a plaquette in $S_{P}$ are unchanged, and those
that belong to two plaquettes are also left unchanged because they
are flipped twice. However, those edges belonging to just one plaquette
in $S_{P}$ are flipped once by $\prod_{p\in S_{P}}\mathcal{K}_{p}$
; we will call these boundary edges. If the operator has boundary
edges then for any state $\mid\psi\rangle$ in our $\sigma^{z}=\pm1$ spin
basis $\langle\psi\mid\prod_{v\in S_{V}}\mathcal{K}_{v}\prod_{p\in S_{P}}\mathcal{K}_{p}\mid\psi\rangle=0$,
because $\prod_{p\in S_{P}}\mathcal{K}_{p}$ will flip at least one
edge in $\mid\psi\rangle$ to an orthogonal state, and
$\prod_{v\in S_{V}}\mathcal{K}_{v}$ cannot undo this. Therefore we
need only worry terms for which $\prod_{p\in S_{P}}\mathcal{K}_{p}$
has no boundary edges, which occurs when $S_{P}=\emptyset$ or $P$. In both cases $\prod_{p\in S_{P}}\mathcal{K}_{p}=1$, therefore

\begin{align}
\text{Tr}\left(\mathcal{P}\right) & =2^{-V-F}\times2\times\sum_{S_{V}\subseteq V}\text{Tr}\left[\prod_{v\in S_{V}}\mathcal{K}_{v}\right]\label{eq:tr_2D_proj3}\end{align}

We use a similar line of reasoning to show that only two terms survive the remaining sum. Note that $\mathcal{K}_{v}$ is -1 if vertex
$v$ has an odd number of 1-strings entering it, and -1 otherwise. Therefore the product $\prod_{v\in S_{V}}\mathcal{K}_{v}$ acts in
the following way. Edges that do not belong to a vertex in $S_{V}$
are unchanged, and those that belong to two vertices in $S_{V}$ are also unchanged because $\left(\sigma^{x}\right)^{2}=1$. However, those boundary edges
belonging to just one vertex in $S_{V}$ are acted on with $\sigma^{x}$.
The fact that $\sigma^{x}$ is traceless implies that $\text{Tr}\left[\prod_{v\in S_{V}}\mathcal{K}_{v}\right]=0$
unless $\prod_{v\in S_{V}}\mathcal{K}_{v}$ has no boundary edges
i.e. unless $S_{V}=\emptyset$ or $V$. In both cases $\prod_{v\in S_{V}}\mathcal{K}_{V}=1$, therefore

\begin{align}
\text{Tr}\left(\mathcal{P}\right) & =2^{2-V-F}\text{Tr}\left[\mathbf{1}\right]\nonumber \\
 & =2^{2-\left(V-E+F\right)}\nonumber \\
 & =2^{2-\chi}\label{eq:tr_2D_proj4}\end{align}

where $\chi$ is the Euler characteristic of the surface that the
lattice tiles, and so the ground state degeneracy of the toric code
is is $2^{2g}$ where the integer $g$ is the genus of the surface. Indeed the
answer is the same in the case of DSem because the $\mathcal{K}_{v}$
are the same and the $\mathcal{K}_{p}=-B_{p}$ still flip edges; the only
non-trivial fact to check is that $\prod_{p\in P}\mathcal{K}_{p}=1$. To
act with $\prod_{p\in P}\mathcal{K}_{p}=\left(-1\right)^{F}\prod_{p\in P}B_{p}$
we can first fuse the $B_{p}$ operators on each edge (which gives
a factor of $-1$ for each edge). All that remain are
the closed $1$-loops on each vertex, which give a factor of $-1$
for each vertex. In total, $\prod_{p\in P}\mathcal{K}_{p}=\left(-1\right)^{V-E+F}=\left(-1\right)^{2g}=1$. Hence we have proved the statement in \eqnref{eq:tr_2D_proj4}.

\subsection{3D ground state degeneracies via trace method}\label{tcDSem3D}
Having calculated the ground state degeneracies of the $2$D models, we now apply similar methods to the 3D toric code and 3DSem on manifolds without boundary. We will confirm the results presented in \secref{sss:TCT^3}: The 3D toric code has degeneracy $2^{b_1}$ where $b_1$ (the first Betti number) is the number of independent non-contractible cycles in the closed 3-manifold, while the semion model has a single ground state.

Following our work in 2D \secref{sec:2DAlg}, we cast the ground state degeneracy in the form

\be\label{eq:3DGSdeg}
\tr \left(\mathcal{P}\right)  =2^{-V-F} \sum_{S_{V}\subseteq V} \sum_{S_{P}\subseteq P}  \tr \left[ \prod_{v\in V}\mathcal{K}_{v}\prod_{p\in P}\mathcal{K}_{p}   \right]\punc{,}
\ee

where $\mathcal{K}_{v}=2\mathcal{P}_{v}-1$ and $\mathcal{K}_{p}=2\mathcal{P}_{p}-1$. We evaluate terms in the above sum, first for the toric code and then for the semion model.

\paragraph{3D toric code:}
 $\mathcal{K}_{p}=B_p$ flips the spin on each edge of plaquette $p$. Therefore the product $\prod_{p\in S_{P}}\mathcal{K}_{p}$ acts in the following way. Edges that do not belong to a plaquette in $S_{P}$ are unchanged, and those that belong to an even number of plaquettes are also left unchanged because they are flipped twice. However, those edges belonging to an odd number of plaquettes in $S_{P}$ are on net flipped once by the operator $\prod_{p\in S_{P}}\mathcal{K}_{p}$; we again call these boundary edges. If the operator has boundary edges then for any state $\mid\psi\rangle$ in our $\sigma^{z}=\pm1$ spin basis

\be\label{eq:3Dtraceless}
 \langle\psi\mid\prod_{v\in S_{V}}\mathcal{K}_{v}\prod_{p\in S_{P}}\mathcal{K}_{p}\mid\psi\rangle=0\punc{,}
 \ee
because $\prod_{p\in S_{P}}\mathcal{K}_{p}$ will flip at least one edge in $\mid\psi\rangle$ to an orthogonal state, and $\prod_{v\in S_{V}}\mathcal{K}_{v}$ cannot undo this. Therefore \eqref{eq:3Dtraceless} is zero, except possibly when $S_P$ has no boundary edges. It can be shown (see below) that there are $2^{b_{1}+C-1}$ sets $S_{P}$ with no boundary edges, where $C$ is the number of cubes on the lattice. Furthermore $\prod_{S_{P}}\mathcal{K}_{p}=1$ for $S_{P}$ with no boundary edges. This leaves us with a ground state degeneracy of

\begin{align}
\tr\left(\mathcal{P}\right) & =2^{-V-F}\times2^{b_{1}+C-1}\times\sum_{S_{V}\subseteq V}\text{Tr}\left[\prod_{v\in S_{V}}\mathcal{K}_{v}\right]\nonumber \\
 & =2^{-V-F}\times2^{b_{1}+C-1}\times2^{1+E}\nonumber \\
 & =2^{b_{1}-\chi}\nonumber\\
 & =2^{b_{1}} \label{eq:tr_3D_proj_tc1}\end{align}

where $\chi=C-F+E-V$ is the Euler characteristic, which is zero for
closed 3-manifolds (Poincar\'{e} duality). To show the second equality
we used the fact that $\prod_{v\in S_{V}}\mathcal{K}_{v}$
is traceless unless $S_{V}=\emptyset$ or $V$ (see \secref{sec:2DAlg}). In both cases $\prod_{v\in S_{V}}\mathcal{K}_{v}=1$.

To justify the factor of $2^{b_{1}+C-1}$ above, note that a set \textcolor{black}{$S_{P}$
}without boundary can be visualized as a set of possibly intersecting
surfaces without boundary. Consider first those\textcolor{black}{{}
$S_{P}$ }with surfaces that can be contracted to a point. Such an $S_{P}$
can be said to form the boundary of one of two sets of cubes: $S_{C}$
or its complement $C-S_{C}$, where $C$ is the set of all cubes.
Therefore those \textcolor{black}{$S_{P}$} whose surfaces can be
contracted to a point can be put into 1:2 correspondence with the set
of subsets of $C$, and so there are $2^{C-1}$ such \textcolor{black}{$S_{P}$.
}But not all \textcolor{black}{$S_{P}$ }have surfaces that can be
contracted to a point. In fact there are $2^{b_{2}}$ topological classes
of surface where $b_{2}$ is the second Betti number
(which for oriented closed three manifolds obeys $b_2=b_{1}$). Each of these classes of surface can be
shown to contain $2^{C-1}$ elements in the same we showed above
for surfaces that could be contracted to a point.

\paragraph{3D semion:}
Having found that the ground state is degenerate for the toric code
on closed 3-manifolds, we now show that the ground state is non-degenerate for
the semion model on any oriented closed 3-manifold. We start again with \eqref{eq:3DGSdeg} recasting it as
\be
\tr\left(\mathcal{P}\right) =2^{-F}\times\sum_{S_{P}\subseteq P}\text{Tr}\left[\prod_{v\in V}\mathcal{P}_{v}\prod_{p\in S_{P}}\mathcal{K}_{p}\right]\punc{.}
\ee
The $\mathcal{K}_{p}=-B_p$ operators in the 3DSem flip each edge of the plaquette $p$ (albeit with possible phases), so we can use the same reasoning as for the toric code to show that $\prod_{p\in S_{P}}\mathcal{K}_{p}$ is traceless unless $S_{P}$ has no boundary edges. We found before that there were $b_{2}$ types of $S_{P}$ without boundary edges by remembering that any such {$S_{P}$ can be visualized as a set of possibly intersecting surfaces without boundary. This has not changed. However we will find that only one of these $b_{1}$ types of surface gives non-zero trace: $\prod_{p\in S_{P}}\mathcal{K}_{p}$ is traceless unless $S_{P}$ has no boundary edges and can be represented as a contractible surface.

To see how this works we recall the result shown in \eqnref{P_{xy}} that $P_{n(S_P)\perp} =\prod_{p\in S_{P}}\db{-B_p}$, where $P_{n(S_P)_\perp}$ is the parity $\pm 1$ of the number of loops crossing the non-contractible closed surface defined by $S_{P}$, which we say has normal $n(S_P)$. Therefore

\be
\tr \left[\prod_{v\in V}\mathcal{P}_{v}\prod_{p\in S_{P}}\mathcal{K}_{p}\right]=\tr\left[\prod_{v\in V}\mathcal{P}_{v} P_{n(S_P)_\perp} \right]
\ee

However, consider the invertible operator $\hat{S}=\prod_{i\in \mathcal{C}_n}\sigma^x_i$, where $\mathcal{C}_n$ is a closed path cutting $S_{P}$ perpendicularly. Notice that $\hat{S}$ commutes with $\proj_v$ but anti-commutes with the parity $P_{n(S_P)_\perp}$ so that

\begin{align}
\tr\left[\prod_{v\in V}\mathcal{P}_{v} P_{n(S_P)_\perp} \right] &= \tr\left[\hat{S}^{-1}\prod_{v\in V}\mathcal{P}_{v} P_{n(S_P)_\perp} \hat{S}\right] \nonumber \\
&=-\tr\left[\prod_{v\in V}\mathcal{P}_{v} P_{n(S_P)_\perp}\right]
\end{align}

Hence, the trace must be zero for $S_{P}$. If $S_{P}$ is contractible, then an odd number of fluxes pierce it only if it encloses vertex violations, but such states are projected out by $\prod_{v\in V}\mathcal{P}_{v}$. Therefore

\begin{align}
\tr\left(\mathcal{P}\right) & =2^{-F}\times\sum_{S_{P}\in\mathcal{E}}\tr\left[\prod_{v\in V}\mathcal{P}_{v}\prod_{p\in S_{P}}\mathcal{K}_{p}\right]\nonumber\\
 & =2^{-F}\times\left|\mathcal{E}\right|\times\text{Tr}\left[\prod_{v\in V}\mathcal{P}_{v}\right]\nonumber\\
 & =2^{-F}\times2^{C-1}\times2^{1-V}\text{Tr}\left[\mathbf{1}\right]\nonumber\\
 & =2^{-\chi}\nonumber\\
 & =1\end{align}

where $\chi$ is the Euler characteristic (which disappears for closed
3-manifolds), and $\mathcal{E}$ is the set of closed and contractible $S_p$'s which has size $2^{C-1}$ (as reasoned in the case of the toric code).


\begin{thebibliography}{10}
\bibitem{fusionmultiplicity}In fuller generality the vertices can also carry an quantum label when the underlying category has `fusion multiplicities'\cite{Bonderson07}. We do not consider this here. In addition, the strings in the categories can be oriented. This makes little difference to the qualitative behaviour of the models.
\bibitem{orientation subtlety} There is an extra complication that this category comes with oriented strings, but we can ignore this subtlety if we choose a convention for orienting the edges of our lattice.
\end{thebibliography}
\end{document}